\documentclass[fleqn,usenatbib]{mnras}

\usepackage{newtxtext,newtxmath}

\usepackage[T1]{fontenc}

\DeclareRobustCommand{\VAN}[3]{#2}
\let\VANthebibliography\thebibliography
\def\thebibliography{\DeclareRobustCommand{\VAN}[3]{##3}\VANthebibliography}

\usepackage{graphicx}	
\usepackage{amsmath}	

\newcommand{\ex}{e.g.~}

\newcommand{\hi}{\text{H{\sc i}}}
\newcommand{\NHI}{N_{\hi}}
\newcommand{\kpc}{\,{\rm kpc}}
\newcommand{\K}{\,{\rm K}}
\newcommand{\Rcirc}{R_{\rm circ}}
\newcommand{\rvir}{r_{\rm vir}}
\newcommand{\Tvir}{T_{\rm vir}}
\newcommand{\msun}{{\rm M}_\odot}

\newcommand{\eagle}{\textsc{EAGLE}~}
\newcommand{\fire}{\textsc{FIRE-2}~}
\newcommand{\tfive}{10^5\K}
\newcommand{\Rdisk}{R_{\rm d}}
\newcommand{\thetam}{\theta_{\rm tilt}}

\newcommand{\vc}{v_{\rm c}}
\newcommand{\vcn}{v_{\rm c, 200}}

\newcommand{\Mdot}{\dot{M}}
\newcommand{\Mdotn}{\Mdot_1}
\newcommand{\Lambdan}{\Lambda_{-22}}
\newcommand{\rn}{r_{20}}
\newcommand{\nH}{n_{\rm H}}
\newcommand{\tcool}{t_{\rm cool}}
\newcommand{\tff}{t_{\rm ff}}
\newcommand{\kmsmath}{\,{\rm km}\,{\rm s}^{-1}}
\newcommand{\Rcircmean}{\bar{R}_{\rm circ}}

\newcommand{\ttorque}{t_{\rm align}}
\renewcommand{\deg}{^\circ}
\newcommand{\fcgm}{f_{\rm cgm}}
\newcommand{\fb}{f_{\rm b}}

\newcommand{\Mvir}{M_{\rm vir}}

\newcommand{\red}[1]{\textcolor{black}{#1}}
\newcommand{\orange}[1]{\textcolor{black}{#1}}
\newcommand{\green}[1]{\textcolor{black}{#1}}

\title[Hot accretion onto extended and warped \hi\ discs]{Hot accretion onto spiral galaxies: the origin of extended and warped \hi\ discs}

\author[Sankar et al.]{Sriram Sankar,$^{1,2}$\thanks{E-mail: sriram.sankar@research.uwa.edu.au}
Jonathan Stern,$^{3}$\thanks{E-mail: sternjon@tauex.tau.ac.il}
Chris Power,$^{1,2}$
Barbara Catinella,$^{1,2}$
Drummond Fielding,$^{4}$\newauthor Claude-André Faucher-Giguère,$^{5}$
Imran Sultan,$^{5}$
Michael Boylan-Kolchin,$^{6,7}$ and
Joss Bland-Hawthorn$^{2,8}$ \\
$^{1}$International Centre for Radio Astronomy Research, The University of Western Australia, Crawley, WA 6009, Australia\\
$^{2}$ARC Centre of Excellence for All Sky Astrophysics in 3 Dimensions (ASTRO 3D), Australia\\
$^{3}$School of Physics and Astronomy, Tel Aviv University, Tel Aviv 69978, Israel\\
$^{4}$Department of Physics, New York University, 726 Broadway, New York, NY, 10003, USA\\
$^{5}$Center for Interdisciplinary Exploration and Research in Astrophysics (CIERA) and Department of Physics and Astronomy, Northwestern University,\\ 1800 Sherman Ave, Evanston, IL 60201, USA\\
$^{6}$Department of Astronomy, The University of Texas at Austin, Austin, TX 78712, USA\\
$^{7}$Cosmic Frontier Center, The University of Texas at Austin, Austin, TX 78712, USA\\
$^{8}$Sydney Institute for Astronomy, School of Physics, A28, The University of Sydney, NSW 2006, Australia\\
}

\date{Accepted 2026 June 2. Received 2026 June 2; in original form 2025 November 5}

\pubyear{2026}

\begin{document}
\label{firstpage}
\pagerange{\pageref{firstpage}--\pageref{lastpage}}
\maketitle

\begin{abstract}
Gas accretion, hot ($\sim 10^6\,{\rm K}$) atmospheres, and a tilt between the rotation axes of the disc and the atmosphere are all common predictions of standard galaxy evolution theory for massive star-forming galaxies at low redshift. Using idealised hydrodynamic simulations, we demonstrate that the central regions of hot galaxy atmospheres continuously condense into cool ($\sim10^4\,{\rm K}$) discs, while being replenished by an inflow from larger scales. The size and orientation of the condensed disc are determined by the angular momentum of the atmosphere, \red{so the condensed disc is expected to often be tilted and more extended than the stellar disc}. \orange{Continuous accretion from hot atmospheres can thus explain the ubiquity of extended and warped \hi\ discs around local spirals, and also potentially provide the necessary fuel for star formation. This hot accretion scenario predicts the absence of significant \hi\ from galaxy halos, consistent with recent  $21\,{\rm cm}$ constraints on nearby spirals (the so-called `\hi\ desert'). Moreover, our analysis indicates that observations of \hi\ warps} can be used to constrain the angular momentum, accretion rate, and gas metallicity of hot galaxy atmospheres, important parameters for disc galaxy evolution that are hard to determine by other means.
\end{abstract}



\section{Introduction}

The neutral hydrogen (\hi) distribution in nearby spiral galaxies typically extends well beyond the stellar disc, providing valuable information about the dark matter halo at radii where stellar emission is hard to detect. Observations by \cite{Broeils97} found a mean \hi\ to optical radius ratio of $R_{\hi}/R_{25}\approx 1.7\pm0.5$ (defined at a surface density/brightness of $1\,\msun\,{\rm pc}^{-2}$ and $25$ ${\rm mag}\,{\rm arcsec}^{-2}$), while recent results from the WALLABY survey indicate that on average, a third of the \hi\ mass resides outside $R_{25}$ \citep{leeWALLABYPilotSurvey2025}. This extended \hi\ typically occupies a tilted plane relative to the inner \hi\ and stellar discs, forming an antisymmetric `integral sign' \hi\ warp \citep[e.g.,][]{briggsRulesBehaviorGalactic1990,sancisiColdGasAccretion2008}. The onset of the warp 
often coincides with the truncation of the stellar disc 
\citep{vanderKruit01,vanderKruit07}. 

Observations of an \hi\ warp in the Milky Way date back to \cite{burkeSystematicDistortionOuter1957} and \cite{kerrMagellanicEffectGalaxy1957}, and surveys of external disc galaxies have since revealed that the warp phenomenon is prevalent. \cite{garcia-ruizNeutralHydrogenOptical2002} identified \hi\ warps in $20$ out of $26$ edge-on galaxies, with warps present in every galaxy where \hi\ extends beyond the stellar disc. The high occurrence of warps suggests that the mechanisms that produce warps are common and long-lasting, and that their formation is independent of external perturbations, which are not as ubiquitous.



Proposed models for warp formation broadly fall into two classes \citep[e.g.,][]{Binney92}: torquing of a pre-existing outer \hi\ disc, and accretion of gas with a misaligned angular momentum vector which circularises into a tilted outer disc. The former disc-torquing-based models require an external mechanism to generate the misalignment between the inner and outer discs, such as companions (\ex \citealt{gomezWarpsWavesStellar2017, willeWarpsInducedSatellites2024, dengPotentialDynamicalOrigin2024}) or tilted dark halos (\ex \citealt{jiangWARPSCosmicInfall1999, shenGalacticWarpsInduced2006, hanTiltedDarkHalo2023, hanTiltedDarkHalos2023}). Tilted dark halos appear to be prevalent in cosmological simulations, sustained either via misaligned accretion of dark matter \citep{jiangWARPSCosmicInfall1999, shenGalacticWarpsInduced2006} or by halo dynamics and merger histories
(\ex \citealt{bailinInternalAlignmentHalos2005, aumerIdealizedModelsGalactic2013, emamiMorphologicalTypesDM2021, shaoTwistedDarkMatter2021, hanTiltedDarkHalos2023, garcia-condeGalactoseismologyCosmologicalSimulations2024}), and thus potentially can explain the ubiquity of warps. \citet{sellwoodInternallyDrivenWarps2022} argued that once a misalignment between the inner and outer discs is established, it can generate internally driven outward-propagating bending waves that sustain the warps for several Gyr.  

In contrast with disc torquing-based models, accretion-based models propose that gas inflows deposit material onto disc outskirts with a misaligned angular momentum vector relative to that of the pre-existing disc \citep[e.g.,][]{lopez-corredoiraGenerationGalacticDisc2002, sancisiColdGasAccretion2008, roskarMisalignedAngularMomentum2010}. Such models can also explain the observed ubiquity and longevity of warps, since accretion is expected to be continuous 
\citep[e.g.,][]{lillyGasRegulationGalaxies2013}, \red{and since cosmological simulations of the circumgalactic medium (CGM) that sources this accretion often indicate a tilted angular momentum vector relative to the disc \citep{Zjupa17, DeFelippis20, Huscher21, Semenov24}. For instance, \citet{DeFelippis20} found a median misalignment angle of $15^\circ$ between the CGM and stellar disc in $z=0$ Milky Way mass galaxies in the \textsc{TNG100} simulations \citep{pillepichFirstResultsTNG502019}, for galaxies in the upper quartile of the angular momentum distribution, and a larger misalignment of $40^\circ$ for galaxies in the lower quartile. \cite{Huscher21} deduced a larger median misalignment angle of $56^\circ$ in the \eagle simulations  \citep{schayeEAGLEProjectSimulating2015}.
} 


\red{The accretion-origin scenario of \hi\ warps could be tested by identifying the accreting gas in the CGM. Such accretion is commonly assumed to proceed via cool ($\sim10^4\,{\rm K}$) clouds or filaments inflowing within a hot ($\sim10^6\,{\rm K}$) and static halo-filling medium, where this class of models includes both `cold flows' as in \cite{Dekel09} and `precipitation' as in \cite{donahueBaryonCyclesBiggest2022}. 
However, it is unclear whether such cool gas can actually reach the galaxy, or rather, is evaporated into the hot phase due to hydrodynamic instabilities \citep{Heitsch09,armillottaEfficiencyGasCooling2016,Mandelker20,Tan23,afruniCloudsAccretingIGM2023a,FGOh23}. Attempts to identify this accreting cool gas in observations have yielded mixed results. On one hand, quasar absorption-line studies have deduced large quantities of cool gas in the halos of star-forming galaxies at $z\sim0.3$ and above \citep[e.g.,][]{Werk13, Lan18, huangCompleteCensusCircumgalactic2021}.} In contrast, recent deep 21~cm observations of nearby spirals in the MHONGOOSE \citep{deblokMHONGOOSEMeerKATNearby2024} and FEASTS surveys \citep{wangFEASTSCombinedInterferometry2024} do not detect cool clouds down to a threshold of $\NHI\sim10^{17}\,{\rm cm}^{-2}$ in the inner CGM \citep[the `\hi\ desert',][]{healyPossibleOriginsAnomalous2024, deblokMHONGOOSEMeerKATNearby2024, Veronese25, Marasco25}. These observations may imply that cool gas is insufficient to sustain star formation in local discs, strengthening similar conclusions from earlier 21cm surveys \citep{sancisiColdGasAccretion2008, Sardone21, Kamphuis22}. A similar lack of cool gas in the inner CGM of the Milky-Way is suggested by the rarity of high velocity clouds (HVCs) beyond $\approx 10\kpc$ \citep{wakkerDistancesGalacticHighVelocity2007, Lehner22}, and by the tight upper limits on cool gas traced by \ion{C}{iv} absorption  ($N_{\rm CIV}<10^{13.4}\,{\rm cm}^{-2}$, \citealt{Bish21}). M31 also exhibits a low cool gas content in its inner CGM with a total Si absorption column of $N_{\rm Si}\lesssim 10^{14}\,{\rm cm}^{-2}$  \citep{Lehner20, lehnerProjectAMIGAInner2025}. Taken together, these observations challenge the cold flow and precipitation models for gas accretion in nearby spirals.

\red{An alternative possibility, that could resolve the challenge posed by observations of the \hi\ desert, is that accretion in the halo occurs \textit{via the hot CGM phase}, while cooling occurs only in the immediate vicinity of the galaxy. In this scenario, the material source of accretion is undetectable by observations of cool halo gas. Such hot inflows are known as `cooling flows' in the intracluster medium literature \citep[e.g.][]{fabianCoolingFlowsClusters1994}, and have been discussed in the context of the CGM by \cite{sternCoolingFlowSolutions2019,sternMaximumAccretionRate2020} and \cite{Sultan25}. 
Note that despite the name, cooling flows remain hot while accreting due to a balance between radiative cooling and compressive heating. Such a hot and rotating CGM inflow will eventually cool at the disc–halo interface, either spontaneously at the radiative cooling rate \citep{hafenHotmodeAccretionPhysics2022, sternAccretionDiskGalaxies2024}, or more rapidly if enhanced by mixing with fountain flows \citep[e.g.,][]{fraternaliGasAccretionCondensation2017,liFountaindrivenGasAccretion2023} or by direct turbulent mixing with disc gas \citep{Lin25,Sharma25}.}

\red{In this work, we demonstrate that hot rotating CGM inflows will create long-lived \hi\ warps when the CGM spin axis is tilted relative to that of the disc. This mechanism can explain the observed ubiquity and longevity of warps: hot CGM are expected to be common in local massive star-forming galaxies \citep{keresHowGalaxiesGet2005, Dekel06, armillottaEfficiencyGasCooling2016,sternMaximumAccretionRate2020,FGOh23}, evolve on cosmological timescales once formed \citep{Bertschinger89, pezzulliAccretionRadialFlows2016, pezzulliAngularMomentumCosmological2017}, and to often spin with a tilted axis relative to the disc as mentioned above. We build on the idealised hydrodynamic setup and results of \cite{sternAccretionDiskGalaxies2024}. They showed that hot inflows spin up and cool into a thin disc at the radius where the flow becomes rotation-supported. Here, we add a tilt between the CGM and disc rotation axes to explore how such inflows form a warp.} 

Our study is timely given recent observations with SKA-mid precursor telescopes such as MeerKAT and FAST, which are enabling detailed studies of extended \hi\ structures at unprecedented resolution (\ex \citealt{wangFEASTSCombinedInterferometry2024, healyPossibleOriginsAnomalous2024, kurapatiUncoveringExtraplanarGas2025, yangFEASTSCombinedInterferometry2025}; see also discussion in \citealt{trappFiguringOutGas2025c}). These observations highlight the power of \hi\ structures, including warps, in tracing physical processes that relate halo dynamics, gas accretion, and disc formation. 


This paper is organised as follows. In section \ref{sec:hotinflows}, we review previous results on the properties of hot rotating CGM inflows that are aligned with the disc, and provide analytic expectations for the misaligned case. In section \ref{sec:sim}, we introduce our numerical simulation setup of a misaligned hot and rotating CGM. We present the results from our simulations and establish the formation of warps under the paradigm of misaligned hot inflows in section \ref{sec:res}. Finally, we discuss the implications and predictions of our work in section \ref{sec:discussion}, and we summarise our key results in section \ref{sec:conclusion}.

\section{Misaligned hot inflows}
\label{sec:hotinflows}

\red{
In this section, we first summarise key aspects of hot rotating inflows (or `rotating cooling flows') derived by previous papers, 
starting with the non-rotating case and then adding rotation. 
We then establish our expectations for cases where the hot CGM rotation axis is tilted (or `misaligned') relative to that of the disc. 
}

\red{The fluid equations of a radiatively cooling, inviscid ideal gas in an external gravitational potential have the following solution, assuming spherical symmetry, steady-state, and a highly subsonic flow \citep[e.g.,][]{fabian84,sternCoolingFlowSolutions2019}:
\begin{eqnarray}
\label{e:rad solution}
  T &=&  2.0\cdot 10^6 \,\vcn^2 \,{\rm K}   \nonumber\\
  \nH &=& 3 \cdot 10^{-4} \, \rn^{-1.5} \vcn \Mdotn^{0.5} \Lambdan^{-0.5}\,{\rm cm}^{-3} \nonumber\\
  -v_r  =  \frac{r}{\tcool} &=& 19 \,\rn^{-0.5}\vcn^{-1}\Mdotn^{0.5} \Lambdan^{0.5}\kmsmath
    \end{eqnarray}
Here, $T,\nH,$ and $v_r$ are the gas temperature, hydrogen density and radial velocity, $\vc=\sqrt{GM(<r)/r}$ is the circular velocity where $M(<r)$ is the mass enclosed in radius $r$, $\Lambda$ is the cooling function, $\Mdot$ is the accretion rate,  $\tcool$ is the cooling time 
\begin{equation}\label{eq:tcool}
t_{\rm cool} = \frac{3}{2}\frac{P}{n_{\rm H}^2\Lambda} = 1.1\,\rn^{1.5}\vcn\Mdotn^{-0.5}\Lambdan^{-0.5}\,{\rm Gyr} ~,
\end{equation}
and $P\approx2.3\,\nH kT$ is the gas pressure. We normalize the variables by values typical of the inner CGM of the Milky-Way, namely $\rn=r/20\,{\rm kpc}$, $\vcn=\vc/200\kmsmath$, 
and $\Lambdan=\Lambda/10^{-22}\,{\rm erg} \,{\rm cm}^3\,{\rm s}^{-1}$, and we normalize also the accretion rate by $\Mdotn=\Mdot/1\,\msun\,{\rm yr}^{-1}$, which is roughly the rate required for sustaining Milky-Way star formation.} 

\red{The analytic solution in eqn.~(\ref{e:rad solution}) captures the key properties of hot inflows. The gas temperature is $\approx\Tvir$, and independent of radius due to a balance between radiative cooling and compressive heating in the inflow. This gas temperature also implies a sound speed $\approx\vc$ and thus the inflow velocity is highly subsonic, as long as $\tcool\gg\tff$ where $\tff$ is the free-fall time
\begin{equation}\label{eq:tff}
    \tff=\sqrt{2}\frac{r}{\vc}=140\,\rn\vcn^{-1}\,{\rm Myr} ~.
\end{equation}
Also, since the accretion timescale is $\approx\tcool$ 
linear thermal instabilities do not have time to develop 
\citep{balbus89,sternCoolingFlowSolutions2019}. 
In a recent paper \cite{Sultan25} showed that such cooling flow solutions provide a good approximation of the hot CGM in \fire cosmological zoom simulations without black hole feedback, for halo masses $\gtrsim10^{12}\,\msun$ at which the $\tcool>\tff$ condition is valid.}





When some initial rotation is taken into account in the hot CGM, the inflow is expected to spin up as it accretes. The hot inflow will then stop and cool at the radius where it becomes rotation supported \citep{Cowie80, sternMaximumAccretionRate2020,sternAccretionDiskGalaxies2024}. This radius is known as the `circularisation radius' $\Rcirc$, and its average value is comparable to the galaxy scale in standard disc formation theory \citep[e.g.,][]{Mo98}. Specifically, \citet{sternAccretionDiskGalaxies2024} showed 
that the hot inflow flattens into a thick hot disc geometry at radii where centrifugal forces become substantial, and then cools from $\sim10^6\K$ to $\sim10^4\K$ at the disc-halo interface when the gas becomes fully rotation supported. 
A main signature of this accretion mode is thus that cooling from $\sim10^6\,{\rm K}$ to $\sim 10^4\,{\rm K}$ in accreting gas occurs near the galaxy disc and is simultaneous with achieving rotation support ($v_\phi\rightarrow\vc$). This signature has been identified in gas accreting onto Milky-Way mass galaxies simulated in \fire \citep{hafenHotmodeAccretionPhysics2022,Sultan26}. 

\red{
How does the mean circularisation radius $\Rcircmean$, at which the gas is expected to cool in hot CGM inflows, compare to the galaxy size? The angular momentum of accreted material is expected to generally increase with cosmic time, yielding inside-out growth of galaxy discs \citep[\ex][]{bouwensInsideoutInfallFormation1997, chiappiniChemicalEvolutionGalaxy1997, pezzulliAccretionRadialFlows2016, pezzulliAngularMomentumCosmological2017}, an effect exacerbated if outflows preferentially remove low-angular momentum material from the disc. We thus expect the value of $\Rcircmean$, which is determined by the angular momentum of current accretion, to be larger than that of the pre-existing disc, which is set by the integrated angular momentum of past accretion. To estimate this relation quantitatively, we compare the angular momentum of hot CGM in simulations with that of observed discs. Using  the spin parameter definition from 
\cite{bullockUniversalAngularMomentum2001}
\begin{equation}\label{eq:spin}
    \lambda_{\rm hot} \equiv \frac{\langle j_{\rm hot}\rangle}{\sqrt{2}v_{\rm vir}\rvir}  ~,
\end{equation}
where $\langle j_{\rm hot}\rangle$ is the mean hot CGM specific angular momentum, 
$\rvir$, $v_{\rm vir}=\sqrt{GM_{\rm vir}/\rvir}$, and $M_{\rm vir}$ are respectively the halo virial radius, virial velocity, and virial mass, and using the following relation between circularisation radius and angular momentum
\begin{equation}\label{eq:Rcirc def}
    \langle j_{\rm hot}\rangle = \vc(\Rcircmean)\Rcircmean ~,
\end{equation}
we get a mean circularisation radius of} 
\begin{equation}\label{eq:Rcirc}
    \Rcircmean = \sqrt{2}\lambda_{\rm hot} f_{\vc}^{-1} \rvir = 0.08\, f_{\vc}^{-1}\left(\frac{\lambda_{\rm hot}}{0.06}\right) \rvir ~,
\end{equation}
where 
$f_{\vc} \equiv \vc(\Rcircmean)/v_{\rm vir} \gtrsim 1$. 
For the average $\lambda_{\rm hot} \approx 0.06-0.08$ found in cosmological simulations \citep[with $\approx0.25\,{\rm dex}$ halo-to-halo dispersion]{Teklu15, Danovich15, stewartHighAngularMomentum2017a,Oppenheimer18,DeFelippis20,Huscher21}, 
we thus get 
$\Rcircmean\approx 20\kpc$ for $\rvir=250\kpc$. 
For comparison, 
typical stellar disc sizes are
\begin{equation}\label{eq:Rd}
    R_{25}\approx4\Rdisk\approx\,0.05\rvir
\end{equation}
where we assumed a factor of four between $R_{25}$ and the disc scale length $\Rdisk$ \citep[e.g.,][]{kregelFlatteningTruncationStellar2002, vanderkruitGalaxyDisks2011,comeronBreaksThinThick2012, martin-navarroUnifiedPictureBreaks2012}, and the relation between  $\Rdisk$ and $\rvir$ from \cite{kravtsovSizeVirialRadiusRelation2013} derived from abundance matching.
Comparison of eqns.~(\ref{eq:Rcirc}) and (\ref{eq:Rd}) implies that $\Rcircmean$ is on average $\approx50\%$ larger than $4\Rdisk$, so most of the accreting gas \red{is expected to circularise and cool beyond the edge of the stellar disc, at a radius consistent with observed \hi\ disc sizes.}\footnote{In \cite{sternAccretionDiskGalaxies2024} we did not account for the high CGM spin relative to that of the dark matter halo as seen in cosmological simulations, so we concluded $\Rcircmean\approx4\Rdisk$ (see section 6.5 there), rather than $\Rcircmean\approx7\Rdisk$ as deduced here. The high dispersion in CGM spin between halos implies that both cases may be relevant, though the latter is likely more common.} A similar argument was made by \cite{Stewart13} and \cite{stewartHighAngularMomentum2017a} in the context of circularisation of cold flows accreting onto $z>1$ galaxies. 


\red{How would a hot rotating inflow cool if the CGM rotation axis is tilted relative to the stellar disc by an angle $\thetam$? If we can disregard disc torques on accreting gas, it will cool and form a disc in the midplane defined by the CGM rotation axis as in the aligned case, and will thus appear as a warp with angle $\thetam$. 
To estimate typical torques by the disc, we use eqn.~(9) from \cite{Binney92}:
\begin{equation}\label{eq:omegapre}
\omega_{\rm pre}(R) \approx 11\,\Omega\,(R/\Rdisk)^{-3}
\end{equation}
}
\red{
where $\omega_{\rm pre}$ is the precession frequency of a moderately inclined orbit at cylindrical radius $R$, the angular frequency of a circular orbit is $\Omega=\vc/R=\sqrt2/\tff$, and this relation is valid at $R>4\Rdisk$ in an axisymmetric potential of an exponential disc. 
The alignment time at the minimum radius of the inflow, which is $\Rcirc\approx7\Rdisk$ (see eqns.~\ref{eq:Rcirc} and \ref{eq:Rd}), is thus
}
\red{
\begin{equation}\label{eq:talign}
\ttorque\equiv\omega_{\rm pre}^{-1}\approx 22\,\tff\left(\frac{R}{7\Rdisk}\right)^3 ~.
\end{equation}
}
\red{
This timescale can be compared to the accretion time, which in a cooling flow is $\approx\tcool$, so
\begin{equation}\label{eq:ttorque to tflow}
    \frac{\ttorque}{t_{\rm accretion}} \approx 2.8\left(\frac{R}{7\Rdisk}\right)^3\left({\frac{\tcool/\tff}{8}}\right)^{-1}
\end{equation}
where we used the characteristic values in eqns.~(\ref{eq:tcool}) and (\ref{eq:tff}) to estimate ${\tcool/\tff}$. The alignment time is thus longer than the accretion time down to the final radius of the inflow, at least for the majority of gas which accretes at radii $7\Rdisk$ or more. We thus expect only mild torquing of accreting gas before it circularises.} In the next section, we test this conclusion using hydrodynamic simulations. 

To conclude this section, we also consider the effect of the dark matter halo on the accretion. Our assumption above of a constant gravitational potential implies the dark matter halo is growing slowly relative to cooling and dynamical times, as expected at low redshift \citep[e.g.,][]{Wechsler02,Diemer13, moTwophaseModelGalaxy2024}. The inner dark matter halo could also have a triaxial shape and principal axes that are tilted with respect to the disc and the outer halo \citep[e.g.,][]{bailinInternalAlignmentHalos2005, velliscigAlignmentShapeDark2015,shaoTwistedDarkMatter2021,emamiMorphologicalTypesDM2021,hanTiltedDarkHalos2023}. The torques on accreting gas by this effect were studied by \cite{trappAngularMomentumTransfer2024} using \fire simulations of $z\sim0$ Milky Way mass galaxies. They found that in some cases the dark matter torque can dominate, which suggests that the hot CGM may cool in a plane which is intermediate between its own and that of the inner halo. Such a plane would still be tilted with respect to that of the disc and thus appear as a warp. 

\section{Simulation Setup}
\label{sec:sim}

\begin{figure} 
\centering
\includegraphics[width=\columnwidth, trim=17.5cm 2cm 16cm 1.5cm,clip]{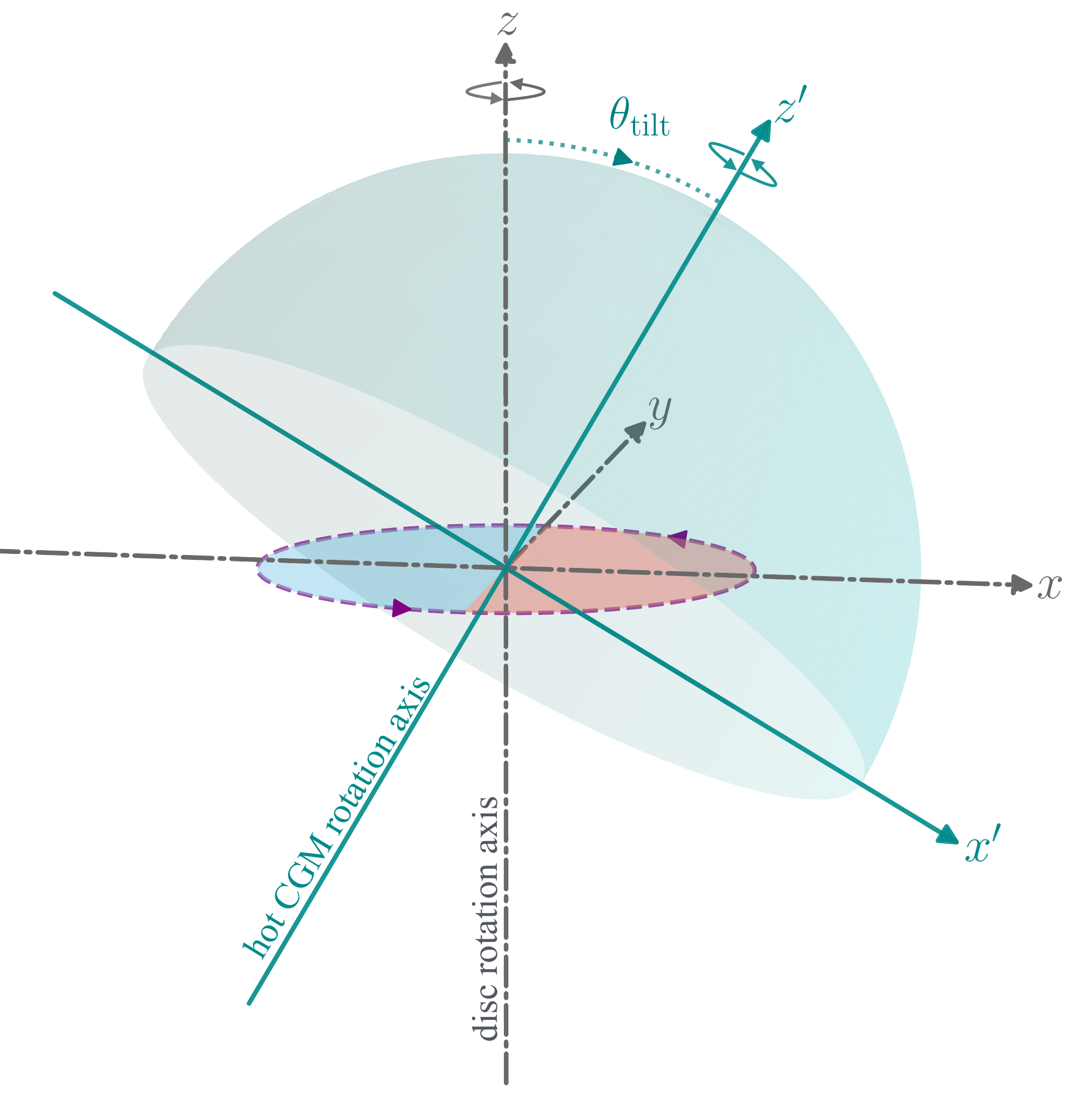}
  \caption{Illustration of the initial conditions in the simulations: a galactic disc surrounded by a hot rotating CGM with a tilted axis. The primed and non-primed coordinate systems respectively describe the CGM and disc orientations, with $z$ and $z^\prime$ along the rotation axes. The $y=y'$ axis is oriented along the line of nodes, and the tilt angle is marked by $\thetam$. }
  \label{fig:angle_convention}
\end{figure}

In this section, we describe the numerical simulation setup we use to study warp formation under the paradigm of misaligned hot inflows. Our simulation setup is similar to that described in \citet{sternAccretionDiskGalaxies2024}, which demonstrated how a thin disc forms out of a hot rotating CGM. We now assume the hot rotating CGM surrounds a pre-existing disc with a tilted rotating axis. 

\subsection{Code and physics}\label{sec:sim physics}

\newcommand{\vco}{v_{\rm c, 0}}

We use the meshless finite-mass (MFM) mode of the GIZMO code \citep{hopkinsNewClassAccurate2015}. MFM is a Lagrangian hydrodynamics solver that combines the advantages of traditional smooth particle hydrodynamics (SPH) and grid-based methods. This allows us to track each resolution element and study its evolution as it inflows from the hot CGM onto the warp. The simulations are initialised with a disc galaxy embedded in a rotating hot CGM as detailed below, where the axis of rotation of the CGM is misaligned with that of the disc by an angle $\thetam$ (see Figure~\ref{fig:angle_convention}). We refer to the disc and halo coordinate frames as ($x, y, z$) and ($x^\prime, y^\prime, z^\prime$), respectively, with $y=y^\prime$ oriented along the line of nodes. The corresponding spherical coordinates are ($r,\theta,\phi$) and ($r^\prime,\theta^\prime,\phi^\prime)$ with $r=r^\prime$. We run four idealised 3D hydrodynamic simulations with $\thetam=0^\circ, 15^\circ, 30^\circ$, and $60^\circ$, where $\thetam=0^\circ$ corresponds to no misalignment. Radiative cooling is calculated using the redshift $z=0$ optically thin cooling tables from \citet{wiersmaEffectPhotoionizationCooling2009} with a gas metallicity $Z=0.3\,{\rm Z}_\odot$ as estimated for the Milky-Way CGM \citep{Bregman18}. Optically thick radiative cooling to temperatures lower than $10^4\,{\rm K}$ is not included,  
since our study is focused on the geometry and thermodynamics of gas outside the ISM, which is largely optically thin at low redshift. 
We also do not include ongoing feedback processes, to focus on the regime where radiative cooling in the CGM dominates over feedback heating. Potential effects of ongoing feedback are addressed in the discussion. `Star formation' is implemented by converting all gas resolution elements with densities $n_{\rm SF} > 10~\mathrm{cm^{-3}}$ into stellar particles. Gravitational forces are calculated including both self-gravity between simulated gas and stars, and an additional spherical acceleration term $-(\vco^2/r)\hat{r}$ with $\vco = 200\kmsmath$ in order to approximate the gravitational field due to the inner dark matter halo. The effects of a non-spherical and/or tilted halo potential (see section~\ref{sec:hotinflows}) are left for future study.  
\red{We run the simulations for $8\,{\rm Gyr}$.}

\begin{figure*}   
    \includegraphics[width=\linewidth]{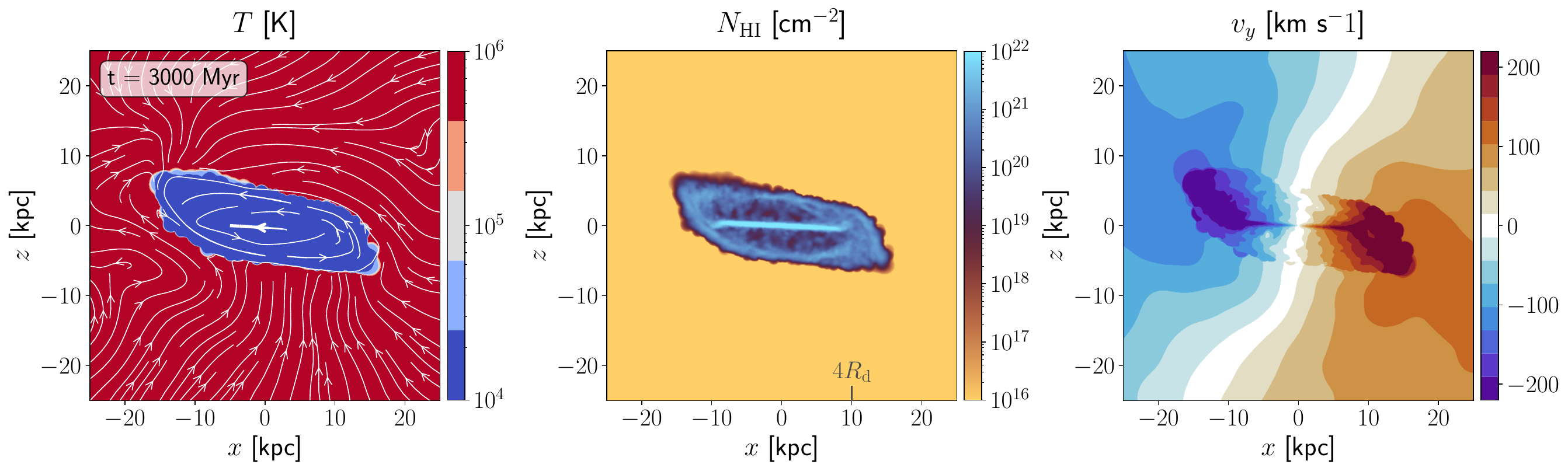}
    \vspace{-0.35cm}
  \caption{An \hi\ warp condensing out of a tilted hot CGM. Panels show edge-on projections of the $\thetam =30^\circ$ simulation at \red{$t = 3000$~Myr}. \hi\ column density is shown in the centre, while \hi-averaged temperature and velocity along the projected axis are shown in the left and right panels. White streamlines in the left panel trace the velocity field in the projected plane. The pre-existing disc is apparent in the $z=0$ plane with a maximum extent of $4\Rdisk=10\kpc$. \red{Newly accreted material is apparent as a tilted ring, oriented in the hot CGM midplane}. The streamlines demonstrate that the hot CGM is inflowing towards the disc and is the source of the accreted material. The cooling of the hot CGM from $\approx10^6\K$ to $\approx10^4\K$ at the warp boundary increases $\NHI$ by orders of magnitude.}
  \label{fig:xz_projection}
\end{figure*}

\begin{figure*}
    \includegraphics[width=\linewidth]{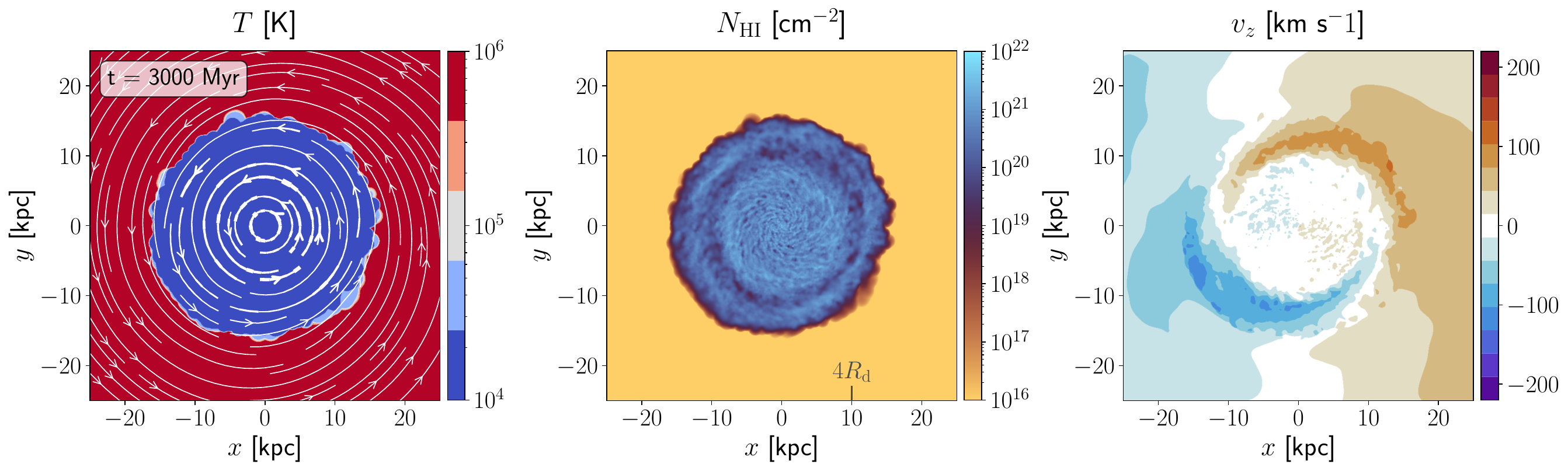}
    \vspace{-0.35cm}
  \caption{Similar to Fig.~\ref{fig:xz_projection}, for a face-on projection. White streamlines on the left demonstrate how the hot CGM spirals inward toward the disc, cooling abruptly from $\approx10^6\K$ to $\approx10^4\K$ at the warp boundary. The right panel shows that \red{the newly accreted gas outside $10\kpc$ is rotating in a tilted plane relative to the pre-existing inner disc.}}
  \label{fig:xy_projection}
\end{figure*}

\subsection{Initialising a galaxy}\label{sec:ics galaxy}

We initialise a galaxy disc using the \textsc{MakeDisk} code \citep{springelModellingFeedbackStars2005}, adopting parameters of a typical main-sequence Milky-Way mass galaxy \citep[e.g.][]{catinellaXGASSTotalCold2018}. The stellar disc is initialized with a mass of $M_{\star} = 4.5 \times 10^{10} \,\msun$, a radial scale length of $\Rdisk = 2.5\,{\rm kpc}$ and a vertical scale length of $0.1\Rdisk$. The stellar disc spans $0.03 - 4\Rdisk$ and connects to a bulge of mass $M_{\rm bulge} = 2 \times 10^8\,\msun$, modelled with a Hernquist profile and a scale length of $0.1\,{\rm kpc}$. The initial gaseous disc has a mass $M_{\rm gas} = f_{\rm gas}M_{\star}$ with $f_{\rm gas} = 0.1$, following the total gas fraction scaling relation from \citet{catinellaXGASSTotalCold2018}, and the same exponential distribution as the stellar disc. \red{We initialise only the region within $4R_{\rm d}$ of the gaseous disc, since one of the aims of our numerical experiment is studying how extended \hi~discs form via cooling from the hot CGM.} 

In the \textsc{MakeDisk} calculation, we include the isothermal gravitational field with circular velocity $\vco=200\kmsmath$ used in our simulations. \red{The resulting $\vc(r=4\Rdisk) \approx 220 \kmsmath$} is consistent with the baryonic Tully-Fisher relation from \citet{lelliBaryonicTullyFisher2019}, \red{where $\vc(r)$ is calculated by including both $\vco$ and the gravitational contribution from all simulated star and gas particles within radius $r$
\begin{equation}\label{eq:vc}
    \vc^2(r) = \vco^2 + \frac{GM(<r)}{r} ~.
\end{equation}}
The mass of the resolution elements is $m_{\rm b} = 8 \times 10^4\,\msun$, yielding a total of $6 \times 10^5$ and $6 \times 10^4$ elements for the initial stellar and gaseous discs, respectively. 

\subsection{Initialising a rotating, inflowing hot CGM}\label{sec:ICs CGM}

\newcommand{\rmax}{r_{\rm max}}

\red{
We initialise the density, temperature, and radial velocity of the CGM using a non-rotating spherical hot inflow solution. This initial condition is approximately equal to the solution in eqn.~(\ref{e:rad solution}) with $\Mdot = 0.5\,\msun\,\mathrm{yr}^{-1}$, $\vc = 200\kmsmath$ and $Z = 0.3\,{\rm Z}_\odot$. In practice, we also account for the dynamic term neglected in the solution in eqn.~(\ref{e:rad solution}), by integrating the 1D spherically symmetric and steady-state flow equations as described in \citet{sternCoolingFlowSolutions2019}, starting at the sonic radius of $r_\mathrm{sonic} = 0.05\,{\rm kpc}$ and proceeding outward. The dynamic term is significant only near $r_{\rm sonic}$, well below the radii of $\gtrsim10\kpc$ we focus on in this study, so eqn.~(\ref{e:rad solution}) provides a good approximation of our initial conditions.}  We then randomly assign initial locations of gas resolution elements in the GIZMO simulation in a way that ensures a radial mass distribution equal to that in the 1D solution. 

\red{
To add rotation to the initial conditions, we assume the hot CGM has an angular momentum profile $j_{\rm hot}(r,\theta^\prime)$, which scales with polar angle as $j_{\rm hot}\propto\sin^2\theta'$ and has only a weak dependence on radius at inner CGM radii. The scaling with polar angle is based on the results of non-radiative
cosmological simulations in \cite{sharmaOriginAngularMomentum2012}. The radial profile is based on the arguments in \cite{pezzulliAngularMomentumCosmological2017}. This profile assumes a truncated exponential angular momentum distribution (AMD) for halo baryons as seen in non-radiative cosmological simulations, with the lower part of the AMD removed, assumed `lost' to the pre-existing disc and to outflows. The exact calculation is detailed in Appendix~A.
The implied value at small radii is \orange{$j_{\rm hot}(r\ll\rvir)=4000\sin^2\theta^\prime \kpc \kmsmath$}, leading to typical circularisation radii of $\Rcirc=j_{\rm hot}/\vc\approx15\kpc$.
This $\Rcirc$ is larger than the initial disc size of $10\kpc$ described in section~\ref{sec:ics galaxy}, and is consistent with the arguments in section~\ref{sec:hotinflows}. Simulations with larger or smaller assumed values of $\Rcircmean$  correspondingly produce larger or smaller warps. }

\red{To avoid rotation velocities above $\vc$ in the initial conditions, at radii $r<4\Rdisk$ we assume $j_{\rm hot}=\vc r\sin^2\theta^\prime$. Gas at these radii cools out quickly ($\ll\,{\rm Gyr}$), so this choice does not affect the simulation beyond the initial transient phase.  }


\begin{figure*}
\includegraphics[width=\textwidth]{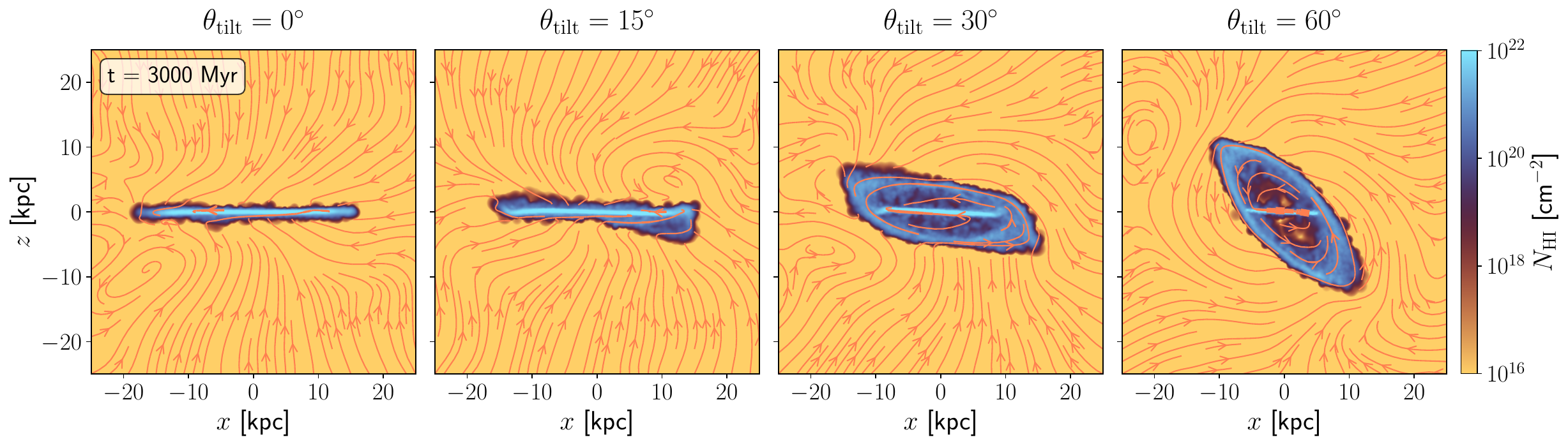}
\vspace{-0.35cm}
  \caption{Formation of \hi\, warps from rotating hot CGM with different tilt angles. Panels show edge-on projections of \hi\ column density in the \red{$t= 3000\,{\rm Myr}$} snapshots of the four simulations, with red streamlines tracing the velocity field in the projected plane. Hot CGM tilt angles are noted on top and are evident from the geometry of the streamlines. \red{The angle of the accreted material traces the hot CGM tilt angle,} creating a \hi\, warp or polar ring when $\thetam\neq0^\circ$. 
  }
  \label{fig:xz_projections_3sims}
\end{figure*}

CGM resolution elements at $r< 200\,{\rm kpc}$ are set with the same mass of $m_{\rm b} = 8 \times 10^4 \,\msun$ as the stars and the gas in the initial disc, implying a characteristic resolution size of $\approx (3m_{\rm b}X/(4\pi n_{\rm H}m_{\rm p}))^{1/3} \approx 0.8\,n_{-3}^{-1/3}\,{\rm kpc}$, where $m_{\rm p}$ is the proton mass, $X$ is the hydrogen mass fraction, and $n_{-3}\equiv n_{\rm H}/10^{-3}\,{\rm cm}^{-3}$. This density scale is typical for hot gas hydrogen densities near the disc \red{(see eqn.~\ref{e:rad solution})}. Characteristic resolution sizes of the $\sim100$ times denser cool gas are $\approx40\,{\rm pc}$. For comparison, the height of the $\approx 10^4\,{\rm K}$ gaseous disc that forms from the cooling of the hot gas is $\approx (10^4\,{\rm K} / T_{\rm hot})^{1/2} \Rcircmean \approx 1\,{\rm kpc}$, where $T_{\rm hot} \approx 2 \cdot 10^6$ K is the temperature of the hot gas \red{(eqn.~\ref{e:rad solution})}. For gas beyond $r=200$~kpc we use a lower mass resolution since it does not participate in the inflow due to its long cooling times. It nevertheless needs to be included in the simulation to confine the gas at smaller radii from expanding outward. The assumed mass resolution becomes coarser by a factor of three every factor of $\sqrt{2}$ in radius beyond $200\,{\rm kpc}$, out to $3.2\,{\rm Mpc}$ where the sound crossing time equals 10 Gyr. In total, the CGM is initialised with $5.5 \times 10^5$ resolution elements.

\red{We emphasise that the initial conditions are not in steady state, as the pressure structure of the integrated 1D hot inflow solution does not account for rotational support or for the gravitational potential of the disc. Cooling flows are expected to relax to a steady state on a cooling timescale \citep[e.g.,][]{Bertschinger89,sternCoolingFlowSolutions2019}, which is $\sim1\,{\rm Gyr}$ at $\Rcirc\approx15\,\kpc$ (see eqn.~\ref{eq:tcool}). Consistent with this expectation, the system reaches a quasi-steady state after $\approx2\,{\rm Gyr}$, which persists until the end of the simulation at $8\,{\rm Gyr}$. This is reflected both in the accretion rate, which stabilises at $0.5-1\,\msun\,{\rm yr}^{-1}$ (see Appendix~B), and in the emergence of a steady warp (see next section). } 

\section{Simulation Results}\label{sec:res}

\subsection{Formation of \hi\ warps from misaligned hot inflows}

Figure \ref{fig:xz_projection} shows edge-on maps of gas temperature ($T$; left panel), neutral hydrogen column density ($\NHI$; middle), and projected velocity ($v_y$; right) in the \red{$t=3000\,{\rm Myr}$} snapshot of the $\thetam=30\deg$ simulation. The projected plane is chosen to be perpendicular to the line of nodes (the $y=y^\prime$ axes), while the snapshot time is chosen to be after the inflow becomes quasi-steady. In the $T$ and $v_y$ panels, the values are \hi-mass weighted averages along the line of sight, and thus dominated by the cool gas in pixels which include the disc and warp. Outside these regions, cool gas is almost completely absent from our simulations, so the hot gas dominates the line-of-sight averages. \orange{We obtain the neutral hydrogen fraction in each resolution element based on its gas density and temperature as described in \citet{hopkinsFIRE2SimulationsPhysics2018}, by assuming equilibrium between recombination, collisional ionisation, and optically-thin photoionisation by the UV background \green{model} from \cite{faucher-giguereCosmicUVXray2020}. We do not distinguish between atomic and molecular hydrogen since the molecular fraction is expected to be small in the extended disc which we focus on in this work.} A cool ($T\approx10^4\,{\rm K}$) `integral sign' warp is apparent in the outer parts of the cool gas disc, where the warp is in the positive $z$ direction on the left and in the negative $z$ direction on the right. The value of $\NHI$ is about $1-2$ orders of magnitude lower in the warp than in the central disc, though still at detectable levels of $\NHI >10^{19}\,{\rm cm}^{-2}$. The warp is also evident in the distinction between the kinematic position angles of the inner and outer discs, as can be seen in the right panel, where the kinematic position angle is defined as the angle of the line that maximises the gradient of the projected velocity \red{field.} The panels show that the warp is roughly perpendicular to the symmetry axis of the velocity field of the hot gas, evident as the $v_y=0\kmsmath$ contour. This indicates that the warp forms in \red{a plane nearly} perpendicular to the hot CGM rotation axis, as further demonstrated below. 

Streamlines in the $v_x,v_z$ plane are plotted as white contours in the left panel of Fig.~\ref{fig:xz_projection}. These streamlines demonstrate that the hot CGM is inflowing towards the disc, as expected in hot gas where radiative cooling dominates over feedback heating. The panel shows that streamlines that are away from the symmetry axis tend to converge onto the warp, indicating that most of the accretion flows onto the warp. Note that although the gas in the cool warp originates from the CGM, the $\NHI$ in the CGM is negligible as the accreting gas is hot ($\gtrsim10^6\,{\rm K}$) and highly ionised prior to joining the warp.  

Figure~\ref{fig:xy_projection} shows face-on projections of the same snapshot shown in Fig.~\ref{fig:xz_projection}. The projected streamlines plotted in the left panel demonstrate that the hot gas is spiralling onto the cool disc, remaining hot until it joins the cool disc edge, as also found by \cite{sternAccretionDiskGalaxies2024} in simulations where the disc and hot CGM are aligned. The right panel of \red{Fig.~\ref{fig:xy_projection} demonstrates that the warp is evident as an asymmetric pattern in projected velocity ($v_z$). }

\begin{figure}
    \centering
    \includegraphics[width=\columnwidth]{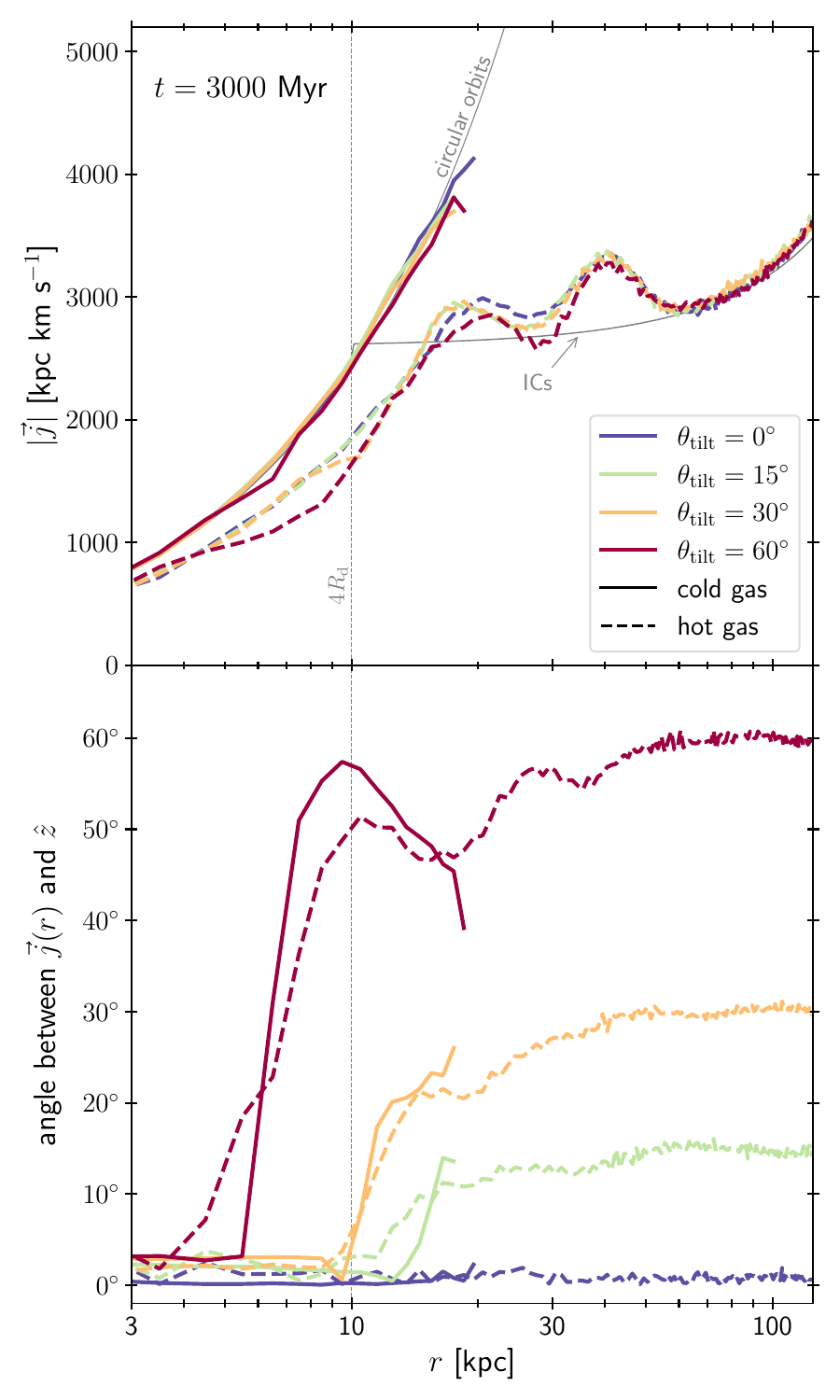} 
\vspace{-0.5cm}
\caption{Angular momentum profiles in hot CGM inflows. Panels show the magnitude (\textit{top}) and angle to the disc axis (\textit{bottom}) of the specific angular momentum vector in spherical shells,  in the \red{$t = 3000 \,{\rm Myr}$} snapshot. Solid lines show cool gas ($T<10^{5}\K$) while dashed lines show hot gas ($T> 10^{5}\K$). Colours denote the four simulations, which differ in CGM tilt angle, as evident in the hot gas angles at large radii. \red{Grey curves in the top panel denote the initial conditions and the angular momentum profile of circular orbits. The $10\kpc$ size of the pre-existing disc is also marked.} The \hi\ warp is evident as an increase in cool gas angle from $\approx 0^\circ$ at $r\lesssim10\kpc$ to \red{$\lesssim\thetam$} at $r\approx 20\kpc$. 
}
\label{fig:theta_j}
\end{figure}

Figure \ref{fig:xz_projections_3sims} shows edge-on $\NHI$ maps and projected streamlines in the \red{$t=3000\,{\rm Myr}$} snapshots of the four simulations. The simulations differ in the tilt angle between the disc and CGM rotation axes, as noted on top of the panels and evident from the directions of the projected streamlines. The tilted nature of accreting gas is also clearly apparent in the $\NHI$ maps of the three simulations with non-zero tilt angles, with the plane of warp roughly perpendicular to the rotation axis of the hot CGM. 
We note also that the accreting warp in the simulation with $\thetam=60^\circ$ may be observationally classified as a polar ring galaxy \citep{schechterNGC2685Spindle1978}, which are observed to have 
an incidence rate of 1-3\% in the local universe \citep{degWALLABYPilotSurvey2023, mosenkovOccurrenceRateGalaxies2024}. 

\red{Our simulations exhibit a non-monotonic $\NHI$ radial profile, where $\NHI$ drops beyond the edge of the pre-existing disc and rises towards the edge of the warp (middle panels of Figs.~\ref{fig:xz_projection}--\ref{fig:xy_projection} and right panels of Fig.~\ref{fig:xz_projections_3sims}). This may reflect the idealised nature of our setup, in which the ISM and CGM angular momentum distributions are initialised separately (sections~\ref{sec:ics galaxy}--\ref{sec:ICs CGM}), whereas in a fully cosmological context they would arise from a common underlying distribution. Moreover, the absence of ISM physics in our simulations likely suppresses radial flows within the disc that would otherwise smooth this distribution. }

The tendency of warps in our simulations to track the midplane defined by the hot CGM rotation axis is demonstrated in Figure~\ref{fig:theta_j}, where we plot specific angular momentum profiles versus spherical radius in the snapshots shown in Fig.~\ref{fig:xz_projections_3sims}. The top and bottom panels show the specific angular momentum magnitude and angle, respectively, while solid and dashed lines show cool ($T< 10^5\,{\rm K}$) and hot gas ($T> 10^5\,{\rm K}$), respectively. Profiles are measured using a mass-weighted average of all resolution elements in spherical shells with thickness $1\,{\rm kpc}$. Different simulations are noted with different colours, with the tilt angles noted in the legend. We also mark the maximum radius of the pre-existing disc ($4\Rdisk=10\,{\rm kpc}$),  the specific angular momentum of a circular orbit $\vc r$, and the specific angular momentum in the initial conditions. \red{The calculation of $\vc$ is done using eqn.~(\ref{eq:vc}) in the $t=3000\,{\rm Myr}$} snapshot shown in the plot, though the difference between different snapshots is minor. 


\red{The top panel of Fig.~\ref{fig:theta_j} shows that the specific angular momentum of cold gas follows that expected for rotation-supported gas (grey line) out to $\approx15\kpc$, including radii beyond the edge of the pre-existing disc at $10\kpc$. At larger radii of $15-20\kpc$, the cool gas rotates slightly slower than circular orbits. Beyond $\approx20\kpc$ there is no cool gas, indicating hot CGM cooling is offset by compressive heating beyond this radius in our simulations.} 

\red{Fig.~\ref{fig:theta_j} also demonstrates that the hot gas specific angular momentum at $r\gtrsim 70\kpc$ is the same as in the initial conditions, both in magnitude (top panel) and in angle (bottom panel). At $20\lesssim r\lesssim70\kpc$ the angular momentum deviates from the initial conditions, due to torques from the disc and due to the inflow that develops at these radii.
The change is however mild, with the rotation angle decreasing to $\gtrsim0.7\,\thetam$ at $r=20\kpc$, and the magnitude increasing by $\lesssim20\%$. This mild change in angular momentum is consistent with the weak torque of the hot inflow by the disc estimated in eqn.~(\ref{eq:ttorque to tflow}). At radii of $r\lesssim20\kpc$ where there is also cool gas, both the angular momentum magnitude and angle decrease significantly relative to the initial conditions. The rotation velocity of the hot gas is notably sub-Keplerian, $\approx0.5\vc$, 
likely since any hot gas that approaches full rotation support immediately cools (see below).}

\begin{figure*}
    \centering
    \includegraphics[width=\textwidth]{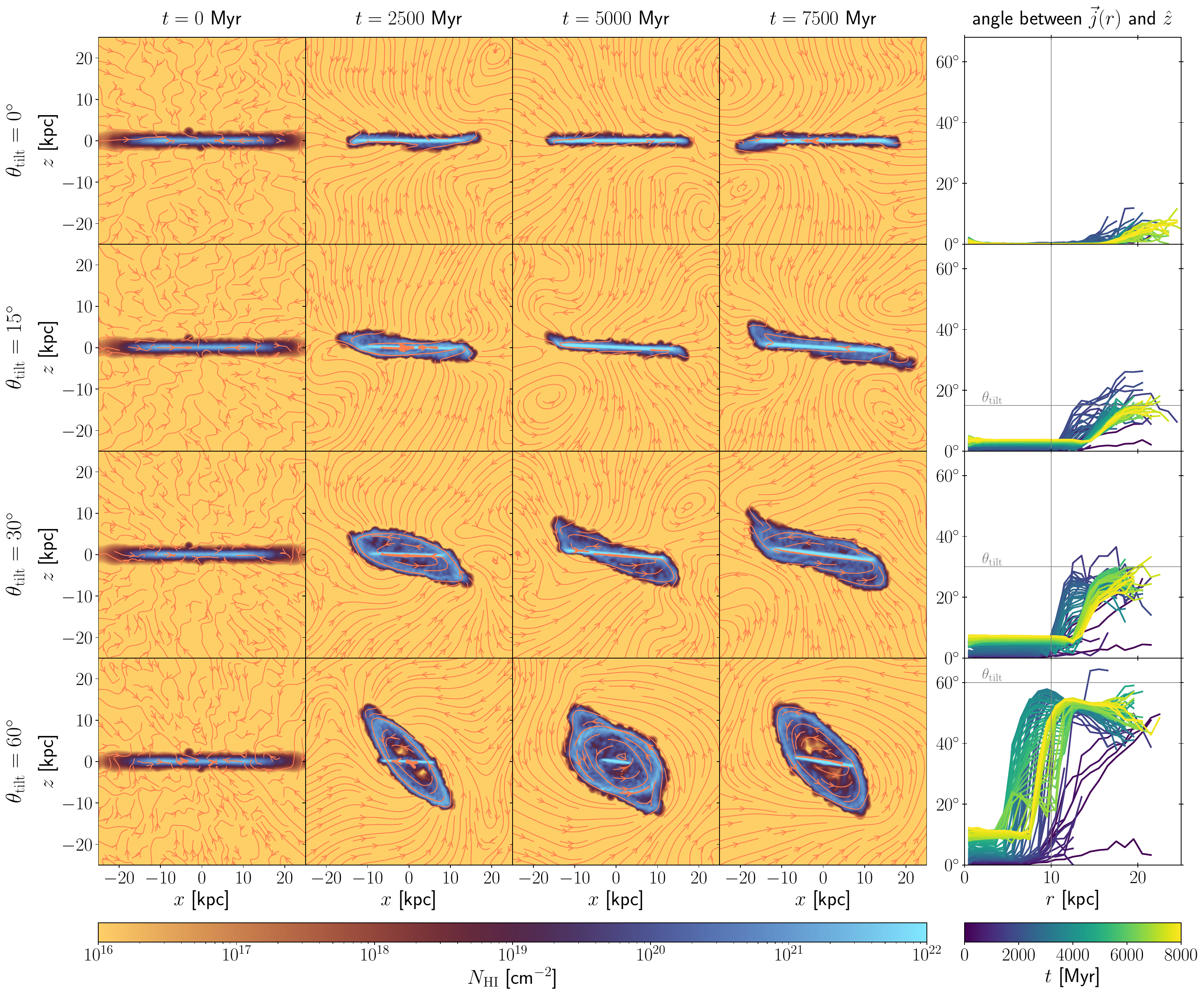}
    \vspace{-0.35cm}
    \caption{The longevity of \hi\, warps condensing out of a rotating hot CGM, for different tilt angles. Each row shows a simulation with different $\thetam$, where the four left panels show the edge-on ($xz$) projections of snapshots at \red{$t=0, 2500, 5000,$ and $7500\,{\rm Myr}$}. The rightmost panel plots the angle of the mean specific angular momentum vector of cool gas versus radius and time. 
    \red{Warps form within $\approx 2000\,{\rm Myr}$, and remain steady afterwards}.
    }
    \label{fig:theta_j_evolution}
\end{figure*}

The bottom panel of Fig.~\ref{fig:theta_j} shows the transition between the inner disc at $r<10\kpc$, where the rotation axis of the cold gas is in the direction of the $z$ axis ($0^\circ$), and the hot CGM at $r>20\kpc$ where the rotation axis of the hot gas is \red{close to} an angle $\thetam$. The warp is evident in the $\thetam=15\deg$ and $30\deg$ simulations at intermediate radii of $\approx10-20\,{\rm kpc}$, where the angle of rotation of the cold gas gradually changes between $0^\circ$ and $\thetam$. The angle of rotation of the hot gas also changes at these radii. 

To demonstrate that \hi\ warps are long-lived in our simulations, Figure~\ref{fig:theta_j_evolution} plots edge-on projections of $\NHI$ at times \red{$t=0, 2500, 5000, 7500\,{\rm Myr}$} for the four simulations. We also plot in the rightmost panels the radial profiles of the angle of the cold gas angular momentum vector at different times, similar to the bottom panel of Fig.~\ref{fig:theta_j}. \red{The panels show that the warps form within $\approx 2000$ Myr and remain roughly steady afterwards. Note also the mild tilt in the inner disc at late times in the simulations, due to accretion of gas from the hot CGM to these inner radii.}
\subsection{Tracking the accretion flow}\label{sec:tracks}

\begin{figure} 
    \includegraphics[width=\linewidth, trim={0 0 0 0}, clip]{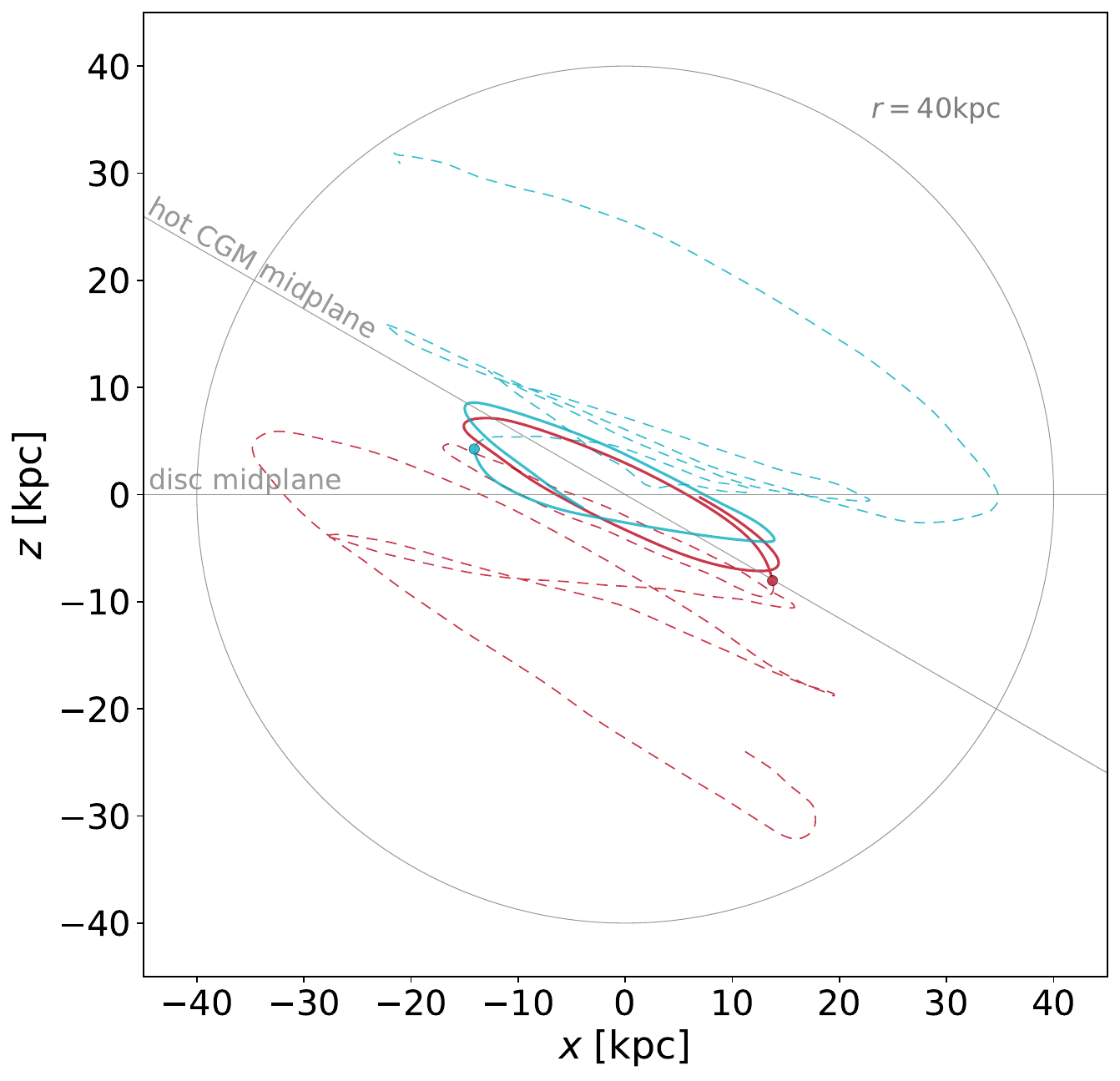}
    \vspace{-0.35cm}
  \caption{\red{Two example trajectories of gas in a hot rotating CGM inflow. We show trajectories that cool at $\phi_{\tfive} \approx 0^\circ$ (red) and at $\phi_{\tfive}\approx180^\circ$ (blue), corresponding to cooling at the maximum vertical offset of the warp in the $xz$-projection. Dashed and solid segments of the tracks correspond to when the inflow is part of the hot CGM and the cool \hi\, warp, respectively. The markers denote the position when the tracks cool. The tracks demonstrate that the hot gas is spiralling in towards the centre, eventually cooling along the midplane of the hot CGM.}}
  \label{fig:helix}
\end{figure}

In this subsection, we analyse the accretion flow and warp formation from a Lagrangian viewpoint, complementing the Eulerian viewpoint explored in the previous subsection. To this end, we track gas resolution elements that form the accretion flow in a manner similar to \cite{hafenHotmodeAccretionPhysics2022} and \cite{sternAccretionDiskGalaxies2024}.  We note that there is no refinement scheme in our simulations, so resolution elements preserve their identity throughout the simulation. 

We track resolution elements that start within a $2\kpc$-thick spherical shell centred at $r_0=40\kpc$ at \red{$t_0 = 2000\,{\rm Myr}$}. Of these, \red{$\sim 96\%$} are subsequently accreted onto the galaxy at any time up to the end of the simulation at \red{$t=8000\,{\rm Myr}$}, where `accreted' implies that the gas has cooled below $T\lesssim 10^5\mathrm{K}$ and either remains cool or forms a star particle. This results in \red{$2225$} tracks for the fiducial $\thetam=30^\circ$ simulation. 
For each track, we identify the time $t_{\tfive}$ when the gas cools to $T=10^5\K$, and measure times in the track relative to $t_{\tfive}$. \red{Other quantities at this time are marked with a subscript `$\tfive$', e.g.\ $R_{\tfive}$ for the cylindrical radius of the track at $t_{\tfive}$. The small minority of tracks which do not accrete are disregarded.} 

\red{We show two example tracks of accreting resolution elements from the $\thetam=30^\circ$ simulation in Figure~\ref{fig:helix}. The tracks are chosen to cool at the average radius of $R_{\tfive}\approx15\kpc$ (see below) and at opposite azimuthal angles of  $\phi_{\tfive}=0^\circ$ (red line) and $\phi_{\tfive}=180^\circ$ (blue line). The dashed curves correspond to where the gas is hot, and the solid line to where the gas is cool, with circular markers denoting the location of cooling. These tracks demonstrate that the hot gas is spiralling in towards the centre and  cools along the midplane of the hot CGM.} 

\begin{figure}
    \centering
    \includegraphics[width=\columnwidth]{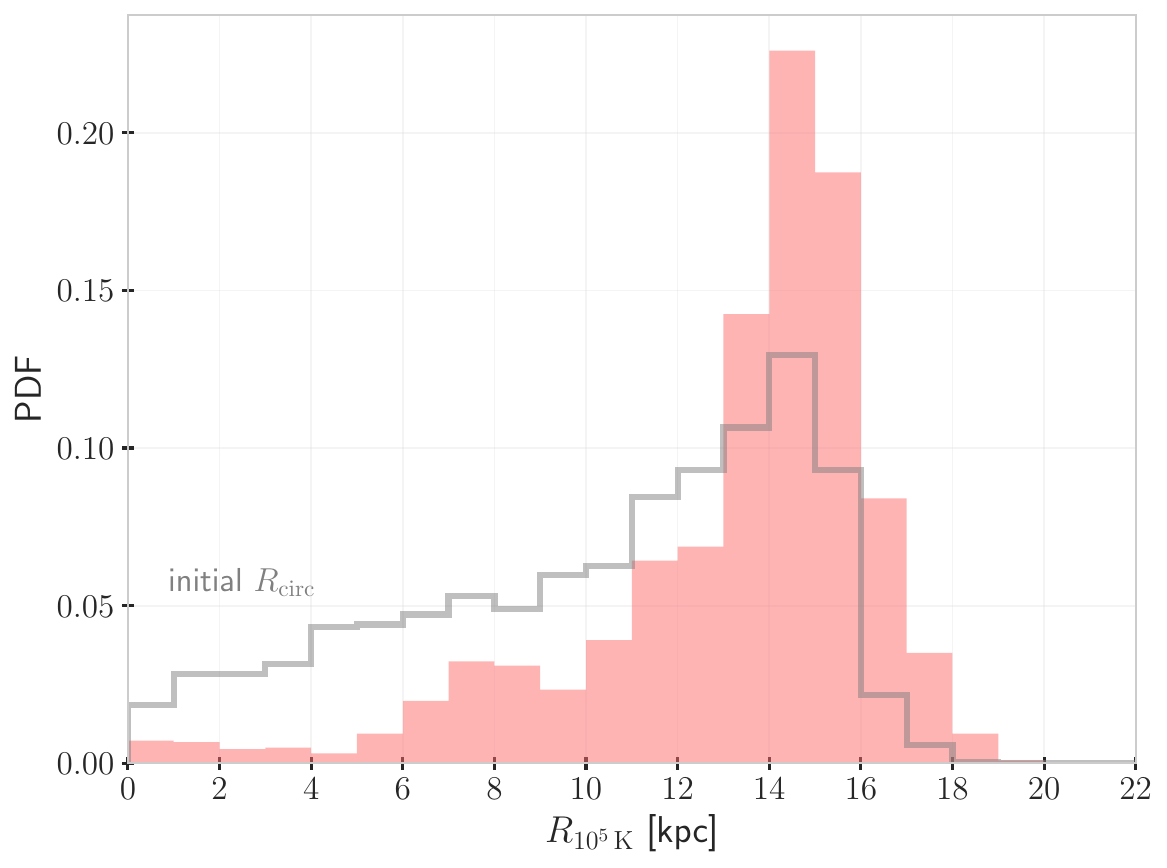}
    \vspace{-0.35cm}
    \caption{\orange{Distribution of cylindrical radii at which the hot inflow cools from $\approx 10^6\K$ to $\approx 10^4\K$, in the $\thetam=30^\circ$ simulation. For comparison, the grey unfilled histogram denotes the PDF of circularisation radii, calculated based on the angular momentum distribution of the accreted gas when it was at $r=40\kpc$. The two distributions are similar, indicating accreting gas cools only when it becomes rotation supported and circularises, while it is hot at larger radii.}
    }
    \label{fig:R_hist}
\end{figure}

\red{Figure \ref{fig:R_hist} shows the distribution of $R_{\tfive}$ in the $\thetam=30^\circ$ simulation, which is the cylindrical radii at which the hot inflow cools.  
Cooling is strongly peaked around $15\kpc$, with a tail towards lower radii. This radius is larger than the pre-existing disc at $<10\kpc$ in our simulation, and thus can explain the nature of extended \hi\ gas.
For comparison, we also plot the distribution of initial circularisation radii of the tracks as a grey line. \orange{The initial $\Rcirc$ are calculated based on the specific angular momentum $j$ of the accreting resolution elements when they are at $r=40\kpc$, using $j=\vc(\Rcirc)\Rcirc$.} 
The figure shows that the accreting gas cools at similar radii as the initial $\Rcirc$ distribution, with a slight offset to higher values. This cooling at $\approx\Rcirc$ has also been found by \cite{hafenHotmodeAccretionPhysics2022} and \cite{sternAccretionDiskGalaxies2024}, and implies that the source of the accreting gas is hot at radii beyond $\Rcirc$, and thus cannot be traced by cool gas observations. }
 
\begin{figure*} 
    \includegraphics[width=\linewidth, trim=1.5cm 2.5cm 2.5cm 2cm, clip]{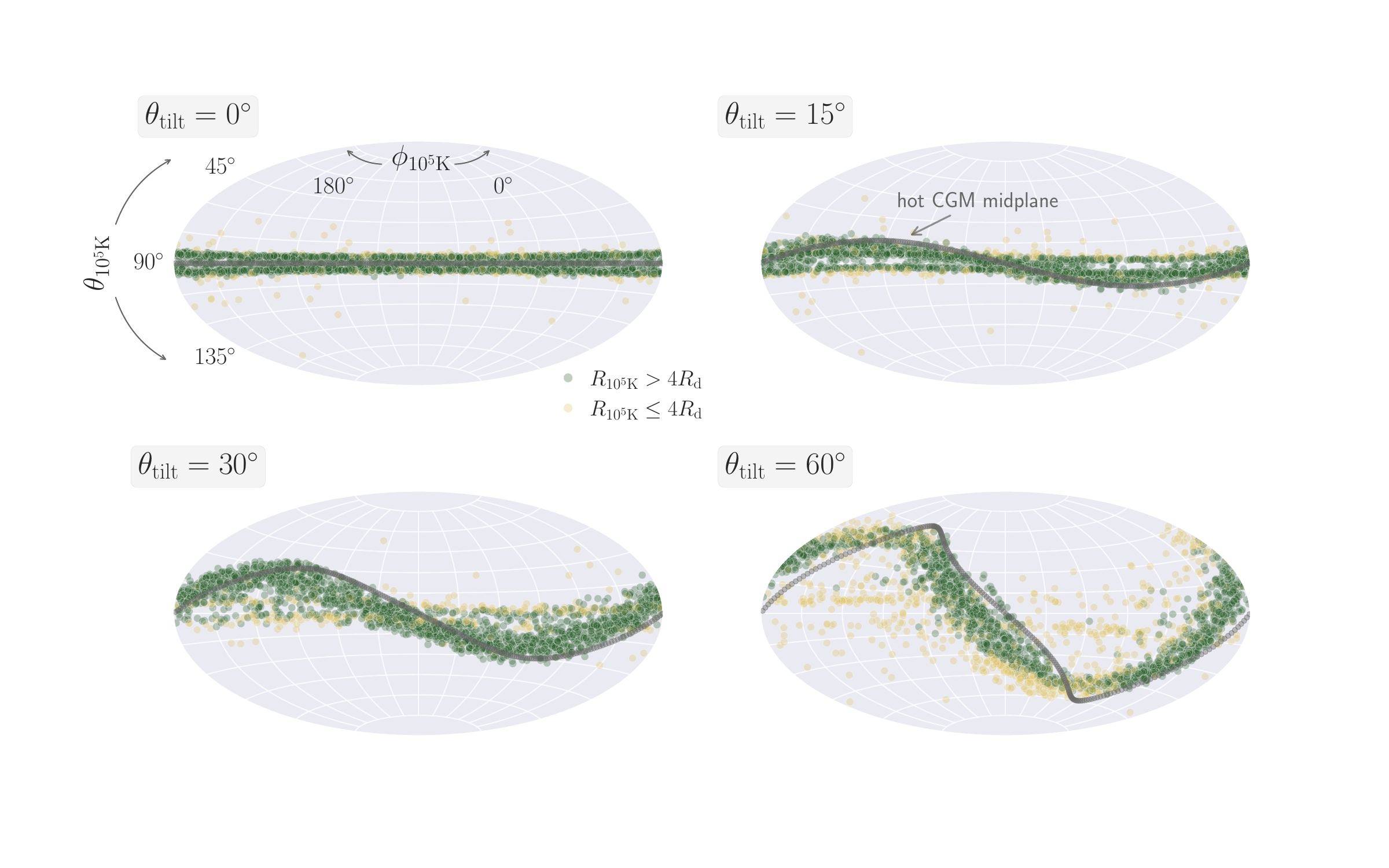}
    \vspace{-0.5cm}
  \caption{The location of cooling in hot inflows traces the hot gas midplane. Panels show the polar and azimuthal angles at which the hot inflow cools from $\approx 10^6\K$ to $\approx 10^4\K$, for the \red{four simulations with $\thetam=0^\circ$ (\textit{top left}), $15^\circ$ (\textit{top right}), $30^\circ$ (\textit{bottom left}), and $60^\circ$ (\textit{bottom right})}. Each point marks the spherical coordinates ($\theta$,$\phi$) of a single accreting resolution element at the time that it cools, with colour denoting the cylindrical radius at cooling. In all simulations, most of the accreting gas cools \red{near} the CGM midplane (marked in the panels as a thick grey curve). A minority of the gas cools near the disc plane ($\theta_{10^5\K}\approx90^\circ$).}
  \label{fig:aitoff_30}
\end{figure*}

In Figure~\ref{fig:aitoff_30} we plot the polar and azimuthal angle distribution of accreting gas at the time of cooling in the four simulations, using Aitoff projections. Each point corresponds to a given track, with colour indicating whether the radius of accretion is smaller or larger than the size of the pre-existing disc ($4\Rdisk$). 
The figure demonstrates that for all four tilt angles, cooling occurs mainly close to the midplane of the CGM (marked with an S-shaped grey band). This cooling near the midplane becomes even more pronounced when considering only gas that cools at $>10\kpc$ (green markers). The \textit{drop in temperature from $\approx10^6\K$ to $\approx 10^4\K$ thus occurs \red{near} the CGM midplane}, suggesting that the balance between radiative cooling and compressive heating, which keeps the accreting gas hot, is maintained at all times until the flowlines are close to the midplane. Observationally, this result, together with the $R_{10^5\K}\approx\Rcirc$ result in Fig.~\ref{fig:R_hist} implies that gas that feeds the warp is hot and thus hard to observe until it joins the warp. 

\begin{figure*}
\centering
    \includegraphics[width=\linewidth]
    {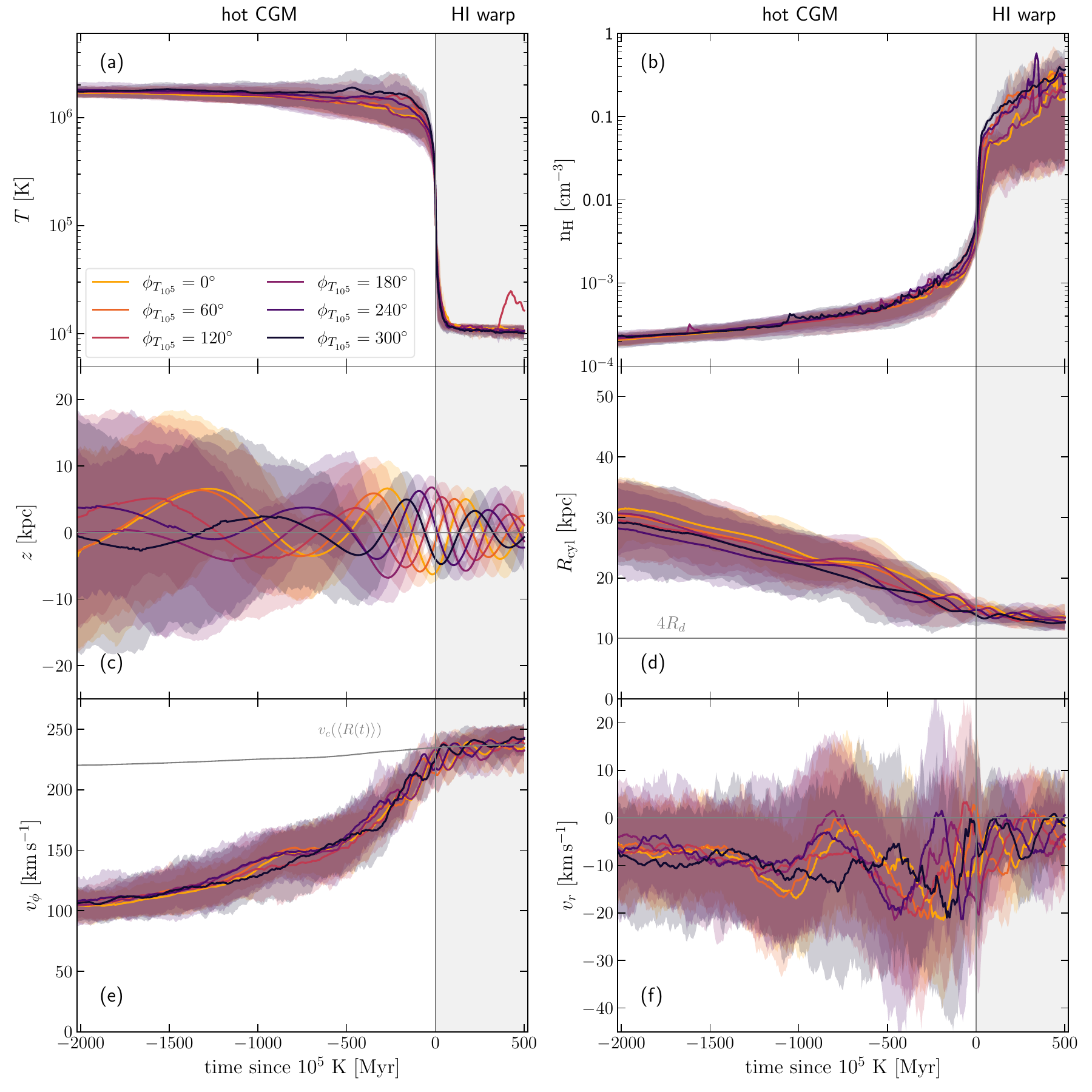} 
    \vspace{-0.35cm}
  \caption{Lagrangian evolution of hot CGM inflows with a tilted rotation axis. Panels show different properties of accreting gas as a function of time relative to cooling, in the $\thetam=30\deg$ simulation. Lines and bands plot the mean and dispersion of properties of all gas that cool at \red{$R_{\tfive}>10\kpc$} (see Fig.~\ref{fig:R_hist}), grouped by azimuthal angle at the time of cooling (see legend in top-left panel). \red{Top row: gas temperature and density. Note that cooling from $10^6\K$ to $10^4\K$ is rapid, within $\lesssim200\,{\rm Myr}$. Second row: $z$ and $R_{\rm cyl}$ coordinates. Note that cooling occurs away from the disc midplane at $R\approx15\kpc$, as also indicated by Figs.~\ref{fig:R_hist} -- \ref{fig:aitoff_30}. Third row: rotation and radial velocities. Note that full rotation support  ($v_\phi=\vc$) is achieved coincidently with cooling, a tell-tale sign of hot rotating inflows. 
  }}
  \label{fig:track_props}
\end{figure*}

Additional locii of points are evident in Fig.~\ref{fig:aitoff_30} just above and below the pre-existing discs. These tracks also have $R_{10^5\K}<10\kpc$ (the tail of the distribution in Fig.~\ref{fig:R_hist}), suggesting that for these tracks, cooling is a result of direct interaction of the accreting gas with the pre-existing discs. These tracks are subdominant, accounting for $10\%-16\%$ of all accretion tracks.
Figure \ref{fig:track_props} demonstrates how the accretion flow proceeds as a function of time relative to the time of cooling, in the $\thetam=30^\circ$ simulation. This figure shows gas which cools at \red{$R_{\tfive}>10\kpc$, corresponding to the peak in Fig.~\ref{fig:R_hist} and including 84\% of all accreting gas, while accretion with $R_{\tfive}<10\kpc$ is shown in Figure~ \ref{fig:plane_track_props} in the appendix}. The accretion tracks are grouped by the azimuthal angle at the time of cooling, with bins spanning $30^\circ$ and centred on the values noted in the legend. Lines and bands plot the mean and 16 -- 84 percentile range of properties. 
Panel \textit{a} demonstrates that cooling from $10^6\K$ to $10^4\K$ occurs within $\lesssim200\,{\rm Myr}$, which is rapid relative to cooling times of $>$ Gyr farther out in the halo. The rapid cooling is made possible by the higher densities of $\nH\approx10^{-3}\,{\rm cm}^{-3}$ at the time of cooling (panel \textit{b}). Panels \textit{c} and \textit{d} show that cooling occurs at $R_{\rm cyl}<20\kpc$ and at a $z$ that depends on $\phi_{10^5\K}$, as expected for cooling in a tilted plane and consistent with the results in Figs.~\ref{fig:R_hist} -- \ref{fig:aitoff_30}. Panel \textit{e} demonstrates that cooling occurs when $v_\phi$ reaches $\vc$, where $\vc$ in the panel is calculated at the mean $R(t)$ shown in panel \textit{d}. This relation between cooling and achieving rotation support is a tell-tale of the rotating hot inflow scenario, and has been identified also in \fire\ simulations of Milky-Way mass galaxies by \cite{hafenHotmodeAccretionPhysics2022} and \cite{Sultan26}.


\red{After rotation support is achieved at $t>t_{\tfive}$, the flow settles on a tilted circular orbit with a mild radial inflow velocity of $\lesssim10\kmsmath$ (panels \textit{c} and \textit{f} in Fig.~\ref{fig:track_props}). Such mild inflows are required to explain how the accreted gas reaches the more inner parts of the galaxy where star formation takes place (see discussion).  We note though that simulating the evolution of gas after cooling requires more realistic disc physics than what we implement in our simulations.  }


Similar tracks for accreting gas with $R_{\tfive}<10\kpc$ are shown in appendix Fig.~\ref{fig:plane_track_props}. This type of accretion is subdominant, constituting $10-16\%$\ of all accreting gas. Panel \textit{b} in Fig.~\ref{fig:plane_track_props} demonstrates that cooling in this case occurs in the disc plane ($z=0\kpc$), as also suggested by Fig.~\ref{fig:aitoff_30}. It thus seems plausible that cooling of this type is a result of direct interaction of a hot flow with the pre-existing disc, in contrast with cooling caused by achieving rotation support, as is the case for accreting gas with $R_{\tfive}>10\kpc$  shown in Fig.~\ref{fig:track_props}.


\section{Discussion}
\label{sec:discussion}


In this work, we demonstrate that hot CGM rotating around an axis tilted with respect to the disc axis create extended \hi\ warps. We show that the warp is the end result of the quasi-steady `rotating cooling flow' process \citep{sternMaximumAccretionRate2020, sternAccretionDiskGalaxies2024, hafenHotmodeAccretionPhysics2022, Sultan25, Sultan26}, wherein the hot CGM phase ($T\approx\Tvir\sim10^6\K$) flows inward while remaining hot and spinning up, eventually cooling to $T\approx10^4\K$ in the plane perpendicular to the rotation axis and at a radius equal to the circularisation radius. 
This model therefore shares the main principle of earlier models that attributed \hi\ warps to misaligned accretion \citep[e.g.,][]{Ostriker89,shenGalacticWarpsInduced2006, sancisiColdGasAccretion2008,roskarMisalignedAngularMomentum2010}, with the main distinction being that accreting gas is hot prior to joining the warp. 

In this section, we discuss how our model compares with observations of low redshift massive spirals (section~\ref{sec:discussion1}) and why our assumption of neglecting feedback heating of the hot CGM is plausible in these systems (section~\ref{sec:discussion2}). We then discuss galaxies at higher redshifts or with lower masses (section~\ref{sec:discussion3}).


\subsection{Comparison to observations of $z\sim0$ spirals}\label{sec:discussion1}

Our results suggest that cooling from the hot CGM is likely to occur in an \hi\ disc that is extended and tilted with respect to the existing ISM, which traces the magnitude and orientation of the hot CGM angular momentum. 
In this subsection, we compare this prediction with available observational constraints.  

The predicted sizes of the extended \hi\ discs in the model are $\approx\Rcircmean$ (Fig.~\ref{fig:R_hist}), which is expected to be somewhat larger than the stellar disc due to the higher spin of the hot gas. This ratio is consistent with the relative extent of \hi\ and stars observed in nearby spirals \citep[e.g.,][]{Broeils97, wangObservationalTheoreticalView2014, wangFEASTSRadialDistribution2025, leeWALLABYPilotSurvey2025}, thus supporting the model. A similar argument on the size of accreted discs was made by \cite{Stewart13} and \cite{stewartHighAngularMomentum2017a}, who posited that large cool discs at $z>1$ are a result of the high spin of the cold flows that feed them. Our results thus demonstrate how the \citeauthor{Stewart13} argument applies at $z\sim0$ when the hot accretion mode dominates, an epoch which is also more accessible to \hi\ observations.  Crucially, in the hot mode, \hi\ gas is not required at larger distances, since accreting gas cool only when reaching the circularisation radius (Fig.~\ref{fig:R_hist}). This is consistent with observations that find a lack of \hi\ in the inner CGM of nearby spirals \citep{healyPossibleOriginsAnomalous2024, deblokMHONGOOSEMeerKATNearby2024, wangFEASTSCombinedInterferometry2024, linFEASTSCombinedInterferometry2025, Veronese25, Marasco25}. 

While observed \hi\ warp sizes appear consistent with hot CGM spins in simulations, observed warp tilt angles may be lower than misalignment angles of simulated hot CGM. \cite{garcia-ruizNeutralHydrogenOptical2002} measured warp angles of $3<\theta_{\rm warp}<25^\circ$ with a median of $12^\circ$ in edge-on galaxies, if we consider only the $16$ out of $26$ objects in their sample with $R_{\hi}/R_{25}>1.2$ where we expect $\theta_{\rm warp}\approx\thetam$ (see Fig.~\ref{fig:theta_j} and  eqn.~\ref{eq:ttorque to tflow}). Somewhat larger warp angles of $15^\circ-20^\circ$ have been measured in seven galaxies from the Local Volume \hi\ survey \citep{wangLocalVolumeSurvey2017}, though they caution that their measurement method may overestimate true warp angles.
For comparison, \citet{DeFelippis20} looked at misalignments between the CGM and stellar disc in $z=0$ MW-mass galaxies in the \textsc{TNG100} simulations \citep{pillepichFirstResultsTNG502019}, and found a median $\thetam=15^\circ$ for galaxies in the upper quartile of the angular momentum distribution, and a larger $\thetam=40^\circ$ for galaxies in the lower quartile. \cite{Huscher21} deduced a larger median $\thetam=56^\circ$ in the \eagle simulations \citep{schayeEAGLEProjectSimulating2015}. The simulated $\theta_{\rm tilt}$ thus appear to be on average larger than observed $\theta_{\rm warp}$. \red{Larger observational surveys and a more careful comparison that accounts for survey selection, together with measuring simulated tilt angles specifically in the inner hot CGM which sources the accretion, will be able to test this conclusion more robustly. }

\red{Since star formation within the warps is inefficient \citep{Bigiel10, wangLocalVolumeSurvey2017}, the metallicity of accreted gas should not change significantly when cooling from the hot CGM onto an \hi\ warp. A prediction of our model is thus that hot CGM and galaxy warps should exhibit similar gas metallicities.} In the Milky Way, \cite{Wenger19} deduced $Z=0.3-0.6\,{\rm Z}_\odot$ from \ion{H}{ii} regions at $R=15\kpc$ (the furthest of their measurements), similar to the $0.4-0.8\,{\rm Z}_\odot$ deduced in the outskirts of MANGA galaxies with similar mass \citep[][at $R=R_{25}$]{Pilyugin19}. This metallicity in the Milky Way outskirts is consistent with the hot CGM metallicity of $\approx0.5\,{\rm Z}_\odot$ deduced from X-ray and dispersion measure observations \citep[with factor of $\approx2$ uncertainty]{Faerman17,Qu18, Bregman18}. A handful of external galaxies also have metallicity estimates for their hot CGM, and show similar values as the Milky Way \citep{Bregman18}. The hot CGM metallicity is thus consistent with the metallicity in galaxy outskirts within the observational errors, further supporting the validity of our hot accretion model.




\green{If local spirals accrete mainly from disc outskirts as suggested by the model herein (Fig.~\ref{fig:R_hist}), then radial inflows are needed to transport gas inward to radii where star formation is observed to occur.} Such radial flows have been suggested by several disc models \citep[\ex][]{krumholzUnifiedModelGalactic2018, Wang22a, Wang22b}, and have beed identified also in some disc simulations \citep{Trapp22, barbaniUnderstandingBaryonCycle2025}. 
Observational evidence for such inflows in local discs remains inconclusive, given the difficulty in identifying the low predicted inflow velocities of $1-5\kmsmath$ in the inner disc \citep[e.g.,][]{Trapp22}. \cite{diTeodoro21} argued that $21\,{\rm cm}$ observations rule out such disc inflows, though this result has been questioned in the outer parts due to known degeneracies between signatures of radial flows and warps \citep{Wang23, zuletaKinematicalSignaturesDistinguishing2024, syloslabiniBreakingDegeneracyWarps2025}. In contrast, an observed correlation between mean gas metallicity and the slope of the disc metallicity gradient has been argued to support the existence of radial inflows \citep{lyuDominantRoleCoplanar2025}. Our simulations exhibit mild mean inflow velocities in the warp of $v_r\approx0-5\kmsmath$ (Fig.~\ref{fig:track_props}, panel \textit{f}), similar to those required to explain star formation in the inner galaxy \orange{if accretion is mainly from the outer edge.
However, we note that our idealised setup focusing on the CGM does not attempt to realistically model the disc, and thus does not accurately model the torques on disc gas which govern disc flows. Specifically, our setup does not include any feedback process or magnetic fields, while star formation is accounted for using a simplified prescription (section~\ref{sec:sim}). This will affect pressure gradients, non-axisymmetric perturbations, and viscous stresses in the disc, all of which will affect the predicted torques \citep[e.g.,][]{hopkinsAnalyticModelAngular2011, krumholzUnifiedModelGalactic2018,Trapp22,Wang22a}.
Incorporating these processes into future simulations of  hot accretion will test whether the resulting disc flows are consistent with observed star formation profiles.
}




Hot CGM that are rotating and inflowing also have specific predictions for how the hot gas entropy, rotation velocity, and temperature depend on radius and polar angle, properties which have observational consequences for hot CGM observations. We refer the reader to previous papers on inflowing and rotating hot CGM for a thorough discussion \citep[and see also \citealt{Sormani18} who discussed hot CGM that are rotating but not inflowing]{sternCoolingFlowSolutions2019,sternAccretionDiskGalaxies2024,Wijers24,Sultan25}.
 
\subsection{Could galaxy feedback disrupt warp formation?}\label{sec:discussion2}

Our calculation above neglects heating of the hot CGM by ongoing galaxy feedback processes. Radiative cooling of the hot CGM is then balanced by compressive heating out in the halo, so the actual drop in temperature occurs as a result of circularisation at a radius $\lesssim\Rcirc\approx0.1\rvir$, rather than as a result of thermal instability in the hot phase further out in the halo. If instead feedback heating offsets or exceeds cooling losses, then the hot CGM would not inflow but rather be radially static or form an outflow \citep{Carr23,Voit24a,Voit24b}. Gas accretion in this latter case would instead proceed via sinking cool clouds formed due to thermal instabilities \citep{McCourt12,Sharma12,Voit17, FGOh23}. We argue here that neglecting ongoing feedback heating of the hot CGM is a plausible assumption in massive spirals at $z\sim0$.  

Main sequence star-forming discs at $z\sim0$ exhibit weak stellar-driven winds \cite[e.g.,][]{Thompson24} which are unlikely to significantly offset  CGM cooling losses. Significant heating by stellar feedback is more likely in starburst galaxies, which exhibit strong winds, though if bursts are separated by more than a cooling time, then the hot inflow would have time to reform. As $\tcool$ is rather short in the inner CGM ($\sim1\,{\rm Gyr}$, see eqn.~\ref{eq:tcool}) this condition is plausibly satisfied. Feedback heating can also result from black hole activity, as is potentially suggested by the existence of the Fermi Bubbles in the Milky Way, though the amount of heating injected into the CGM by this process is highly uncertain \citep[see review of Fermi bubble models in][]{Sarkar24}. Even if significant, feedback by the black hole mostly affects the hot CGM near the rotation axis \citep[e.g.,][]{Truong21}, so it is unclear if it can disrupt the formation of \hi\, warps near the disc edge as discussed here. We thus conclude that feedback heating is plausibly small near the midplane of hot CGM of $z\sim0$ main-sequence galaxies, supporting our calculation above.  

We note that the relatively weak stellar-driven winds seen in local spirals, known as `fountain flows', could also affect the hot CGM by \textit{promoting} cooling at the disc-halo interface, due to mixing of the cool winds with the hot gas \citep{marascoSupernovadrivenGasAccretion2012, fraternaliGasAccretionCondensation2017, liFountaindrivenGasAccretion2023}. 
\red{Our simulations do not account for fountain flows and result in most of the hot inflow accreting beyond the disc edge (Fig.~\ref{fig:R_hist}), rather than at inner galaxy radii above star formation sites where fountain flows are observed. 
It would thus be useful to run simulations which account for both fountain-driven accretion and spontaneous hot CGM cooling, in order to understand the conditions under which each of these accretion modes dominates.}



\subsection{At which masses and redshifts are \hi\ warps formed via hot accretion?}\label{sec:discussion3}

Condensation of the hot CGM directly onto the warp requires that the hot CGM is stable and long-lived down to the circularisation radius $\Rcirc\approx0.1\rvir$. It has long been argued that a stable hot CGM can exist only if its cooling time $\tcool$ is longer than the halo dynamical or free-fall time $\tff$, a condition satisfied for mean CGM densities above a halo mass threshold of order $\sim 10^{11.5}\,\msun$ \citep{White78,birnboim03,keresHowGalaxiesGet2005,Fielding17,Pandya23}, with \red{the exact threshold depending} on CGM mass and metallicity. \cite{sternMaximumAccretionRate2020} showed that at inner halo radii near $\Rcirc$, the higher density and shorter cooling time imply a higher threshold mass than the outer CGM. This threshold is found to be $\sim 10^{12}\,\msun$ for a CGM mass equal to half the halo baryon budget and CGM metallicity of $0.3~{\rm Z}_\odot$ (see also fig.~2 in \citealt{goldnerAccretionDrivenTurbulenceCircumgalactic2025}). This predicted `outside-in' hot CGM formation scenario has been identified in the FIRE cosmological zoom simulations \citep{stern21a, stern21b, Gurvich23,Kakoly25} and potentially also in TNG50 \citep{Semenov24}. As the warp forms in the inner CGM, it is this higher mass threshold that is required to be surpassed for the warp to condense directly out of the hot CGM.  
Furthermore, even if a stable hot phase exists down to $\Rcirc$, it may be subdominant in terms of accretion to cold flows penetrating the halo and reaching the disc, as may be expected at $z>1$ \citep{Dekel06}.   
\red{We thus expect \hi\ warps condensing directly out of the hot CGM mainly in Milky-Way mass spirals at $z<1$. If CGM have been significantly depleted by feedback, hot CGM can extend down to $\Rcirc$ and form warps also at lower halo masses.} 

In principle, warps could also form due to misaligned accretion of cool gas, which is expected at lower masses and higher redshifts than hot accretion. 
However, observations of warped local spirals indicate discs which are sufficiently thin and kinematically cold that even mild warp angles of $\approx3^\circ$ are identifiable \citep{garcia-ruizNeutralHydrogenOptical2002}. This suggests that local warped spirals prefer a gentle and smooth accretion process, such as the hot accretion mode discussed herein.
Cold accretion, in contrast, can enhance disc turbulence, so the disc would become thicker and the warps would be less apparent. This can occur either directly by cold flows driving turbulence in the disc \citep{Ginzburg22}, or indirectly if cold accretion affects feedback properties and disc gas fractions \citep{stern21a, Gurvich23, Sun25} -- characteristics which set disc turbulence levels \cite[e.g.][]{bland-hawthornTurbulentGasrichDisks2024,bland-hawthornTurbulentGasrichDiscs2025}.  


\red{\subsection{Caveats and future directions}}

\red{The importance of the \hi\, warp formation model discussed herein can be further tested using cosmological simulations, which include several physical mechanisms that are absent in our idealised setup. These simulations include CGM spins and CGM-disc tilts that emerge self-consistently in a $\Lambda$CDM framework rather than being \orange{implemented ad-hoc} as we do here. \orange{Additionally,  explicit feedback models in the simulations may affect CGM and warp properties, and dark matter potentials that are tilted and/or not perfectly spherical may induce additional torques. These simulations could thus be used to test the conditions under which the CGM inflow and warp-formation mechanisms explored here dominate over alternative scenarios} \citep[e.g.,][]{shenGalacticWarpsInduced2006, roskarMisalignedAngularMomentum2010, hanTiltedDarkHalo2023, binneyDiscDistortionRevisited2024}.
We expect our model to be valid in simulations in which accretion is dominated by inflows in the hot CGM, such as the \fire zoom simulations of Milky Way-mass galaxies \citep{hafenHotmodeAccretionPhysics2022,Sultan25,Sultan26}. }

\red{
Simulations which account for ISM physics could also allow studying additional properties of warps in the context of our model and whether they are consistent with observations. These include radial disc inflows (see section~\ref{sec:discussion1}), bending waves \citep[e.g.,][]{sellwoodInternallyDrivenWarps2022}, corrugation patterns \citep[e.g.,][]{bland-hawthornGalacticSeismologyEvolving2021}, and line of nodes patterns \citep[e.g.,][]{briggsRulesBehaviorGalactic1990, jozsaKinematicModellingDisk2007a}. From the observational side, a more extensive sample of observed \hi\ warps and their orientations with respect to the inner disc would be useful for comparison against the hot CGM angular momentum properties found in cosmological simulations. Such observations will be within reach with the upcoming SKA-Mid telescope, as already demonstrated by sensitive observations with precursor telescopes \citep[e.g.,][]{deblokMHONGOOSEMeerKATNearby2024, wangFEASTSCombinedInterferometry2024}. 
}

\vspace{0.5cm}
\section{Summary}
\label{sec:conclusion}

In this work, we demonstrate that an extended and warped \hi\ disc forms out of hot CGM that are rotating with a tilted axis relative to that of the stellar disc. Our simulation setup uses the \textsc{GIZMO} code and includes a Milky Way-like pre-existing disc, a radiatively cooling and rotating hot CGM, and an analytic gravitational profile mimicking the dark matter halo potential. This setup is similar to \cite{sternAccretionDiskGalaxies2024}, with the main distinction that the CGM and disc rotation axes are misaligned. 
Our study further differs in its focus on the formed \hi\ discs, in contrast with predicted hot CGM properties discussed in \cite{sternAccretionDiskGalaxies2024}.

Our main findings can be summarised as follows.
\begin{enumerate}
    \item We demonstrate that extended \hi\ discs can form through continuous and smooth cooling of the hot CGM, \red{which occurs near} the midplane defined by the CGM rotation axis (Fig.~\ref{fig:aitoff_30}) and at radii equal to the circularisation radius (Fig.~\ref{fig:R_hist}). The continuous and smooth nature of this accretion mode is consistent with observations of regular and low-turbulent \hi\ discs in the local universe \citep{Marasco25}, which suggest gentle accretion and feedback processes are at play.

    \item Accreting gas in our simulations is hot and invisible to cool gas observations prior to joining the extended \hi\, disc (Figs.~\ref{fig:xz_projection}--\ref{fig:xz_projections_3sims}), consistent with recent deep $21\,{\rm cm}$ observations of local spirals that indicate CGM \hi\ is insufficient to explain gas accretion \citep{deblokMHONGOOSEMeerKATNearby2024, wangFEASTSCombinedInterferometry2024, linFEASTSCombinedInterferometry2025}. 
    \item Observed sizes of \hi\ discs in the local Universe, which are roughly twice the stellar disc size \citep[e.g.][]{Broeils97,leeWALLABYPilotSurvey2025}, are consistent with the ratio of hot CGM and stellar disc spins seen in cosmological simulations \citep[e.g.][]{stewartHighAngularMomentum2017a,DeFelippis20,Huscher21}, supporting our model. Observed warp angles appear smaller than hot CGM-disc misalignment angles found in simulations (section~\ref{sec:discussion1}).  


    \item Warps are long-lived in our simulations, remaining steady at $2-8\,{\rm Gyr}$ (Fig.~\ref{fig:theta_j_evolution}, Fig~\ref{fig:accretion_rate}). We note that our simulations assume constant hot CGM properties, but these are expected to evolve on a timescale of several Gyrs once the hot CGM has formed.
    \end{enumerate}

\section*{Acknowledgements}

\red{We thank the anonymous referee for rigorous and insightful comments that greatly improved the paper.} SS was supported by the Australian Research Council Centre of Excellence for All Sky Astrophysics in 3 Dimensions (ASTRO 3D), through project number CE170100013. JS was supported by the Israel Science Foundation (grant No.\ 2584/21) and by a grant from the United States-Israel Binational Science Foundation (BSF), Jerusalem, Israel. CAFG was supported by NSF through grants AST-2108230 and AST-2307327; by NASA through grants 21-ATP21-0036 and 23-ATP23-0008; and by STScI through grant JWST-AR-03252.001-A. MBK acknowledges support from NSF grants AST-1910346, AST-2108962, and AST-2408247; NASA grant 80NSSC22K0827; HST-GO-16686, HST-AR-17028, HST-AR-17043, JWST-GO-03788, and JWST-AR-06278 from the Space Telescope Science Institute, which is operated by AURA, Inc., under NASA contract NAS5-26555; and from the Samuel T. and Fern Yanagisawa Regents Professorship in Astronomy at UT Austin. This work was supported by resources provided by the \hyperlink{https://doi.org/10.48569/18sb-8s43}{Pawsey Supercomputing Research Centre’s Setonix Supercomputer}, with funding from the Australian Government and the Government of Western Australia.

\section*{Data Availability}
The simulation data underlying this article will be shared on reasonable request to the corresponding authors. A public version of the GIZMO simulation code is available at \hyperlink{https://github.com/pfhopkins/gizmo-public}{https://github.com/pfhopkins/gizmo-public}.


\bibliographystyle{mnras}
\bibliography{tau}

@article{krumholzUnifiedModelGalactic2018,
  title = {A Unified Model for Galactic Discs: Star Formation, Turbulence Driving, and Mass Transport},
  shorttitle = {A Unified Model for Galactic Discs},
  author = {Krumholz, Mark R. and Burkhart, Blakesley and Forbes, John C. and Crocker, Roland M.},
  year = 2018,
  month = jun,
  journal = {Monthly Notices of the Royal Astronomical Society},
  volume = {477},
  number = {2},
  pages = {2716--2740},
  issn = {0035-8711},
  doi = {10.1093/mnras/sty852},
  urldate = {2025-11-20},
  langid = {english}
}

@article{hopkinsAnalyticModelAngular2011,
  title = {An Analytic Model of Angular Momentum Transport by Gravitational Torques: From Galaxies to Massive Black Holes},
  shorttitle = {An Analytic Model of Angular Momentum Transport by Gravitational Torques},
  author = {Hopkins, Philip F. and Quataert, Eliot},
  year = 2011,
  month = aug,
  journal = {Monthly Notices of the Royal Astronomical Society},
  volume = {415},
  number = {2},
  pages = {1027--1050},
  issn = {0035-8711},
  doi = {10.1111/j.1365-2966.2011.18542.x},
  urldate = {2026-05-17},
  langid = {english}
}

@article{Sultan26,
	adsurl = {https://ui.adsabs.harvard.edu/abs/2026arXiv260414273S},
	archiveprefix = {arXiv},
	author = {{Sultan}, Imran and {Faucher-Gigu{\`e}re}, Claude-Andr{\'e} and {Stern}, Jonathan and {Sun}, Guochao},
	doi = {10.48550/arXiv.2604.14273},
	eid = {arXiv:2604.14273},
	eprint = {2604.14273},
	journal = {arXiv e-prints},
	month = apr,
	pages = {arXiv:2604.14273},
	primaryclass = {astro-ph.GA},
	title = {{Cold vs. Hot Gas Accretion and Angular Momentum in FIRE Simulations: From Halo to Galaxy Scales}},
	year = 2026}

@article{Wechsler02,
	adsurl = {https://ui.adsabs.harvard.edu/abs/2002ApJ...568...52W},
	archiveprefix = {arXiv},
	author = {{Wechsler}, Risa H. and {Bullock}, James S. and {Primack}, Joel R. and {Kravtsov}, Andrey V. and {Dekel}, Avishai},
	doi = {10.1086/338765},
	eprint = {astro-ph/0108151},
	journal = {\apj},
	month = mar,
	number = {1},
	pages = {52-70},
	primaryclass = {astro-ph},
	title = {{Concentrations of Dark Halos from Their Assembly Histories}},
	volume = {568},
	year = 2002}

@article{Mandelker20,
	adsurl = {https://ui.adsabs.harvard.edu/abs/2020MNRAS.494.2641M},
	archiveprefix = {arXiv},
	author = {{Mandelker}, Nir and {Nagai}, Daisuke and {Aung}, Han and {Dekel}, Avishai and {Birnboim}, Yuval and {van den Bosch}, Frank C.},
	doi = {10.1093/mnras/staa812},
	eprint = {1910.05344},
	journal = {\mnras},
	month = may,
	number = {2},
	pages = {2641-2663},
	primaryclass = {astro-ph.GA},
	title = {{Instability of supersonic cold streams feeding galaxies - IV. Survival of radiatively cooling streams}},
	volume = {494},
	year = 2020}

@article{Dekel09,
	adsurl = {https://ui.adsabs.harvard.edu/abs/2009ApJ...703..785D},
	archiveprefix = {arXiv},
	author = {{Dekel}, Avishai and {Sari}, Re'em and {Ceverino}, Daniel},
	doi = {10.1088/0004-637X/703/1/785},
	eprint = {0901.2458},
	journal = {\apj},
	month = sep,
	number = {1},
	pages = {785-801},
	primaryclass = {astro-ph.GA},
	title = {{Formation of Massive Galaxies at High Redshift: Cold Streams, Clumpy Disks, and Compact Spheroids}},
	volume = {703},
	year = 2009}

@article{Thompson24,
	adsurl = {https://ui.adsabs.harvard.edu/abs/2024ARA&A..62..529T},
	archiveprefix = {arXiv},
	author = {{Thompson}, Todd A. and {Heckman}, Timothy M.},
	doi = {10.1146/annurev-astro-041224-011924},
	eprint = {2406.08561},
	journal = {\araa},
	month = sep,
	number = {1},
	pages = {529-591},
	primaryclass = {astro-ph.GA},
	title = {{Theory and Observation of Winds from Star-Forming Galaxies}},
	volume = {62},
	year = 2024}

@article{faucher-giguereCosmicUVXray2020,
	author = {{Faucher-Gigu{\`e}re}, Claude-Andr{\'e}},
	doi = {10.1093/mnras/staa302},
	issn = {0035-8711},
	journal = {Monthly Notices of the Royal Astronomical Society},
	langid = {english},
	month = apr,
	number = {2},
	pages = {1614--1632},
	title = {A Cosmic {{UV}}/{{X-ray}} Background Model Update},
	urldate = {2026-04-07},
	volume = {493},
	year = 2020}

@article{huangCompleteCensusCircumgalactic2021,
	author = {Huang, Yun-Hsin and Chen, Hsiao-Wen and Shectman, Stephen A. and Johnson, Sean D. and Zahedy, Fakhri S. and Helsby, Jennifer E. and Gauthier, Jean-Ren{\'e} and Thompson, Ian B.},
	doi = {10.1093/mnras/stab360},
	issn = {0035-8711},
	journal = {Monthly Notices of the Royal Astronomical Society, Volume 502, Issue 4, pp.4743-4761},
	langid = {english},
	month = apr,
	number = {4},
	pages = {4743},
	title = {A Complete Census of Circumgalactic {{Mg II}} at Redshift z {$\lessequivlnt$} 0.5},
	urldate = {2026-03-31},
	volume = {502},
	year = 2021}

@article{Lin25,
	adsurl = {https://ui.adsabs.harvard.edu/abs/2025ApJ...982..151L},
	archiveprefix = {arXiv},
	author = {{Lin}, Xuchen and {Wang}, Jing and {Staveley-Smith}, Lister and {Ji}, Suoqing and {Yang}, Dong and {Chen}, Xinkai and {Walter}, Fabian and {Chen}, Hsiao-Wen and {Ho}, Luis C. and {Jiang}, Peng and {Mandelker}, Nir and {Oh}, Se-Heon and {Peng}, Bo and {P{\'e}roux}, C{\'e}line and {Qu}, Zhijie and {Wang}, Q. Daniel},
	doi = {10.3847/1538-4357/adb718},
	eid = {151},
	eprint = {2502.10672},
	journal = {\apj},
	month = apr,
	number = {2},
	pages = {151},
	primaryclass = {astro-ph.GA},
	title = {{FEASTS Combined with Interferometry. III. The Low Column Density H I Around M51 and Possibility of Turbulent-mixing Gas Accretion}},
	volume = {982},
	year = 2025}

@article{Sharma25,
	adsurl = {https://ui.adsabs.harvard.edu/abs/2025arXiv250903802S},
	archiveprefix = {arXiv},
	author = {{Sharma}, Prateek and {Kumar}, Arnav and {Datta}, Dipayan and {Babul}, Arif and {Das}, Rishita and {Aditya}, Konduri},
	doi = {10.48550/arXiv.2509.03802},
	eid = {arXiv:2509.03802},
	eprint = {2509.03802},
	journal = {arXiv e-prints},
	month = sep,
	pages = {arXiv:2509.03802},
	primaryclass = {astro-ph.GA},
	title = {{Universal Structure of Turbulent Radiative Mixing Layers}},
	year = 2025}

@article{binneyDiscDistortionRevisited2024,
	author = {Binney, James},
	doi = {10.1093/mnras/stae2481},
	issn = {0035-8711},
	journal = {Monthly Notices of the Royal Astronomical Society},
	langid = {english},
	month = dec,
	number = {2},
	pages = {1898--1912},
	title = {Disc Distortion Revisited},
	urldate = {2026-03-26},
	volume = {535},
	year = 2024}

@article{trappFiguringOutGas2025c,
	author = {Trapp, Cameron W. and Peeples, Molly S. and Tumlinson, Jason and O'Shea, Brian W. and Wright, Anna C. and Acharyya, Ayan and Smith, Britton D. and Saeedzadeh, Vida and Augustin, Ramona},
	doi = {10.48550/arXiv.2511.00159},
	journal = {arXiv e-prints},
	langid = {english},
	month = oct,
	pages = {arXiv:2511.00159},
	title = {Figuring {{Out Gas}} \&amp; {{Galaxies In Enzo}} ({{FOGGIE}}). {{XIII}}. {{On}} the {{Observability}} of {{Extended HI Disks}} and {{Warps}}},
	urldate = {2026-03-26},
	year = 2025}

@article{zuletaKinematicalSignaturesDistinguishing2024,
	author = {Zuleta, A. and Birnstiel, T. and Teague, R.},
	doi = {10.1051/0004-6361/202451145},
	issn = {0004-6361},
	journal = {Astronomy \&amp; Astrophysics, Volume 692, id.A56, 14 pp.},
	langid = {english},
	month = dec,
	pages = {A56},
	shorttitle = {Kinematical Signatures},
	title = {Kinematical Signatures: {{Distinguishing}} between Warps and Radial Flows},
	urldate = {2025-05-21},
	volume = {692},
	year = 2024}

@article{velliscigAlignmentShapeDark2015,
	author = {Velliscig, Marco and Cacciato, Marcello and Schaye, Joop and Crain, Robert A. and Bower, Richard G. and {van Daalen}, Marcel P. and Dalla Vecchia, Claudio and Frenk, Carlos S. and Furlong, Michelle and McCarthy, I. G. and Schaller, Matthieu and Theuns, Tom},
	doi = {10.1093/mnras/stv1690},
	issn = {0035-8711},
	journal = {Monthly Notices of the Royal Astronomical Society},
	langid = {english},
	month = oct,
	number = {1},
	pages = {721--738},
	title = {The Alignment and Shape of Dark Matter, Stellar, and Hot Gas Distributions in the {{EAGLE}} and Cosmo-{{OWLS}} Simulations},
	urldate = {2026-03-24},
	volume = {453},
	year = 2015}

@misc{goldnerAccretionDrivenTurbulenceCircumgalactic2025,
	archiveprefix = {arXiv},
	author = {Goldner, Roy and Stern, Jonathan and Fielding, Drummond and {Faucher-Gigu{\`e}re}, Claude-Andr{\'e} and Faerman, Yakov and Kakoly, Aharon},
	doi = {10.48550/arXiv.2510.27678},
	eprint = {2510.27678},
	month = oct,
	number = {arXiv:2510.27678},
	primaryclass = {astro-ph},
	publisher = {arXiv},
	title = {Accretion-{{Driven Turbulence}} in the {{Circumgalactic Medium}}},
	urldate = {2025-11-05},
	year = 2025}

@article{wangFEASTSRadialDistribution2025,
	author = {Wang, Jing and Yang, Dong and Lin, Xuchen and Huang, Qifeng and Qu, Zhijie and Chen, Hsiao-wen and Guo, Hong and Ho, Luis C. and Jiang, Peng and Liang, Zezhong and P{\'e}roux, C{\'e}line and {Staveley-Smith}, Lister and Weng, Simon},
	doi = {10.3847/1538-4357/ada95a},
	issn = {0004-637X},
	journal = {The Astrophysical Journal},
	langid = {english},
	month = feb,
	number = {1},
	pages = {25},
	shorttitle = {{{FEASTS}}},
	title = {{{FEASTS}}: {{Radial Distribution}} of {{H I Surface Densities Down}} to 0.01 {{M}}{\textsubscript{{$\odot$}}} Pc{\textsuperscript{-2}} of 35 {{Nearby Galaxies}}},
	urldate = {2025-04-26},
	volume = {980},
	year = 2025}

@article{wangObservationalTheoreticalView2014,
	author = {Wang, Jing and Fu, Jian and Aumer, Michael and Kauffmann, Guinevere and J{\'o}zsa, Gyula I. G. and Serra, Paolo and Huang, Mei-ling and Brinchmann, Jarle and {van der Hulst}, Thijs and Bigiel, Frank},
	doi = {10.1093/mnras/stu649},
	issn = {0035-8711},
	journal = {Monthly Notices of the Royal Astronomical Society, Volume 441, Issue 3, p.2159-2172},
	langid = {english},
	month = jul,
	number = {3},
	pages = {2159},
	title = {An Observational and Theoretical View of the Radial Distribution of {{H I}} Gas in Galaxies},
	urldate = {2025-05-21},
	volume = {441},
	year = 2014}

@article{pezzulliAngularMomentumCosmological2017,
	author = {Pezzulli, Gabriele and Fraternali, Filippo and Binney, James},
	doi = {10.1093/mnras/stx029},
	issn = {0035-8711},
	journal = {Monthly Notices of the Royal Astronomical Society},
	langid = {english},
	month = may,
	number = {1},
	pages = {311--329},
	title = {The Angular Momentum of Cosmological Coronae and the Inside-out Growth of Spiral Galaxies},
	urldate = {2025-12-02},
	volume = {467},
	year = 2017}

@article{Kakoly25,
	adsurl = {https://ui.adsabs.harvard.edu/abs/2025MNRAS.543.3345K},
	archiveprefix = {arXiv},
	author = {{Kakoly}, Aharon and {Stern}, Jonathan and {Faucher-Gigu{\`e}re}, Claude-Andr{\'e} and {Fielding}, Drummond B. and {Goldner}, Roy and {Sun}, Guochao and {Hummels}, Cameron B.},
	doi = {10.1093/mnras/staf1516},
	eprint = {2504.17001},
	journal = {\mnras},
	month = nov,
	number = {4},
	pages = {3345-3366},
	primaryclass = {astro-ph.GA},
	title = {{Turbulence-dominated CGM: the origin of UV absorbers with equivalent widths of {\ensuremath{\sim}}1 {\r{A}}}},
	volume = {543},
	year = 2025}

@article{FGOh23,
	adsurl = {https://ui.adsabs.harvard.edu/abs/2023ARA&A..61..131F},
	archiveprefix = {arXiv},
	author = {{Faucher-Gigu{\`e}re}, Claude-Andr{\'e} and {Oh}, S. Peng},
	doi = {10.1146/annurev-astro-052920-125203},
	eprint = {2301.10253},
	journal = {\araa},
	month = aug,
	pages = {131-195},
	primaryclass = {astro-ph.GA},
	title = {{Key Physical Processes in the Circumgalactic Medium}},
	volume = {61},
	year = 2023}

@article{Diemer13,
	adsurl = {https://ui.adsabs.harvard.edu/abs/2013ApJ...766...25D},
	archiveprefix = {arXiv},
	author = {{Diemer}, Benedikt and {More}, Surhud and {Kravtsov}, Andrey V.},
	doi = {10.1088/0004-637X/766/1/25},
	eid = {25},
	eprint = {1207.0816},
	journal = {\apj},
	month = mar,
	number = {1},
	pages = {25},
	primaryclass = {astro-ph.CO},
	title = {{The Pseudo-evolution of Halo Mass}},
	volume = {766},
	year = 2013}

@article{moTwophaseModelGalaxy2024,
	author = {Mo, Houjun and Chen, Yangyao and Wang, Huiyuan},
	doi = {10.1093/mnras/stae1727},
	issn = {0035-8711},
	journal = {Monthly Notices of the Royal Astronomical Society},
	langid = {english},
	month = aug,
	number = {4},
	pages = {3808--3838},
	shorttitle = {A Two-Phase Model of Galaxy Formation},
	title = {A Two-Phase Model of Galaxy Formation: {{I}}. {{The}} Growth of Galaxies and Supermassive Black Holes},
	urldate = {2025-11-02},
	volume = {532},
	year = 2024}

@article{Carr23,
	adsurl = {https://ui.adsabs.harvard.edu/abs/2023ApJ...949...21C},
	archiveprefix = {arXiv},
	author = {{Carr}, Christopher and {Bryan}, Greg L. and {Fielding}, Drummond B. and {Pandya}, Viraj and {Somerville}, Rachel S.},
	doi = {10.3847/1538-4357/acc4c7},
	eid = {21},
	eprint = {2211.05115},
	journal = {\apj},
	month = may,
	number = {1},
	pages = {21},
	primaryclass = {astro-ph.GA},
	title = {{Regulation of Star Formation by a Hot Circumgalactic Medium}},
	volume = {949},
	year = 2023}

@article{Voit24b,
	adsurl = {https://ui.adsabs.harvard.edu/abs/2024ApJ...976..151V},
	archiveprefix = {arXiv},
	author = {{Voit}, G. Mark and {Carr}, Christopher and {Fielding}, Drummond B. and {Pandya}, Viraj and {Bryan}, Greg L. and {Donahue}, Megan and {Oppenheimer}, Benjamin D. and {Somerville}, Rachel S.},
	doi = {10.3847/1538-4357/ad81d5},
	eid = {151},
	eprint = {2406.07632},
	journal = {\apj},
	month = dec,
	number = {2},
	pages = {151},
	primaryclass = {astro-ph.GA},
	title = {{Equilibrium States of Galactic Atmospheres. II. Interpretation and Implications}},
	volume = {976},
	year = 2024}

@article{Voit24a,
	adsurl = {https://ui.adsabs.harvard.edu/abs/2024ApJ...976..150V},
	archiveprefix = {arXiv},
	author = {{Voit}, G. Mark and {Pandya}, Viraj and {Fielding}, Drummond B. and {Bryan}, Greg L. and {Carr}, Christopher and {Donahue}, Megan and {Oppenheimer}, Benjamin D. and {Somerville}, Rachel S.},
	doi = {10.3847/1538-4357/ad81d6},
	eid = {150},
	eprint = {2406.07631},
	journal = {\apj},
	month = dec,
	number = {2},
	pages = {150},
	primaryclass = {astro-ph.GA},
	title = {{Equilibrium States of Galactic Atmospheres. I. The Flip Side of Mass Loading}},
	volume = {976},
	year = 2024}

@article{Sun25,
	adsurl = {https://ui.adsabs.harvard.edu/abs/2025arXiv250804768S},
	archiveprefix = {arXiv},
	author = {{Sun}, Guochao and {Faucher-Gigu{\`e}re}, Claude-Andr{\'e} and {Stern}, Jonathan},
	eid = {arXiv:2508.04768},
	eprint = {2508.04768},
	journal = {arXiv e-prints},
	month = aug,
	pages = {arXiv:2508.04768},
	primaryclass = {astro-ph.GA},
	title = {{A Turbulent Framework for Star Formation in High-Redshift Galaxies}},
	year = 2025}

@article{Ginzburg22,
	adsurl = {https://ui.adsabs.harvard.edu/abs/2022MNRAS.513.6177G},
	archiveprefix = {arXiv},
	author = {{Ginzburg}, Omri and {Dekel}, Avishal and {Mandelker}, Nir and {Krumholz}, Mark R.},
	doi = {10.1093/mnras/stac1324},
	eprint = {2202.12331},
	journal = {\mnras},
	month = jul,
	number = {4},
	pages = {6177-6195},
	primaryclass = {astro-ph.GA},
	title = {{The evolution of turbulent galactic discs: gravitational instability, feedback, and accretion}},
	volume = {513},
	year = 2022}

@article{bland-hawthornTurbulentGasrichDiscs2025,
	author = {{Bland-Hawthorn}, Joss and {Tepper-Garcia}, Thor and Agertz, Oscar and Federrath, Christoph and Haywood, Misha and {di Matteo}, Paola and Bedding, Timothy R. and Tsukui, Takafumi and Wisnioski, Emily and Ness, Melissa and Freeman, Ken},
	doi = {10.48550/arXiv.2502.01895},
	journal = {arXiv e-prints},
	langid = {english},
	month = feb,
	pages = {arXiv:2502.01895},
	shorttitle = {Turbulent Gas-Rich Discs at High Redshift},
	title = {Turbulent Gas-Rich Discs at High Redshift: Origin of Thick Stellar Discs through {{3D}} 'Baryon Sloshing'},
	urldate = {2025-10-26},
	year = 2025}

@article{bland-hawthornTurbulentGasrichDisks2024,
	author = {{Bland-Hawthorn}, Joss and {Tepper-Garcia}, Thor and Agertz, Oscar and Federrath, Christoph},
	doi = {10.3847/1538-4357/ad4118},
	issn = {0004-637X},
	journal = {The Astrophysical Journal},
	langid = {english},
	month = jun,
	number = {2},
	pages = {86},
	shorttitle = {Turbulent {{Gas-rich Disks}} at {{High Redshift}}},
	title = {Turbulent {{Gas-rich Disks}} at {{High Redshift}}: {{Bars}} and {{Bulges}} in a {{Radial Shear Flow}}},
	urldate = {2025-10-26},
	volume = {968},
	year = 2024}

@article{schechterNGC2685Spindle1978,
	author = {Schechter, P. L. and Gunn, J. E.},
	doi = {10.1086/112324},
	issn = {0004-6256},
	journal = {Astronomical Journal, Vol. 83, p. 1360-1362 (1978)},
	langid = {english},
	month = nov,
	pages = {1360},
	shorttitle = {{{NGC}} 2685},
	title = {{{NGC}} 2685: Spindle or Pancake?},
	urldate = {2025-10-23},
	volume = {83},
	year = 1978}

@article{degWALLABYPilotSurvey2023,
	author = {Deg, N. and Palleske, R. and Spekkens, K. and Wang, J. and Jarrett, T. and English, J. and Lin, X. and Yeung, J. and Mould, J. R. and Catinella, B. and D{\'e}nes, H. and Elagali, A. and For, B.-Q. and Kamphuis, P. and Koribalski, B. S. and {Lee-Waddell}, K. and Murugeshan, C. and Oh, S. and Rhee, J. and Serra, P. and Westmeier, T. and Wong, O. I. and Bekki, K. and Bosma, A. and Carignan, C. and Holwerda, B. W. and Yu, N.},
	doi = {10.1093/mnras/stad2312},
	issn = {0035-8711},
	journal = {Monthly Notices of the Royal Astronomical Society},
	langid = {english},
	month = nov,
	number = {3},
	pages = {4663--4684},
	shorttitle = {{{WALLABY}} Pilot Survey},
	title = {{{WALLABY}} Pilot Survey: The Potential Polar Ring Galaxies {{NGC}} 4632 and {{NGC}} 6156},
	urldate = {2025-10-23},
	volume = {525},
	year = 2023}

@article{mosenkovOccurrenceRateGalaxies2024,
	author = {Mosenkov, A. V. and Bahr, S. K. H. and Reshetnikov, V. P. and Shakespear, Z. and Smirnov, D. V.},
	doi = {10.1051/0004-6361/202348494},
	issn = {0004-6361},
	journal = {Astronomy \&amp; Astrophysics, Volume 681, id.L15, 5 pp.},
	langid = {english},
	month = jan,
	pages = {L15},
	title = {The Occurrence Rate of Galaxies with Polar Structures May Be Significantly Underestimated},
	urldate = {2025-10-23},
	volume = {681},
	year = 2024}

@article{bland-hawthornGalacticSeismologyEvolving2021,
	author = {{Bland-Hawthorn}, Joss and {Tepper-Garc{\'\i}a}, Thor},
	doi = {10.1093/mnras/stab704},
	issn = {0035-8711},
	journal = {Monthly Notices of the Royal Astronomical Society},
	langid = {english},
	month = jul,
	number = {3},
	pages = {3168--3186},
	shorttitle = {Galactic Seismology},
	title = {Galactic Seismology: The Evolving 'phase Spiral' after the {{Sagittarius}} Dwarf Impact},
	urldate = {2025-10-23},
	volume = {504},
	year = 2021}

@article{Pandya23,
	adsurl = {https://ui.adsabs.harvard.edu/abs/2023ApJ...956..118P},
	archiveprefix = {arXiv},
	author = {{Pandya}, Viraj and {Fielding}, Drummond B. and {Bryan}, Greg L. and {Carr}, Christopher and {Somerville}, Rachel S. and {Stern}, Jonathan and {Faucher-Gigu{\`e}re}, Claude-Andr{\'e} and {Hafen}, Zachary and {Angl{\'e}s-Alc{\'a}zar}, Daniel and {Forbes}, John C.},
	doi = {10.3847/1538-4357/acf3ea},
	eid = {118},
	eprint = {2211.09755},
	journal = {\apj},
	month = oct,
	number = {2},
	pages = {118},
	primaryclass = {astro-ph.GA},
	title = {{A Unified Model for the Coevolution of Galaxies and Their Circumgalactic Medium: The Relative Roles of Turbulence and Atomic Cooling Physics}},
	volume = {956},
	year = 2023}

@article{McCourt12,
	adsurl = {https://ui.adsabs.harvard.edu/abs/2012MNRAS.419.3319M},
	archiveprefix = {arXiv},
	author = {{McCourt}, Michael and {Sharma}, Prateek and {Quataert}, Eliot and {Parrish}, Ian J.},
	doi = {10.1111/j.1365-2966.2011.19972.x},
	eprint = {1105.2563},
	journal = {\mnras},
	month = feb,
	number = {4},
	pages = {3319-3337},
	primaryclass = {astro-ph.CO},
	title = {{Thermal instability in gravitationally stratified plasmas: implications for multiphase structure in clusters and galaxy haloes}},
	volume = {419},
	year = 2012}

@article{Tan23,
	adsurl = {https://ui.adsabs.harvard.edu/abs/2023MNRAS.520.2571T},
	archiveprefix = {arXiv},
	author = {{Tan}, Brent and {Oh}, S. Peng and {Gronke}, Max},
	doi = {10.1093/mnras/stad236},
	eprint = {2210.06493},
	journal = {\mnras},
	month = apr,
	number = {2},
	pages = {2571-2592},
	primaryclass = {astro-ph.GA},
	title = {{Cloudy with a chance of rain: accretion braking of cold clouds}},
	volume = {520},
	year = 2023}

@article{Heitsch09,
	adsurl = {https://ui.adsabs.harvard.edu/abs/2009ApJ...698.1485H},
	archiveprefix = {arXiv},
	author = {{Heitsch}, Fabian and {Putman}, Mary E.},
	doi = {10.1088/0004-637X/698/2/148510.48550/arXiv.0904.1995},
	eprint = {0904.1995},
	journal = {\apj},
	month = jun,
	number = {2},
	pages = {1485-1496},
	primaryclass = {astro-ph.GA},
	title = {{The Fate of High-Velocity Clouds: Warm or Cold Cosmic Rain?}},
	volume = {698},
	year = 2009}

@article{vanderKruit01,
	adsurl = {https://ui.adsabs.harvard.edu/abs/2001A&A...379..374V},
	archiveprefix = {arXiv},
	author = {{van der Kruit}, P.~C. and {Jim{\'e}nez-Vicente}, J. and {Kregel}, M. and {Freeman}, K.~C.},
	doi = {10.1051/0004-6361:20011311},
	eprint = {astro-ph/0109477},
	journal = {\aap},
	month = nov,
	pages = {374-383},
	primaryclass = {astro-ph},
	title = {{Kinematics and dynamics of the ``superthin'' edge-on disk galaxy IC 5249}},
	volume = {379},
	year = 2001}

@article{Sarkar24,
	adsurl = {https://ui.adsabs.harvard.edu/abs/2024A&ARv..32....1S},
	archiveprefix = {arXiv},
	author = {{Sarkar}, Kartick C.},
	doi = {10.1007/s00159-024-00152-1},
	eid = {1},
	eprint = {2403.09824},
	journal = {\aapr},
	month = mar,
	number = {1},
	pages = {1},
	primaryclass = {astro-ph.HE},
	title = {{The Fermi/eROSITA bubbles: a look into the nuclear outflow from the Milky Way}},
	volume = {32},
	year = 2024}

@article{Truong21,
	adsurl = {https://ui.adsabs.harvard.edu/abs/2021MNRAS.508.1563T},
	archiveprefix = {arXiv},
	author = {{Truong}, Nhut and {Pillepich}, Annalisa and {Nelson}, Dylan and {Werner}, Norbert and {Hernquist}, Lars},
	doi = {10.1093/mnras/stab2638},
	eprint = {2109.06884},
	journal = {\mnras},
	month = dec,
	number = {2},
	pages = {1563-1581},
	primaryclass = {astro-ph.GA},
	title = {{Predictions for anisotropic X-ray signatures in the circumgalactic medium: imprints of supermassive black hole driven outflows}},
	volume = {508},
	year = 2021}

@article{Lan18,
	author = {Ting-Wen Lan and Houjun Mo},
	doi = {https://doi.org/10.3847/1538-4357/aadc08},
	journal = {The Astrophysical Journal, 866, 36},
	note = {14 pages, 8 figures, matched the accepted version},
	title = {The Circumgalactic Medium of eBOSS Emission Line Galaxies: Signatures of Galactic Outflows in Gas Distribution and Kinematics},
	year = {2018}}

@article{Sormani18,
	adsurl = {https://ui.adsabs.harvard.edu/abs/2018MNRAS.481.3370S},
	archiveprefix = {arXiv},
	author = {{Sormani}, Mattia C. and {Sobacchi}, Emanuele and {Pezzulli}, Gabriele and {Binney}, James and {Klessen}, Ralf S.},
	doi = {10.1093/mnras/sty2500},
	eprint = {1809.03437},
	journal = {\mnras},
	month = dec,
	number = {3},
	pages = {3370-3381},
	primaryclass = {astro-ph.GA},
	title = {{Models of rotating coronae}},
	volume = {481},
	year = 2018}

@article{Birnboim03,
	adsurl = {https://ui.adsabs.harvard.edu/abs/2003MNRAS.345..349B},
	archiveprefix = {arXiv},
	author = {{Birnboim}, Yuval and {Dekel}, Avishai},
	doi = {10.1046/j.1365-8711.2003.06955.x},
	eprint = {astro-ph/0302161},
	journal = {\mnras},
	month = oct,
	number = {1},
	pages = {349-364},
	primaryclass = {astro-ph},
	title = {{Virial shocks in galactic haloes?}},
	volume = {345},
	year = 2003}

@article{White78,
	adsurl = {https://ui.adsabs.harvard.edu/abs/1978MNRAS.183..341W},
	author = {{White}, S.~D.~M. and {Rees}, M.~J.},
	doi = {10.1093/mnras/183.3.341},
	journal = {\mnras},
	month = may,
	pages = {341-358},
	title = {{Core condensation in heavy halos: a two-stage theory for galaxy formation and clustering.}},
	volume = {183},
	year = 1978}

@article{Fielding17,
	adsurl = {https://ui.adsabs.harvard.edu/abs/2017MNRAS.466.3810F},
	archiveprefix = {arXiv},
	author = {{Fielding}, Drummond and {Quataert}, Eliot and {McCourt}, Michael and {Thompson}, Todd A.},
	doi = {10.1093/mnras/stw3326},
	eprint = {1606.06734},
	journal = {\mnras},
	month = apr,
	number = {4},
	pages = {3810-3826},
	primaryclass = {astro-ph.GA},
	title = {{The impact of star formation feedback on the circumgalactic medium}},
	volume = {466},
	year = 2017}

@article{Qu18,
	adsurl = {https://ui.adsabs.harvard.edu/abs/2018ApJ...856....5Q},
	archiveprefix = {arXiv},
	author = {{Qu}, Zhijie and {Bregman}, Joel N.},
	doi = {10.3847/1538-4357/aaafd4},
	eid = {5},
	eprint = {1801.05833},
	journal = {\apj},
	month = mar,
	number = {1},
	pages = {5},
	primaryclass = {astro-ph.GA},
	title = {{The Mass and Absorption Columns of Galactic Gaseous Halos}},
	volume = {856},
	year = 2018}

@article{Faerman17,
	adsurl = {https://ui.adsabs.harvard.edu/abs/2017ApJ...835...52F},
	archiveprefix = {arXiv},
	author = {{Faerman}, Yakov and {Sternberg}, Amiel and {McKee}, Christopher F.},
	doi = {10.3847/1538-4357/835/1/52},
	eid = {52},
	eprint = {1602.00689},
	journal = {\apj},
	month = jan,
	number = {1},
	pages = {52},
	primaryclass = {astro-ph.GA},
	title = {{Massive Warm/Hot Galaxy Coronae as Probed by UV/X-Ray Oxygen Absorption and Emission. I. Basic Model}},
	volume = {835},
	year = 2017}

@article{Pilyugin19,
	adsurl = {https://ui.adsabs.harvard.edu/abs/2019A&A...623A.122P},
	archiveprefix = {arXiv},
	author = {{Pilyugin}, L.~S. and {Grebel}, E.~K. and {Zinchenko}, I.~A. and {Nefedyev}, Y.~A. and {V{\'\i}lchez}, J.~M.},
	doi = {10.1051/0004-6361/201834239},
	eid = {A122},
	eprint = {1901.11001},
	journal = {\aap},
	month = mar,
	pages = {A122},
	primaryclass = {astro-ph.GA},
	title = {{Relations between abundance characteristics and rotation velocity for star-forming MaNGA galaxies}},
	volume = {623},
	year = 2019}

@article{Stewart13,
	adsurl = {https://ui.adsabs.harvard.edu/abs/2013ApJ...769...74S},
	archiveprefix = {arXiv},
	author = {{Stewart}, Kyle R. and {Brooks}, Alyson M. and {Bullock}, James S. and {Maller}, Ariyeh H. and {Diemand}, J{\"u}rg and {Wadsley}, James and {Moustakas}, Leonidas A.},
	doi = {10.1088/0004-637X/769/1/74},
	eid = {74},
	eprint = {1301.3143},
	journal = {\apj},
	month = may,
	number = {1},
	pages = {74},
	primaryclass = {astro-ph.CO},
	title = {{Angular Momentum Acquisition in Galaxy Halos}},
	volume = {769},
	year = 2013}

@article{Bigiel10,
	adsurl = {https://ui.adsabs.harvard.edu/abs/2010AJ....140.1194B},
	archiveprefix = {arXiv},
	author = {{Bigiel}, F. and {Leroy}, A. and {Walter}, F. and {Blitz}, L. and {Brinks}, E. and {de Blok}, W.~J.~G. and {Madore}, B.},
	doi = {10.1088/0004-6256/140/5/1194},
	eprint = {1007.3498},
	journal = {\aj},
	month = nov,
	number = {5},
	pages = {1194-1213},
	primaryclass = {astro-ph.CO},
	title = {{Extremely Inefficient Star Formation in the Outer Disks of Nearby Galaxies}},
	volume = {140},
	year = 2010}

@article{Bregman18,
	adsurl = {https://ui.adsabs.harvard.edu/abs/2018ApJ...862....3B},
	archiveprefix = {arXiv},
	author = {{Bregman}, Joel N. and {Anderson}, Michael E. and {Miller}, Matthew J. and {Hodges-Kluck}, Edmund and {Dai}, Xinyu and {Li}, Jiang-Tao and {Li}, Yunyang and {Qu}, Zhijie},
	doi = {10.3847/1538-4357/aacafe},
	eid = {3},
	eprint = {1803.08963},
	journal = {\apj},
	month = jul,
	number = {1},
	pages = {3},
	primaryclass = {astro-ph.GA},
	title = {{The Extended Distribution of Baryons around Galaxies}},
	volume = {862},
	year = 2018}

@article{Wenger19,
	adsurl = {https://ui.adsabs.harvard.edu/abs/2019ApJ...887..114W},
	archiveprefix = {arXiv},
	author = {{Wenger}, Trey V. and {Balser}, Dana S. and {Anderson}, L.~D. and {Bania}, T.~M.},
	doi = {10.3847/1538-4357/ab53d3},
	eid = {114},
	eprint = {1910.14605},
	journal = {\apj},
	month = dec,
	number = {2},
	pages = {114},
	primaryclass = {astro-ph.GA},
	title = {{Metallicity Structure in the Milky Way Disk Revealed by Galactic H II Regions}},
	volume = {887},
	year = 2019}

@article{Broeils97,
	adsurl = {https://ui.adsabs.harvard.edu/abs/1997A&A...324..877B},
	author = {{Broeils}, A.~H. and {Rhee}, M. -H.},
	journal = {\aap},
	month = aug,
	pages = {877-887},
	title = {{Short 21-cm WSRT observations of spiral and irregular galaxies. HI properties.}},
	volume = {324},
	year = 1997}

@article{Wijers24,
	adsurl = {https://ui.adsabs.harvard.edu/abs/2024ApJ...973...99W},
	archiveprefix = {arXiv},
	author = {{Wijers}, Nastasha A. and {Faucher-Gigu{\`e}re}, Claude-Andr{\'e} and {Stern}, Jonathan and {Byrne}, Lindsey and {Sultan}, Imran},
	doi = {10.3847/1538-4357/ad63a0},
	eid = {99},
	eprint = {2401.08776},
	journal = {\apj},
	month = oct,
	number = {2},
	pages = {99},
	primaryclass = {astro-ph.GA},
	title = {{Ne VIII in the Warm-hot Circumgalactic Medium of FIRE Simulations and in Observations}},
	volume = {973},
	year = 2024}

@article{Sultan25,
	adsurl = {https://ui.adsabs.harvard.edu/abs/2025MNRAS.540.1017S},
	archiveprefix = {arXiv},
	author = {{Sultan}, Imran and {Faucher-Gigu{\`e}re}, Claude-Andr{\'e} and {Stern}, Jonathan and {Rotshtein}, Shaked and {Byrne}, Lindsey and {Wijers}, Nastasha},
	doi = {10.1093/mnras/staf786},
	eprint = {2410.16359},
	journal = {\mnras},
	month = jun,
	number = {1},
	pages = {1017-1041},
	primaryclass = {astro-ph.GA},
	title = {{Cooling flows as a reference solution for the hot circumgalactic medium}},
	volume = {540},
	year = 2025}

@article{Teklu15,
	adsurl = {https://ui.adsabs.harvard.edu/abs/2015ApJ...812...29T},
	archiveprefix = {arXiv},
	author = {{Teklu}, Adelheid F. and {Remus}, Rhea-Silvia and {Dolag}, Klaus and {Beck}, Alexander M. and {Burkert}, Andreas and {Schmidt}, Andreas S. and {Schulze}, Felix and {Steinborn}, Lisa K.},
	doi = {10.1088/0004-637X/812/1/29},
	eid = {29},
	eprint = {1503.03501},
	journal = {\apj},
	month = oct,
	number = {1},
	pages = {29},
	primaryclass = {astro-ph.GA},
	title = {{Connecting Angular Momentum and Galactic Dynamics: The Complex Interplay between Spin, Mass, and Morphology}},
	volume = {812},
	year = 2015}

@article{Oppenheimer18,
	adsurl = {https://ui.adsabs.harvard.edu/abs/2018MNRAS.480.2963O},
	archiveprefix = {arXiv},
	author = {{Oppenheimer}, Benjamin D.},
	doi = {10.1093/mnras/sty1918},
	eprint = {1801.00788},
	journal = {\mnras},
	month = nov,
	number = {3},
	pages = {2963-2975},
	primaryclass = {astro-ph.GA},
	title = {{Deviations from hydrostatic equilibrium in the circumgalactic medium: spinning hot haloes and accelerating flows}},
	volume = {480},
	year = 2018}

@article{Semenov24,
	adsurl = {https://ui.adsabs.harvard.edu/abs/2024ApJ...972...73S},
	archiveprefix = {arXiv},
	author = {{Semenov}, Vadim A. and {Conroy}, Charlie and {Chandra}, Vedant and {Hernquist}, Lars and {Nelson}, Dylan},
	doi = {10.3847/1538-4357/ad57ba},
	eid = {73},
	eprint = {2306.13125},
	journal = {\apj},
	month = sep,
	number = {1},
	pages = {73},
	primaryclass = {astro-ph.GA},
	title = {{Formation of Galactic Disks. II. The Physical Drivers of Disk Spin-up}},
	volume = {972},
	year = 2024}

@article{Huscher21,
	adsurl = {https://ui.adsabs.harvard.edu/abs/2021MNRAS.500.1476H},
	archiveprefix = {arXiv},
	author = {{Huscher}, Ezra and {Oppenheimer}, Benjamin D. and {Lonardi}, Alice and {Crain}, Robert A. and {Richings}, Alexander J. and {Schaye}, Joop},
	doi = {10.1093/mnras/staa3203},
	eprint = {2005.06310},
	journal = {\mnras},
	month = jan,
	number = {1},
	pages = {1476-1490},
	primaryclass = {astro-ph.GA},
	title = {{The changing circumgalactic medium over the last 10 Gyr - I. Physical and dynamical properties}},
	volume = {500},
	year = 2021}

@article{Zjupa17,
	adsurl = {https://ui.adsabs.harvard.edu/abs/2017MNRAS.466.1625Z},
	archiveprefix = {arXiv},
	author = {{Zjupa}, Jolanta and {Springel}, Volker},
	doi = {10.1093/mnras/stw2945},
	eprint = {1608.01323},
	journal = {\mnras},
	month = apr,
	number = {2},
	pages = {1625-1647},
	primaryclass = {astro-ph.CO},
	title = {{Angular momentum properties of haloes and their baryon content in the Illustris simulation}},
	volume = {466},
	year = 2017}

@article{comeronBreaksThinThick2012,
	author = {Comer{\'o}n, S{\'e}bastien and Elmegreen, Bruce G. and Salo, Heikki and Laurikainen, Eija and Athanassoula, E. and Bosma, Albert and Knapen, Johan H. and Gadotti, Dimitri A. and Sheth, Kartik and Hinz, Joannah L. and Regan, Michael W. and {Gil de Paz}, Armando and {Mu{\~n}oz-Mateos}, Juan Carlos and {Men{\'e}ndez-Delmestre}, Kar{\'\i}n and Seibert, Mark and Kim, Taehyun and Mizusawa, Trisha and Laine, Jarkko and Ho, Luis C. and Holwerda, Benne},
	doi = {10.1088/0004-637X/759/2/98},
	issn = {0004-637X},
	journal = {The Astrophysical Journal, Volume 759, Issue 2, article id. 98, 29 pp. (2012).},
	langid = {english},
	month = nov,
	number = {2},
	pages = {98},
	title = {Breaks in {{Thin}} and {{Thick Disks}} of {{Edge-on Galaxies Imaged}} in the {{Spitzer Survey Stellar Structure}} in {{Galaxies}} ({{S}}{\textsuperscript{4}}{{G}})},
	urldate = {2025-09-16},
	volume = {759},
	year = {2012}}

@article{kregelFlatteningTruncationStellar2002,
	author = {Kregel, Michiel and {van der Kruit}, Pieter C. and {de Grijs}, Richard},
	doi = {10.1046/j.1365-8711.2002.05556.x},
	issn = {0035-8711},
	journal = {Monthly Notices of the Royal Astronomical Society, Volume 334, Issue 3, pp. 646-668.},
	langid = {english},
	month = aug,
	number = {3},
	pages = {646},
	title = {Flattening and Truncation of Stellar Discs in Edge-on Spiral Galaxies},
	urldate = {2025-09-16},
	volume = {334},
	year = {2002}}

@article{martin-navarroUnifiedPictureBreaks2012,
	author = {{Mart{\'\i}n-Navarro}, Ignacio and Bakos, Judit and Trujillo, Ignacio and Knapen, Johan H. and Athanassoula, E. and Bosma, Albert and Comer{\'o}n, S{\'e}bastien and Elmegreen, Bruce G. and {Erroz-Ferrer}, Santiago and Gadotti, Dimitri A. and {Gil de Paz}, Armando and Hinz, Joannah L. and Ho, Luis C. and Holwerda, Benne W. and Kim, Taehyun and Laine, Jarkko and Laurikainen, Eija and {Men{\'e}ndez-Delmestre}, Kar{\'\i}n and Mizusawa, Trisha and {Mu{\~n}oz-Mateos}, Juan-Carlos and Regan, Michael W. and Salo, Heikki and Seibert, Mark and Sheth, Kartik},
	doi = {10.1111/j.1365-2966.2012.21929.x},
	issn = {0035-8711},
	journal = {Monthly Notices of the Royal Astronomical Society, Volume 427, Issue 2, pp. 1102-1134.},
	langid = {english},
	month = dec,
	number = {2},
	pages = {1102},
	title = {A Unified Picture of Breaks and Truncations in Spiral Galaxies from {{SDSS}} and {{S}}{\textsuperscript{4}}{{G}} Imaging},
	urldate = {2025-09-16},
	volume = {427},
	year = {2012}}

@article{wakkerDistancesGalacticHighVelocity2007,
	author = {Wakker, B. P. and York, D. G. and Howk, J. C. and Barentine, J. C. and Wilhelm, R. and Peletier, R. F. and {van Woerden}, H. and Beers, T. C. and Ivezi{\'c}, {\v Z} and Richter, P. and Schwarz, U. J.},
	doi = {10.1086/524222},
	issn = {0004-637X},
	journal = {The Astrophysical Journal, Volume 670, Issue 2, pp. L113-L116.},
	langid = {english},
	month = dec,
	number = {2},
	pages = {L113},
	shorttitle = {Distances to {{Galactic High-Velocity Clouds}}},
	title = {Distances to {{Galactic High-Velocity Clouds}}: {{Complex C}}},
	urldate = {2025-09-16},
	volume = {670},
	year = {2007}}

@article{mo98,
	adsurl = {https://ui.adsabs.harvard.edu/abs/1998MNRAS.295..319M},
	archiveprefix = {arXiv},
	author = {{Mo}, H.~J. and {Mao}, Shude and {White}, Simon D.~M.},
	doi = {10.1046/j.1365-8711.1998.01227.x},
	eprint = {astro-ph/9707093},
	journal = {\mnras},
	month = apr,
	number = {2},
	pages = {319-336},
	primaryclass = {astro-ph},
	title = {{The formation of galactic discs}},
	volume = {295},
	year = 1998}

@article{Fabian84,
	adsurl = {https://ui.adsabs.harvard.edu/abs/1984Natur.310..733F},
	author = {{Fabian}, A.~C. and {Nulsen}, P.~E.~J. and {Canizares}, C.~R.},
	doi = {10.1038/310733a0},
	journal = {\nat},
	month = aug,
	number = {5980},
	pages = {733-740},
	title = {{Cooling flows in clusters of galaxies}},
	volume = {310},
	year = 1984}

@article{Cowie80,
	adsurl = {https://ui.adsabs.harvard.edu/abs/1980MNRAS.191..399C},
	author = {{Cowie}, L.~L. and {Fabian}, A.~C. and {Nulsen}, P.~E.~J.},
	doi = {10.1093/mnras/191.2.399},
	journal = {\mnras},
	month = may,
	pages = {399-410},
	title = {{NGC 1275 and the Perseus cluster - The formation of optical filaments in cooling gas in X-ray clusters}},
	volume = {191},
	year = 1980}

@article{Binney92,
	adsurl = {https://ui.adsabs.harvard.edu/abs/1992ARA&A..30...51B},
	author = {{Binney}, James},
	doi = {10.1146/annurev.aa.30.090192.000411},
	journal = {\araa},
	month = jan,
	pages = {51-74},
	title = {{Warps.}},
	volume = {30},
	year = 1992}

@article{diTeodoro21,
	adsurl = {https://ui.adsabs.harvard.edu/abs/2021ApJ...923..220D},
	archiveprefix = {arXiv},
	author = {{Di Teodoro}, Enrico M. and {Peek}, J.~E.~G.},
	doi = {10.3847/1538-4357/ac2cbd},
	eid = {220},
	eprint = {2110.01618},
	journal = {\apj},
	month = dec,
	number = {2},
	pages = {220},
	primaryclass = {astro-ph.GA},
	title = {{Radial Motions and Radial Gas Flows in Local Spiral Galaxies}},
	volume = {923},
	year = 2021}

@article{Wang23,
	adsurl = {https://ui.adsabs.harvard.edu/abs/2023ApJ...944..143W},
	archiveprefix = {arXiv},
	author = {{Wang}, Enci and {Lilly}, Simon J.},
	doi = {10.3847/1538-4357/acaf31},
	eid = {143},
	eprint = {2205.04215},
	journal = {\apj},
	month = feb,
	number = {2},
	pages = {143},
	primaryclass = {astro-ph.GA},
	title = {{Similar Signatures of Coplanar Gas Inflow and Disk Warps in Galactic Gas Kinematic Maps}},
	volume = {944},
	year = 2023}

@article{Wang22b,
	adsurl = {https://ui.adsabs.harvard.edu/abs/2022ApJ...929...95W},
	archiveprefix = {arXiv},
	author = {{Wang}, Enci and {Lilly}, Simon J.},
	doi = {10.3847/1538-4357/ac5e31},
	eid = {95},
	eprint = {2201.04151},
	journal = {\apj},
	month = apr,
	number = {1},
	pages = {95},
	primaryclass = {astro-ph.GA},
	title = {{Gas-phase Metallicity Profiles of Star-forming Galaxies in the Modified Accretion Disk Framework}},
	volume = {929},
	year = 2022}

@article{Wang22a,
	adsurl = {https://ui.adsabs.harvard.edu/abs/2022ApJ...927..217W},
	archiveprefix = {arXiv},
	author = {{Wang}, Enci and {Lilly}, Simon J.},
	doi = {10.3847/1538-4357/ac49ed},
	eid = {217},
	eprint = {2201.04148},
	journal = {\apj},
	month = mar,
	number = {2},
	pages = {217},
	primaryclass = {astro-ph.GA},
	title = {{The Origin of Exponential Star-forming Disks}},
	volume = {927},
	year = 2022}

@article{Lehner22,
	adsurl = {https://ui.adsabs.harvard.edu/abs/2022MNRAS.513.3228L},
	archiveprefix = {arXiv},
	author = {{Lehner}, Nicolas and {Howk}, J. Christopher and {Marasco}, Antonino and {Fraternali}, Filippo},
	doi = {10.1093/mnras/stac987},
	eprint = {2202.05848},
	journal = {\mnras},
	month = jul,
	number = {3},
	pages = {3228-3240},
	primaryclass = {astro-ph.GA},
	title = {{Intermediate- and high-velocity clouds in the Milky Way - I. Covering factors and vertical heights}},
	volume = {513},
	year = 2022}

@article{Marasco25,
	adsurl = {https://ui.adsabs.harvard.edu/abs/2025A&A...697A..86M},
	archiveprefix = {arXiv},
	author = {{Marasco}, A. and {de Blok}, W.~J.~G. and {Maccagni}, F.~M. and {Fraternali}, F. and {Oman}, K.~A. and {Oosterloo}, T. and {Combes}, F. and {McGaugh}, S.~S. and {Kamphuis}, P. and {Spekkens}, K. and {Kleiner}, D. and {Veronese}, S. and {Amram}, P. and {Chemin}, L. and {Brinks}, E.},
	doi = {10.1051/0004-6361/202453172},
	eid = {A86},
	eprint = {2503.03818},
	journal = {\aap},
	month = may,
	pages = {A86},
	primaryclass = {astro-ph.GA},
	title = {{HI within and around observed and simulated galaxy discs: Comparing MeerKAT observations with mock data from TNG50 and FIRE-2}},
	volume = {697},
	year = 2025}

@article{Veronese25,
	adsurl = {https://ui.adsabs.harvard.edu/abs/2025A&A...693A..97V},
	archiveprefix = {arXiv},
	author = {{Veronese}, S. and {de Blok}, W.~J.~G. and {Healy}, J. and {Kleiner}, D. and {Marasco}, A. and {Maccagni}, F.~M. and {Kamphuis}, P. and {Brinks}, E. and {Holwerda}, B.~W. and {Zabel}, N. and {Chemin}, L. and {Adams}, E.~A.~K. and {Kurapati}, S. and {Sorgho}, A. and {Spekkens}, K. and {Combes}, F. and {Pisano}, D.~J. and {Walter}, F. and {Amram}, P. and {Bigiel}, F. and {Wong}, O.~I. and {Athanassoula}, E.},
	doi = {10.1051/0004-6361/202452085},
	eid = {A97},
	eprint = {2411.11584},
	journal = {\aap},
	month = jan,
	pages = {A97},
	primaryclass = {astro-ph.GA},
	title = {{Searching for HI around MHONGOOSE galaxies via spectral stacking}},
	volume = {693},
	year = 2025}

@article{vanderKruit07,
	adsurl = {https://ui.adsabs.harvard.edu/abs/2007A&A...466..883V},
	archiveprefix = {arXiv},
	author = {{van der Kruit}, P.~C.},
	doi = {10.1051/0004-6361:20066941},
	eprint = {astro-ph/0702486},
	journal = {\aap},
	month = may,
	number = {3},
	pages = {883-893},
	primaryclass = {astro-ph},
	title = {{Truncations of stellar disks and warps of HI-layers in edge-on spiral galaxies}},
	volume = {466},
	year = 2007}

@article{Ostriker89,
	adsurl = {https://ui.adsabs.harvard.edu/abs/1989MNRAS.237..785O},
	author = {{Ostriker}, E.~C. and {Binney}, J.~J.},
	doi = {10.1093/mnras/237.3.785},
	journal = {\mnras},
	month = apr,
	pages = {785-798},
	title = {{Warped and tilted galactic discs}},
	volume = {237},
	year = 1989}

@article{Trapp22,
	adsurl = {https://ui.adsabs.harvard.edu/abs/2022MNRAS.509.4149T},
	archiveprefix = {arXiv},
	author = {{Trapp}, Cameron W. and {Kere{\v{s}}}, Du{\v{s}}an and {Chan}, Tsang Keung and {Escala}, Ivanna and {Hummels}, Cameron and {Hopkins}, Philip F. and {Faucher-Gigu{\`e}re}, Claude-Andr{\'e} and {Murray}, Norman and {Quataert}, Eliot and {Wetzel}, Andrew},
	doi = {10.1093/mnras/stab3251},
	eprint = {2105.11472},
	journal = {\mnras},
	month = jan,
	number = {3},
	pages = {4149-4170},
	primaryclass = {astro-ph.GA},
	title = {{Gas infall and radial transport in cosmological simulations of milky way-mass discs}},
	volume = {509},
	year = 2022}

@article{Gurvich23,
	author = {Alexander B. Gurvich and Jonathan Stern and Claude-Andr{\'e} Faucher-Gigu{\`e}re and Philip F. Hopkins and Andrew Wetzel and Jorge Moreno and Christopher C. Hayward and Alexander J. Richings and Zachary Hafen},
	doi = {https://doi.org/10.1093/mnras/stac3712},
	journal = {MNRAS, 519, 2},
	title = {Rapid disc settling and the transition from bursty to steady star formation in Milky Way-mass galaxies},
	year = {2023}}

@article{stern21a,
  title = {Virialization of the {{Inner CGM}} in the {{FIRE Simulations}} and {{Implications}} for {{Galaxy Disks}}, {{Star Formation}}, and {{Feedback}}},
  author = {Stern, Jonathan and {Faucher-Gigu{\`e}re}, Claude-Andr{\'e} and Fielding, Drummond and Quataert, Eliot and Hafen, Zachary and Gurvich, Alexander B. and Ma, Xiangcheng and Byrne, Lindsey and {El-Badry}, Kareem and {Angl{\'e}s-Alc{\'a}zar}, Daniel and Chan, T. K. and Feldmann, Robert and Kere{\v s}, Du{\v s}an and Wetzel, Andrew and Murray, Norman and Hopkins, Philip F.},
  year = 2021,
  month = apr,
  journal = {The Astrophysical Journal},
  volume = {911},
  number = {2},
  pages = {88},
  publisher = {The American Astronomical Society},
  issn = {0004-637X},
  doi = {10.3847/1538-4357/abd776},
  langid = {english}
}

@article{stern21b,
  title = {Neutral {{CGM}} as Damped {{Ly}} {$\alpha$} Absorbers at High Redshift},
  author = {Stern, Jonathan and Sternberg, Amiel and {Faucher-Gigu{\`e}re}, Claude-Andr{\'e} and Hafen, Zachary and Fielding, Drummond and Quataert, Eliot and Wetzel, Andrew and {Angl{\'e}s-Alc{\'a}zar}, Daniel and {El-Badry}, Kareem and Kere{\v s}, Du{\v s}an and Hopkins, Philip F.},
  year = 2021,
  month = oct,
  journal = {Monthly Notices of the Royal Astronomical Society, Volume 507, Issue 2, pp.2869-2884},
  volume = {507},
  number = {2},
  pages = {2869},
  issn = {0035-8711},
  doi = {10.1093/mnras/stab2240},
  langid = {english}
}

@article{Werk13,
	author = {Jessica K. Werk and J. Xavier Prochaska and Christopher Thom and Jason Tumlinson and Todd M. Tripp and John M. O'Meara and Molly S. Peeples},
	doi = {https://doi.org/10.1088/0067-0049/204/2/17},
	journal = {The Astrophysical Journal Supplement, 204, 17},
	note = {19 pages, 15 figures, 7 Tables -- re-submitted to ApJS; Based on HST/COS and Keck/HIRES observations of quasars within 160 kpc of target galaxies},
	title = {The COS-Halos Survey: An Empirical Description of Metal-line Absorption in the Low-redshift Circumgalactic Medium},
	year = {2013}}

@article{Bish21,
	adsurl = {https://ui.adsabs.harvard.edu/abs/2021ApJ...912....8B},
	archiveprefix = {arXiv},
	author = {{Bish}, Hannah V. and {Werk}, Jessica K. and {Peek}, Joshua and {Zheng}, Yong and {Putman}, Mary},
	doi = {10.3847/1538-4357/abeb6b},
	eid = {8},
	eprint = {2010.03610},
	journal = {\apj},
	month = may,
	number = {1},
	pages = {8},
	primaryclass = {astro-ph.GA},
	title = {{The QuaStar Survey: Detecting Hidden Low-velocity Gas in the Milky Way's Circumgalactic Medium}},
	volume = {912},
	year = 2021}

@article{Lehner20,
	author = {Nicolas Lehner and Samantha C. Berek and J. Christopher Howk and Bart P. Wakker and Jason Tumlinson and Edward B. Jenkins and J. Xavier Prochaska and Ramona Augustin and Suoqing Ji and Claude-Andr{\'e} Faucher-Gigu{\`e}re and Zachary Hafen and Molly S. Peeples and Kat A. Barger and Michelle A. Berg and Rongmon Bordoloi and Thomas M. Brown and Andrew J. Fox and Karoline M. Gilbert and Puragra Guhathakurta and Jason S. Kalirai and others},
	doi = {https://doi.org/10.3847/1538-4357/aba49c},
	journal = {The Astrophysical Journal, 900, 9},
	note = {Submitted to the Astrophysical Journal; Comments welcome},
	title = {Project AMIGA: The Circumgalactic Medium of Andromeda},
	year = {2020}}

@article{Kamphuis22,
	adsurl = {https://ui.adsabs.harvard.edu/abs/2022A&A...668A.182K},
	archiveprefix = {arXiv},
	author = {{Kamphuis}, P. and {J{\"u}tte}, E. and {Heald}, G.~H. and {Herrera Ruiz}, N. and {J{\'o}zsa}, G.~I.~G. and {de Blok}, W.~J.~G. and {Serra}, P. and {Marasco}, A. and {Dettmar}, R. -J. and {Pingel}, N.~M. and {Oosterloo}, T. and {Rand}, R.~J. and {Walterbos}, R.~A.~M. and {van der Hulst}, J.~M.},
	doi = {10.1051/0004-6361/202140704},
	eid = {A182},
	eprint = {2210.09383},
	journal = {\aap},
	month = dec,
	pages = {A182},
	primaryclass = {astro-ph.GA},
	title = {{HALOGAS: Strong constraints on the neutral gas reservoir and accretion rate in nearby spiral galaxies}},
	volume = {668},
	year = 2022}

@article{Sardone21,
	adsurl = {https://ui.adsabs.harvard.edu/abs/2021ApJ...910...69S},
	archiveprefix = {arXiv},
	author = {{Sardone}, Amy and {Pisano}, D.~J. and {Pingel}, N.~M. and {Sorgho}, A. and {Carignan}, Claude and {de Blok}, W.~J.~G.},
	doi = {10.3847/1538-4357/abde45},
	eid = {69},
	eprint = {2101.07822},
	journal = {\apj},
	month = mar,
	number = {1},
	pages = {69},
	primaryclass = {astro-ph.GA},
	title = {{A Census of the Extended Neutral Hydrogen around 18 MHONGOOSE Galaxies}},
	volume = {910},
	year = 2021}

@article{Dekel06,
	adsurl = {https://ui.adsabs.harvard.edu/abs/2006MNRAS.368....2D},
	archiveprefix = {arXiv},
	author = {{Dekel}, Avishai and {Birnboim}, Yuval},
	doi = {10.1111/j.1365-2966.2006.10145.x},
	eprint = {astro-ph/0412300},
	journal = {\mnras},
	month = may,
	number = {1},
	pages = {2-20},
	primaryclass = {astro-ph},
	title = {{Galaxy bimodality due to cold flows and shock heating}},
	volume = {368},
	year = 2006}

@article{Voit17,
	adsurl = {https://ui.adsabs.harvard.edu/abs/2017ApJ...845...80V},
	archiveprefix = {arXiv},
	author = {{Voit}, G. Mark and {Meece}, Greg and {Li}, Yuan and {O'Shea}, Brian W. and {Bryan}, Greg L. and {Donahue}, Megan},
	doi = {10.3847/1538-4357/aa7d04},
	eid = {80},
	eprint = {1607.02212},
	journal = {\apj},
	month = aug,
	number = {1},
	pages = {80},
	primaryclass = {astro-ph.GA},
	title = {{A Global Model for Circumgalactic and Cluster-core Precipitation}},
	volume = {845},
	year = 2017}

@article{Sharma12,
	adsurl = {https://ui.adsabs.harvard.edu/abs/2012MNRAS.420.3174S},
	archiveprefix = {arXiv},
	author = {{Sharma}, Prateek and {McCourt}, Michael and {Quataert}, Eliot and {Parrish}, Ian J.},
	doi = {10.1111/j.1365-2966.2011.20246.x},
	eprint = {1106.4816},
	journal = {\mnras},
	month = mar,
	number = {4},
	pages = {3174-3194},
	primaryclass = {astro-ph.CO},
	title = {{Thermal instability and the feedback regulation of hot haloes in clusters, groups and galaxies}},
	volume = {420},
	year = 2012}

@article{Bertschinger89,
	adsurl = {https://ui.adsabs.harvard.edu/abs/1989ApJ...340..666B},
	author = {{Bertschinger}, Edmund},
	doi = {10.1086/167428},
	journal = {\apj},
	month = may,
	pages = {666},
	title = {{The Evolution of Cooling Flows: Self-similar Cooling Waves}},
	volume = {340},
	year = 1989}

@article{DeFelippis20,
	adsurl = {https://ui.adsabs.harvard.edu/abs/2020ApJ...895...17D},
	archiveprefix = {arXiv},
	author = {{DeFelippis}, Daniel and {Genel}, Shy and {Bryan}, Greg L. and {Nelson}, Dylan and {Pillepich}, Annalisa and {Hernquist}, Lars},
	doi = {10.3847/1538-4357/ab8a4a},
	eid = {17},
	eprint = {2004.07846},
	journal = {\apj},
	month = may,
	number = {1},
	pages = {17},
	primaryclass = {astro-ph.GA},
	title = {{The Angular Momentum of the Circumgalactic Medium in the TNG100 Simulation}},
	volume = {895},
	year = 2020}

@article{armillottaEfficiencyGasCooling2016,
	author = {Armillotta, L. and Fraternali, F. and Marinacci, F.},
	doi = {10.1093/mnras/stw1930},
	issn = {0035-8711},
	journal = {Monthly Notices of the Royal Astronomical Society},
	month = nov,
	pages = {4157--4170},
	publisher = {OUP},
	title = {Efficiency of Gas Cooling and Accretion at the Disc-Corona Interface},
	urldate = {2024-07-04},
	volume = {462},
	year = {2016}}

@article{briggsRulesBehaviorGalactic1990,
	author = {Briggs, F. H.},
	doi = {10.1086/168512},
	issn = {0004-637X},
	journal = {The Astrophysical Journal},
	month = mar,
	pages = {15},
	title = {Rules of {{Behavior}} for {{Galactic WARPS}}},
	urldate = {2023-10-02},
	volume = {352},
	year = {1990}}

@misc{deblokMHONGOOSEMeerKATNearby2024,
	archiveprefix = {arXiv},
	author = {{de Blok}, W. J. G. and Healy, J. and Maccagni, F. M. and Pisano, D. J. and Bosma, A. and English, J. and Jarrett, T. and Marasco, A. and Meurer, G. R. and Veronese, S. and Bigiel, F. and Chemin, L. and Fraternali, F. and Holwerda, B. W. and Kamphuis, P. and Kl{\"o}ckner, H. R. and Kleiner, D. and Leroy, A. K. and Mogotsi, M. and Oman, K. A. and Schinnerer, E. and {Verdes-Montenegro}, L. and Westmeier, T. and Wong, O. I. and Zabel, N. and Amram, P. and Carignan, C. and Combes, F. and Brinks, E. and Dettmar, R. J. and Gibson, B. K. and Jozsa, G. I. G. and Koribalski, B. S. and McGaugh, S. S. and Oosterloo, T. A. and Spekkens, K. and Schr{\"o}der, A. C. and Adams, E. A. K. and Athanassoula, E. and Bershady, M. A. and Beswick, R. J. and Blyth, S. and Elson, E. C. and Frank, B. S. and Heald, G. and Henning, P. A. and Kurapati, S. and Loubser, S. I. and Lucero, D. and Meyer, M. and Namumba, B. and Oh, S.-H. and Sardone, A. and Sheth, K. and Smith, M. W. L. and Sorgho, A. and Walter, F. and Williams, T. and Woudt, P. A. and Zijlstra, A.},
	eprint = {2404.01774},
	langid = {english},
	month = jun,
	number = {arXiv:2404.01774},
	primaryclass = {astro-ph},
	publisher = {arXiv},
	title = {{{MHONGOOSE}} -- {{A MeerKAT Nearby Galaxy HI Survey}}},
	urldate = {2024-06-27},
	year = {2024}}

@misc{dengPotentialDynamicalOrigin2024,
	archiveprefix = {arXiv},
	author = {Deng, Mingji and Du, Cuihua and Yang, Yanbin and Liao, Jiwei and Ye, Dashuang},
	doi = {10.48550/arXiv.2409.03264},
	eprint = {2409.03264},
	month = sep,
	number = {arXiv:2409.03264},
	primaryclass = {astro-ph},
	publisher = {arXiv},
	shorttitle = {A {{Potential Dynamical Origin}} of {{The Galactic Disk Warp}}},
	title = {A {{Potential Dynamical Origin}} of {{The Galactic Disk Warp}}: {{The Gaia-Sausage-Enceladus Major Merger}}},
	urldate = {2024-12-27},
	year = {2024}}

@article{fabianCoolingFlowsClusters1994,
	author = {Fabian, A. C.},
	doi = {10.1146/annurev.aa.32.090194.001425},
	issn = {0066-4146, 1545-4282},
	journal = {Annual Review of Astronomy and Astrophysics},
	langid = {english},
	month = sep,
	number = {Volume 32, 1994},
	pages = {277--318},
	publisher = {Annual Reviews},
	title = {Cooling {{Flows}} in {{Clusters}} of {{Galaxies}}},
	urldate = {2024-03-28},
	volume = {32},
	year = {1994}}

@article{hafenHotmodeAccretionPhysics2022,
	author = {Hafen, Zachary and Stern, Jonathan and Bullock, James and Gurvich, Alexander B. and Yu, Sijie and {Faucher-Gigu{\`e}re}, Claude-Andr{\'e} and Fielding, Drummond B. and {Angl{\'e}s-Alc{\'a}zar}, Daniel and Quataert, Eliot and Wetzel, Andrew and Starkenburg, Tjitske and {Boylan-Kolchin}, Michael and Moreno, Jorge and Feldmann, Robert and {El-Badry}, Kareem and Chan, T. K. and Trapp, Cameron and Kere{\v s}, Du{\v s}an and Hopkins, Philip F.},
	doi = {10.1093/mnras/stac1603},
	issn = {0035-8711},
	journal = {Monthly Notices of the Royal Astronomical Society},
	month = aug,
	pages = {5056--5073},
	title = {Hot-Mode Accretion and the Physics of Thin-Disc Galaxy Formation},
	urldate = {2023-03-20},
	volume = {514},
	year = {2022}}

@misc{hanTiltedDarkHalo2023,
	author = {Han, Jiwon Jesse and Conroy, Charlie and Hernquist, Lars},
	doi = {10.1038/s41550-023-02076-9},
	howpublished = {https://arxiv.org/abs/2309.07209v1},
	journal = {arXiv.org},
	langid = {english},
	month = sep,
	title = {A {{Tilted Dark Halo Origin}} of the {{Galactic Disk Warp}} and {{Flare}}},
	urldate = {2023-12-29},
	year = {2023}}

@article{healyPossibleOriginsAnomalous2024,
	author = {Healy, J. and De Blok, W. J. G. and Maccagni, F. M. and Amram, P. and Chemin, L. and Combes, F. and Holwerda, B. W. and Kamphuis, P. and Pisano, D. J. and Schinnerer, E. and Spekkens, K. and {Verdes-Montenegro}, L. and Walter, F. and Adams, E. A. K. and Gibson, B. K. and Kleiner, D. and Veronese, S. and Zabel, N. and English, J. and Carignan, C.},
	copyright = {https://creativecommons.org/licenses/by/4.0},
	doi = {10.1051/0004-6361/202347475},
	issn = {0004-6361, 1432-0746},
	journal = {Astronomy \& Astrophysics},
	langid = {english},
	month = jul,
	pages = {A254},
	title = {Possible Origins of Anomalous {{H I}} Gas around {{MHONGOOSE}} Galaxy, {{NGC}} 5068},
	urldate = {2024-08-07},
	volume = {687},
	year = {2024}}

@article{hopkinsNewClassAccurate2015,
	author = {Hopkins, Philip F.},
	doi = {10.1093/mnras/stv195},
	issn = {0035-8711},
	journal = {Monthly Notices of the Royal Astronomical Society},
	month = jun,
	pages = {53--110},
	title = {A New Class of Accurate, Mesh-Free Hydrodynamic Simulation Methods},
	urldate = {2023-07-27},
	volume = {450},
	year = {2015}}

@article{kravtsovSizeVirialRadiusRelation2013,
	author = {Kravtsov, Andrey V.},
	doi = {10.1088/2041-8205/764/2/L31},
	issn = {0004-637X},
	journal = {The Astrophysical Journal},
	month = feb,
	pages = {L31},
	publisher = {IOP},
	title = {The {{Size-Virial Radius Relation}} of {{Galaxies}}},
	urldate = {2024-08-07},
	volume = {764},
	year = {2013}}

@article{pezzulliAccretionRadialFlows2016,
	author = {Pezzulli, Gabriele and Fraternali, Filippo},
	doi = {10.1093/mnras/stv2397},
	issn = {0035-8711},
	journal = {Monthly Notices of the Royal Astronomical Society},
	month = jan,
	pages = {2308--2322},
	publisher = {OUP},
	title = {Accretion, Radial Flows and Abundance Gradients in Spiral Galaxies},
	urldate = {2024-07-04},
	volume = {455},
	year = {2016}}

@article{roskarMisalignedAngularMomentum2010,
	author = {Ro{\v s}kar, Rok and Debattista, Victor P. and Brooks, Alyson M. and Quinn, Thomas R. and Brook, Chris B. and Governato, Fabio and Dalcanton, Julianne J. and Wadsley, James},
	doi = {10.1111/j.1365-2966.2010.17178.x},
	issn = {0035-8711},
	journal = {Monthly Notices of the Royal Astronomical Society},
	month = oct,
	pages = {783--796},
	shorttitle = {Misaligned Angular Momentum in Hydrodynamic Cosmological Simulations},
	title = {Misaligned Angular Momentum in Hydrodynamic Cosmological Simulations: Warps, Outer Discs and Thick Discs},
	urldate = {2023-07-27},
	volume = {408},
	year = {2010}}

@article{sellwoodInternallyDrivenWarps2022,
	author = {Sellwood, J. A. and Debattista, Victor P.},
	doi = {10.1093/mnras/stab3433},
	issn = {0035-8711},
	journal = {Monthly Notices of the Royal Astronomical Society},
	month = feb,
	pages = {1375--1382},
	title = {Internally Driven Warps in Disc Galaxies},
	urldate = {2024-03-06},
	volume = {510},
	year = {2022}}

@article{sharmaOriginAngularMomentum2012,
	author = {Sharma, Sanjib and Steinmetz, Matthias and {Bland-Hawthorn}, Joss},
	doi = {10.1088/0004-637X/750/2/107},
	issn = {0004-637X},
	journal = {The Astrophysical Journal},
	month = may,
	pages = {107},
	title = {On the {{Origin}} of the {{Angular Momentum Properties}} of {{Gas}} and {{Dark Matter}} in {{Galactic Halos}} and {{Its Implications}}},
	urldate = {2023-07-29},
	volume = {750},
	year = {2012}}

@article{springelModellingFeedbackStars2005,
	author = {Springel, Volker and Di Matteo, Tiziana and Hernquist, Lars},
	doi = {10.1111/j.1365-2966.2005.09238.x},
	issn = {0035-8711},
	journal = {Monthly Notices of the Royal Astronomical Society},
	month = aug,
	pages = {776--794},
	publisher = {OUP},
	title = {Modelling Feedback from Stars and Black Holes in Galaxy Mergers},
	urldate = {2024-07-06},
	volume = {361},
	year = {2005}}

@article{sternAccretionDiskGalaxies2024,
	author = {Stern, Jonathan and Fielding, Drummond and Hafen, Zachary and Su, Kung-Yi and Naor, Nadav and {Faucher-Gigu{\`e}re}, Claude-Andr{\'e} and Quataert, Eliot and Bullock, James},
	doi = {10.1093/mnras/stae824},
	issn = {0035-8711},
	journal = {Monthly Notices of the Royal Astronomical Society},
	month = mar,
	publisher = {OUP},
	title = {Accretion onto Disk Galaxies via Hot and Rotating {{CGM}} Inflows},
	urldate = {2024-04-17},
	year = {2024}}

@article{sternCoolingFlowSolutions2019,
	author = {Stern, Jonathan and Fielding, Drummond and {Faucher-Gigu{\`e}re}, Claude-Andr{\'e} and Quataert, Eliot},
	doi = {10.1093/mnras/stz1859},
	issn = {0035-8711},
	journal = {Monthly Notices of the Royal Astronomical Society},
	month = sep,
	pages = {2549--2572},
	title = {Cooling Flow Solutions for the Circumgalactic Medium},
	urldate = {2023-03-21},
	volume = {488},
	year = {2019}}

@article{sternMaximumAccretionRate2020,
	author = {Stern, Jonathan and Fielding, Drummond and {Faucher-Gigu{\`e}re}, Claude-Andr{\'e} and Quataert, Eliot},
	doi = {10.1093/mnras/staa198},
	issn = {0035-8711},
	journal = {Monthly Notices of the Royal Astronomical Society},
	month = mar,
	pages = {6042--6058},
	title = {The Maximum Accretion Rate of Hot Gas in Dark Matter Haloes},
	urldate = {2023-03-21},
	volume = {492},
	year = {2020}}

@misc{trappAngularMomentumTransfer2024,
	author = {Trapp, Cameron W. and Kere{\v s}, Du{\v s}an and Hopkins, Philip F. and {Faucher-Gigu{\`e}re}, Claude-Andr{\'e}},
	doi = {10.48550/arXiv.2405.01632},
	journal = {arXiv e-prints},
	month = may,
	title = {Angular Momentum Transfer in Cosmological Simulations of {{Milky Way-mass}} Discs},
	urldate = {2024-06-20},
	year = {2024}}

@article{vanderkruitGalaxyDisks2011,
	author = {{van der Kruit}, P. C. and Freeman, K. C.},
	doi = {10.1146/annurev-astro-083109-153241},
	issn = {0066-4146},
	journal = {Annual Review of Astronomy and Astrophysics, vol. 49, issue 1, pp. 301-371},
	langid = {english},
	month = sep,
	number = {1},
	pages = {301},
	title = {Galaxy {{Disks}}},
	urldate = {2022-06-15},
	volume = {49},
	year = {2011}}

@article{bouwensInsideoutInfallFormation1997,
	author = {Bouwens, Rychard J. and Cay{\'o}n, Laura and Silk, Joseph},
	doi = {10.1086/310969},
	issn = {0004-637X},
	journal = {The Astrophysical Journal},
	langid = {english},
	month = nov,
	number = {1},
	pages = {L21-L24},
	shorttitle = {Inside-out {{Infall Formation}} of {{Disk Galaxies}}},
	title = {Inside-out {{Infall Formation}} of {{Disk Galaxies}}: {{Do Predictions Differ}} from {{Models}} without {{Size Evolution}}?},
	urldate = {2026-04-23},
	volume = {489},
	year = 1997}

@article{chiappiniChemicalEvolutionGalaxy1997,
	author = {Chiappini, C. and Matteucci, F. and Gratton, R.},
	doi = {10.1086/303726},
	issn = {0004-637X},
	journal = {The Astrophysical Journal},
	langid = {english},
	month = mar,
	number = {2},
	pages = {765--780},
	shorttitle = {The {{Chemical Evolution}} of the {{Galaxy}}},
	title = {The {{Chemical Evolution}} of the {{Galaxy}}: {{The Two-Infall Model}}},
	urldate = {2026-04-23},
	volume = {477},
	year = 1997}

@article{wangFEASTSCombinedInterferometry2024,
	author = {Wang, Jing and Lin, Xuchen and Yang, Dong and {Staveley-Smith}, Lister and Walter, Fabian and Wang, Q. Daniel and Wang, Ran and Battisti, A. J. and Catinella, Barbara and Chen, Hsiao-Wen and Cortese, Luca and Fisher, D. B. and Ho, Luis C. and Ji, Suoqing and Jiang, Peng and Kauffmann, Guinevere and Kong, Xu and Liu, Ziming and Shao, Li and Wang, Jie and Wang, Lile and Wang, Shun},
	doi = {10.3847/1538-4357/ad3e61},
	issn = {0004-637X},
	journal = {The Astrophysical Journal},
	langid = {english},
	month = jun,
	number = {1},
	pages = {48},
	title = {{{FEASTS Combined}} with {{Interferometry}}. {{I}}. {{Overall Properties}} of {{Diffuse H I}} and {{Implications}} for {{Gas Accretion}} in {{Nearby Galaxies}}},
	urldate = {2026-04-23},
	volume = {968},
	year = 2024}

@article{wiersmaEffectPhotoionizationCooling2009,
	author = {Wiersma, Robert P. C. and Schaye, Joop and Smith, Britton D.},
	doi = {10.1111/j.1365-2966.2008.14191.x},
	issn = {0035-8711},
	journal = {Monthly Notices of the Royal Astronomical Society},
	month = feb,
	number = {1},
	pages = {99--107},
	title = {The Effect of Photoionization on the Cooling Rates of Enriched, Astrophysical Plasmas},
	urldate = {2024-12-27},
	volume = {393},
	year = {2009}}

@misc{willeWarpsInducedSatellites2024,
	archiveprefix = {arXiv},
	author = {Wille, Andressa and Machado, Rubens E. G.},
	eprint = {2408.09932},
	langid = {english},
	month = aug,
	number = {arXiv:2408.09932},
	primaryclass = {astro-ph},
	publisher = {arXiv},
	title = {Warps Induced by Satellites on Barred and Non-Barred Galaxies},
	urldate = {2024-08-20},
	year = {2024}}

@article{bailinInternalAlignmentHalos2005,
	author = {Bailin, Jeremy and Kawata, Daisuke and Gibson, Brad K. and Steinmetz, Matthias and Navarro, Julio F. and Brook, Chris B. and Gill, Stuart P. D. and Ibata, Rodrigo A. and Knebe, Alexander and Lewis, Geraint F. and Okamoto, Takashi},
	doi = {10.1086/432157},
	issn = {0004-637X},
	journal = {The Astrophysical Journal},
	month = jul,
	pages = {L17-L20},
	publisher = {IOP},
	title = {Internal {{Alignment}} of the {{Halos}} of {{Disk Galaxies}} in {{Cosmological Hydrodynamic Simulations}}},
	urldate = {2025-03-28},
	volume = {627},
	year = {2005}}

@article{jiangWARPSCosmicInfall1999,
	author = {Jiang, Ing-Guey and Binney, James},
	doi = {10.1046/j.1365-8711.1999.02333.x},
	issn = {0035-8711},
	journal = {Monthly Notices of the Royal Astronomical Society},
	month = feb,
	pages = {L7-L10},
	publisher = {OUP},
	title = {{{WARPS}} and Cosmic Infall},
	urldate = {2025-03-26},
	volume = {303},
	year = {1999}}

@article{jozsaKinematicModellingDisk2007a,
	author = {J{\'o}zsa, G. I. G.},
	doi = {10.1051/0004-6361:20066165},
	issn = {0004-6361},
	journal = {Astronomy and Astrophysics},
	month = jun,
	pages = {903--917},
	title = {Kinematic Modelling of Disk Galaxies. {{II}}. {{A}} Case-Study of Symmetrically Warped Galaxy Disks},
	urldate = {2025-03-26},
	volume = {468},
	year = {2007}}

@article{aumerIdealizedModelsGalactic2013,
	author = {Aumer, Michael and White, Simon D. M.},
	doi = {10.1093/mnras/sts083},
	issn = {0035-8711},
	journal = {Monthly Notices of the Royal Astronomical Society},
	langid = {english},
	month = jan,
	number = {2},
	pages = {1055--1076},
	title = {Idealized Models for Galactic Disc Formation and Evolution in `realistic' {{$\Lambda$CDM}} Haloes},
	urldate = {2025-04-18},
	volume = {428},
	year = {2013}}

@article{garcia-condeGalactoseismologyCosmologicalSimulations2024,
	author = {{Garc{\'\i}a-Conde}, B. and Antoja, T. and {Roca-F{\`a}brega}, S. and G{\'o}mez, F. and Ramos, P. and {Garavito-Camargo}, N. and {G{\'o}mez-Flechoso}, M. A.},
	doi = {10.1051/0004-6361/202347446},
	issn = {0004-6361},
	journal = {Astronomy and Astrophysics},
	langid = {english},
	month = mar,
	pages = {A47},
	title = {Galactoseismology in Cosmological Simulations. {{Vertical}} Perturbations by Dark Matter, Satellite Galaxies, and Gas},
	urldate = {2025-04-18},
	volume = {683},
	year = {2024}}

@article{gomezWarpsWavesStellar2017,
	author = {G{\'o}mez, Facundo A. and White, Simon D. M. and Grand, Robert J. J. and Marinacci, Federico and Springel, Volker and Pakmor, R{\"u}diger},
	doi = {10.1093/mnras/stw2957},
	issn = {0035-8711},
	journal = {Monthly Notices of the Royal Astronomical Society},
	langid = {english},
	month = mar,
	number = {3},
	pages = {3446--3460},
	title = {Warps and Waves in the Stellar Discs of the {{Auriga}} Cosmological Simulations},
	urldate = {2025-04-18},
	volume = {465},
	year = {2017}}

@article{hanTiltedDarkHalos2023,
	author = {Han, Jiwon Jesse and Semenov, Vadim and Conroy, Charlie and Hernquist, Lars},
	doi = {10.3847/2041-8213/ad0641},
	issn = {0004-637X},
	journal = {The Astrophysical Journal},
	langid = {english},
	month = nov,
	number = {2},
	pages = {L24},
	title = {Tilted {{Dark Halos Are Common}} and {{Long-lived}}, and {{Can Warp Galactic Disks}}},
	urldate = {2025-04-18},
	volume = {957},
	year = {2023}}

@article{lopez-corredoiraGenerationGalacticDisc2002,
	author = {{L{\'o}pez-Corredoira}, M. and {Betancort-Rijo}, J. and Beckman, J. E.},
	doi = {10.1051/0004-6361:20020229},
	issn = {0004-6361, 1432-0746},
	journal = {Astronomy \& Astrophysics},
	langid = {english},
	month = apr,
	number = {1},
	pages = {169--186},
	title = {Generation of Galactic Disc Warps Due to Intergalactic Accretion Flows onto the Disc},
	urldate = {2025-04-19},
	volume = {386},
	year = {2002}}

@article{syloslabiniBreakingDegeneracyWarps2025,
	author = {Sylos Labini, Francesco and De Marzo, Giordano and Straccamore, Matteo},
	doi = {10.48550/arXiv.2503.22306},
	journal = {arXiv e-prints},
	langid = {english},
	month = mar,
	pages = {arXiv:2503.22306},
	title = {Breaking the Degeneracy between Warps and Radial Flows in External Galaxies},
	urldate = {2025-04-18},
	year = {2025}}

@article{garcia-ruizNeutralHydrogenOptical2002,
	author = {{Garc{\'\i}a-Ruiz}, I. and Sancisi, R. and Kuijken, K.},
	doi = {10.1051/0004-6361:20020976},
	issn = {0004-6361},
	journal = {Astronomy and Astrophysics, v.394, p.769-789 (2002)},
	langid = {english},
	month = nov,
	pages = {769},
	shorttitle = {Neutral Hydrogen and Optical Observations of Edge-on Galaxies},
	title = {Neutral Hydrogen and Optical Observations of Edge-on Galaxies: {{Hunting}} for Warps},
	urldate = {2025-04-18},
	volume = {394},
	year = {2002}}

@article{shenGalacticWarpsInduced2006,
	author = {Shen, Juntai and Sellwood, J. A.},
	doi = {10.1111/j.1365-2966.2006.10477.x},
	issn = {0035-8711},
	journal = {Monthly Notices of the Royal Astronomical Society, Volume 370, Issue 1, pp. 2-14.},
	langid = {english},
	month = jul,
	number = {1},
	pages = {2},
	title = {Galactic Warps Induced by Cosmic Infall},
	urldate = {2025-04-22},
	volume = {370},
	year = {2006}}

@article{afruniCloudsAccretingIGM2023a,
	author = {Afruni, Andrea and Pezzulli, Gabriele and Fraternali, Filippo and Gr{\o}nnow, Asger},
	doi = {10.1093/mnras/stad1963},
	issn = {0035-8711},
	journal = {Monthly Notices of the Royal Astronomical Society},
	langid = {english},
	month = sep,
	number = {2},
	pages = {2351--2367},
	title = {Clouds Accreting from the {{IGM}} Are Not Able to Feed the Star Formation of Low-Redshift Disc Galaxies},
	urldate = {2025-04-23},
	volume = {524},
	year = {2023}}

@article{keresHowGalaxiesGet2005,
	author = {Kere{\v s}, Du{\v s}an and Katz, Neal and Weinberg, David H. and Dav{\'e}, Romeel},
	doi = {10.1111/j.1365-2966.2005.09451.x},
	issn = {0035-8711},
	journal = {Monthly Notices of the Royal Astronomical Society},
	langid = {english},
	month = oct,
	number = {1},
	pages = {2--28},
	title = {How Do Galaxies Get Their Gas?},
	urldate = {2025-04-23},
	volume = {363},
	year = {2005}}

@article{sancisiColdGasAccretion2008,
	author = {Sancisi, Renzo and Fraternali, Filippo and Oosterloo, Tom and {van der Hulst}, Thijs},
	doi = {10.1007/s00159-008-0010-0},
	issn = {0935-4956},
	journal = {The Astronomy and Astrophysics Review, Volume 15, Issue 3, pp.189-223},
	langid = {english},
	month = jun,
	number = {3},
	pages = {189},
	title = {Cold Gas Accretion in Galaxies},
	urldate = {2022-08-04},
	volume = {15},
	year = {2008}}

@incollection{fraternaliGasAccretionCondensation2017,
	address = {Cham},
	author = {Fraternali, Filippo},
	booktitle = {Gas {{Accretion}} onto {{Galaxies}}},
	doi = {10.1007/978-3-319-52512-9_14},
	editor = {Fox, Andrew and Dav{\'e}, Romeel},
	isbn = {978-3-319-52512-9},
	langid = {english},
	pages = {323--353},
	publisher = {Springer International Publishing},
	series = {Astrophysics and {{Space Science Library}}},
	title = {Gas {{Accretion}} via {{Condensation}} and {{Fountains}}},
	urldate = {2023-11-09},
	year = {2017}}

@article{stewartHighAngularMomentum2017a,
	author = {Stewart, Kyle R. and Maller, Ariyeh H. and O{\~n}orbe, Jose and Bullock, James S. and Joung, M. Ryan and Devriendt, Julien and Ceverino, Daniel and Kere{\v s}, Du{\v s}an and Hopkins, Philip F. and {Faucher-Gigu{\`e}re}, Claude-Andr{\'e}},
	doi = {10.3847/1538-4357/aa6dff},
	issn = {0004-637X},
	journal = {The Astrophysical Journal, Volume 843, Issue 1, article id. 47, 15 pp. (2017).},
	langid = {english},
	month = jul,
	number = {1},
	pages = {47},
	shorttitle = {High {{Angular Momentum Halo Gas}}},
	title = {High {{Angular Momentum Halo Gas}}: {{A Feedback}} and {{Code-independent Prediction}} of {{LCDM}}},
	urldate = {2025-04-26},
	volume = {843},
	year = {2017}}

@article{wangLocalVolumeSurvey2017,
	author = {Wang, Jing and Koribalski, B{\"a}rbel S. and Jarrett, Tom H. and Kamphuis, Peter and Li, Zhao-Yu and Ho, Luis C. and Westmeier, Tobias and Shao, Li and Lagos, Claudia del P. and Wong, Oiwei Ivy and Serra, Paolo and {Staveley-Smith}, Lister and J{\'o}zsa, Gyula and {van der Hulst}, Thijs and {L{\'o}pez-S{\'a}nchez}, {\'A} R.},
	doi = {10.1093/mnras/stx2073},
	issn = {0035-8711},
	journal = {Monthly Notices of the Royal Astronomical Society, Volume 472, Issue 3, p.3029-3057},
	langid = {english},
	month = dec,
	number = {3},
	pages = {3029},
	shorttitle = {The {{Local Volume H I Survey}}},
	title = {The {{Local Volume H I Survey}}: Star Formation Properties},
	urldate = {2025-04-26},
	volume = {472},
	year = {2017}}

@article{yangFEASTSCombinedInterferometry2025,
	author = {Yang, Dong and Wang, Jing and Qu, Zhijie and Liang, Zezhong and Lin, Xuchen and Weng, Simon and Chen, Xinkai and Catinella, Barbara and Cortese, Luca and Fisher, D. B. and Ho, Luis C. and Jing, Yingjie and Jiang, Fangzhou and Jiang, Peng and Liu, Ziming and P{\'e}roux, C{\'e}line and Shao, Li and {Staveley-Smith}, Lister and Wang, Q. Daniel and Wang, Jie},
	doi = {10.48550/arXiv.2502.16926},
	journal = {arXiv e-prints},
	langid = {english},
	month = feb,
	pages = {arXiv:2502.16926},
	shorttitle = {{{FEASTS Combined}} with {{Interferometry}} ({{IV}})},
	title = {{{FEASTS Combined}} with {{Interferometry}} ({{IV}}): {{Mapping HI Emission}} to a Limit of \${{N}}\_\{{\textbackslash}text\{\vphantom{\}\}}{{HI}}\vphantom\{\}\vphantom\{\}=10{\textasciicircum}\{17.7\} {\textbackslash}text\{cm\}{\textasciicircum}\{-2\}\$ in {{Seven Edge-on Galaxies}}},
	urldate = {2025-04-25},
	year = {2025}}

@misc{kurapatiUncoveringExtraplanarGas2025,
	archiveprefix = {arXiv},
	author = {Kurapati, Sushma and Pisano, D. J. and de Blok, W. J. G. and Kamphuis, Peter and Zabel, Nikki and de Villiers, Mikhail and Healy, Julia and Maccagni, Filippo M. and Kleiner, Dane and Adams, Elizabeth A. K. and Amram, Philippe and Athanassoula, E. and Bigiel, Frank and Bosma, Albert and Brinks, Elias and Chemin, Laurent and Combes, Francoise and Dettmar, Ralf-J{\"u}rgen and J{\'o}zsa, Gyula and Koribalski, Baerbel and Marasco, Antonino and Meurer, Gerhardt and Mogotsi, Moses and Mohapatra, Abhisek and Rajohnson, Sambatriniaina H. A. and Schinnerer, Eva and Sorgho, Amidou and Spekkens, Kristine and {Verdes-Montenegro}, Lourdes and Veronese, Simone and Walter, Fabian},
	doi = {10.1093/mnras/staf387},
	eprint = {2503.03483},
	month = mar,
	primaryclass = {astro-ph},
	title = {Uncovering {{Extraplanar Gas}} in {{UGCA}} 250 with the {{Ultra-deep MHONGOOSE Survey}}},
	urldate = {2025-03-07},
	year = {2025}}

@misc{linFEASTSCombinedInterferometry2025,
	archiveprefix = {arXiv},
	author = {Lin, Xuchen and Wang, Jing and {Staveley-Smith}, Lister and Ji, Suoqing and Yang, Dong and Chen, Xinkai and Walter, Fabian and Chen, Hsiao-Wen and Ho, Luis C. and Jiang, Peng and Mandelker, Nir and Oh, Se-Heon and Peng, Bo and P{\'e}roux, C{\'e}line and Qu, Zhijie and Wang, Q. Daniel},
	doi = {10.48550/arXiv.2502.10672},
	eprint = {2502.10672},
	month = feb,
	number = {arXiv:2502.10672},
	primaryclass = {astro-ph},
	publisher = {arXiv},
	title = {{{FEASTS Combined}} with {{Interferometry}}. {{III}}. {{The Low-column-density HI Around M51}} and {{Possibility}} of {{Turbulent-mixing Gas Accretion}}},
	urldate = {2025-02-22},
	year = {2025}}

@article{burkeSystematicDistortionOuter1957,
	author = {Burke, Bernard F.},
	doi = {10.1086/107463},
	issn = {0004-6256},
	journal = {Astronomical Journal, Vol. 62, p. 90},
	langid = {english},
	month = may,
	pages = {90},
	title = {Systematic Distortion of the Outer Regions of the Galaxy.},
	urldate = {2025-04-26},
	volume = {62},
	year = {1957}}

@article{kerrMagellanicEffectGalaxy1957,
	author = {Kerr, F. J.},
	doi = {10.1086/107466},
	issn = {0004-6256},
	journal = {Astronomical Journal, Vol. 62, p. 93-93},
	langid = {english},
	month = may,
	pages = {93},
	title = {A {{Magellanic}} Effect on the Galaxy.},
	urldate = {2025-04-26},
	volume = {62},
	year = {1957}}

@article{donahueBaryonCyclesBiggest2022,
	archiveprefix = {arXiv},
	author = {Donahue, Megan and Voit, G. Mark},
	doi = {10.1016/j.physrep.2022.04.005},
	eprint = {2204.08099},
	issn = {03701573},
	journal = {Physics Reports},
	langid = {english},
	month = aug,
	pages = {1--109},
	primaryclass = {astro-ph},
	title = {Baryon {{Cycles}} in the {{Biggest Galaxies}}},
	urldate = {2023-03-19},
	volume = {973},
	year = {2022}}

@article{catinellaXGASSTotalCold2018,
	author = {Catinella, Barbara and Saintonge, Am{\'e}lie and Janowiecki, Steven and Cortese, Luca and Dav{\'e}, Romeel and Lemonias, Jenna J. and Cooper, Andrew P. and Schiminovich, David and Hummels, Cameron B. and Fabello, Silvia and Ger{\'e}b, Katinka and Kilborn, Virginia and Wang, Jing},
	doi = {10.1093/mnras/sty089},
	issn = {0035-8711},
	journal = {Monthly Notices of the Royal Astronomical Society},
	langid = {english},
	month = may,
	number = {1},
	pages = {875--895},
	shorttitle = {{{xGASS}}},
	title = {{{xGASS}}: Total Cold Gas Scaling Relations and Molecular-to-Atomic Gas Ratios of Galaxies in the Local {{Universe}}},
	urldate = {2025-06-03},
	volume = {476},
	year = {2018}}

@article{lelliBaryonicTullyFisher2019,
	author = {Lelli, Federico and McGaugh, Stacy S and Schombert, James M and Desmond, Harry and Katz, Harley},
	doi = {10.1093/mnras/stz205},
	issn = {0035-8711, 1365-2966},
	journal = {Monthly Notices of the Royal Astronomical Society},
	langid = {english},
	month = apr,
	number = {3},
	pages = {3267--3278},
	title = {The Baryonic {{Tully}}--{{Fisher}} Relation for Different Velocity Definitions and Implications for Galaxy Angular Momentum},
	urldate = {2022-11-28},
	volume = {484},
	year = {2019}}

@article{hopkinsFIRE2SimulationsPhysics2018,
	author = {Hopkins, Philip F. and Wetzel, Andrew and Kere{\v s}, Du{\v s}an and {Faucher-Gigu{\`e}re}, Claude-Andr{\'e} and Quataert, Eliot and {Boylan-Kolchin}, Michael and Murray, Norman and Hayward, Christopher C. and {Garrison-Kimmel}, Shea and Hummels, Cameron and Feldmann, Robert and Torrey, Paul and Ma, Xiangcheng and {Angl{\'e}s-Alc{\'a}zar}, Daniel and Su, Kung-Yi and Orr, Matthew and Schmitz, Denise and Escala, Ivanna and Sanderson, Robyn and Grudi{\'c}, Michael Y. and Hafen, Zachary and Kim, Ji-Hoon and Fitts, Alex and Bullock, James S. and Wheeler, Coral and Chan, T. K. and Elbert, Oliver D. and Narayanan, Desika},
	doi = {10.1093/mnras/sty1690},
	issn = {0035-8711},
	journal = {Monthly Notices of the Royal Astronomical Society},
	month = oct,
	pages = {800--863},
	publisher = {OUP},
	shorttitle = {{{FIRE-2}} Simulations},
	title = {{{FIRE-2}} Simulations: Physics versus Numerics in Galaxy Formation},
	urldate = {2025-01-06},
	volume = {480},
	year = {2018}}

@article{bullockUniversalAngularMomentum2001,
	author = {Bullock, J. S. and Dekel, A. and Kolatt, T. S. and Kravtsov, A. V. and Klypin, A. A. and Porciani, C. and Primack, J. R.},
	doi = {10.1086/321477},
	issn = {0004-637X},
	journal = {The Astrophysical Journal, Volume 555, Issue 1, pp. 240-257.},
	langid = {english},
	month = jul,
	number = {1},
	pages = {240},
	title = {A {{Universal Angular Momentum Profile}} for {{Galactic Halos}}},
	urldate = {2025-07-27},
	volume = {555},
	year = {2001}}

@article{leeWALLABYPilotSurvey2025,
	author = {Lee, Seona and Catinella, Barbara and Westmeier, Tobias and Cortese, Luca and Wang, Jing and Spekkens, Kristine and Deg, Nathan and D{\'e}nes, Helga and Elagali, Ahmed and Koribalski, B{\"a}rbel S. and {Lee-Waddell}, Karen and Murugeshan, Chandrashekar and Rhee, Jonghwan and {Staveley-Smith}, Lister and Wong, O. Ivy and Holwerda, Benne W.},
	doi = {10.1017/pasa.2025.30},
	issn = {1323-3580},
	journal = {Publications of the Astronomical Society of Australia},
	langid = {english},
	month = may,
	pages = {e046},
	shorttitle = {{{WALLABY}} Pilot Survey},
	title = {{{WALLABY}} Pilot Survey: {{Spatially}} Resolved Gas Scaling Relations within the Stellar Discs of Nearby Galaxies},
	urldate = {2025-06-02},
	volume = {42},
	year = {2025}}

@article{emamiMorphologicalTypesDM2021,
	author = {Emami, Razieh and Genel, Shy and Hernquist, Lars and Alcock, Charles and Bose, Sownak and Weinberger, Rainer and Vogelsberger, Mark and Marinacci, Federico and Loeb, Abraham and Torrey, Paul and Forbes, John C.},
	doi = {10.3847/1538-4357/abf147},
	issn = {0004-637X},
	journal = {The Astrophysical Journal},
	langid = {english},
	month = may,
	number = {1},
	pages = {36},
	shorttitle = {Morphological {{Types}} of {{DM Halos}} in {{Milky Way-like Galaxies}} in the {{TNG50 Simulation}}},
	title = {Morphological {{Types}} of {{DM Halos}} in {{Milky Way-like Galaxies}} in the {{TNG50 Simulation}}: {{Simple}}, {{Twisted}}, or {{Stretched}}},
	urldate = {2025-08-17},
	volume = {913},
	year = {2021}}

@article{shaoTwistedDarkMatter2021,
	author = {Shao, Shi and Cautun, Marius and Deason, Alis and Frenk, Carlos S.},
	doi = {10.1093/mnras/staa3883},
	issn = {0035-8711},
	journal = {Monthly Notices of the Royal Astronomical Society},
	langid = {english},
	month = jul,
	number = {4},
	pages = {6033--6048},
	title = {The Twisted Dark Matter Halo of the {{Milky Way}}},
	urldate = {2025-08-17},
	volume = {504},
	year = {2021}}

@article{pillepichFirstResultsTNG502019,
	author = {Pillepich, Annalisa and Nelson, Dylan and Springel, Volker and Pakmor, R{\"u}diger and Torrey, Paul and Weinberger, Rainer and Vogelsberger, Mark and Marinacci, Federico and Genel, Shy and {van der Wel}, Arjen and Hernquist, Lars},
	doi = {10.1093/mnras/stz2338},
	issn = {0035-8711},
	journal = {Monthly Notices of the Royal Astronomical Society},
	langid = {english},
	month = dec,
	number = {3},
	pages = {3196--3233},
	shorttitle = {First Results from the {{TNG50}} Simulation},
	title = {First Results from the {{TNG50}} Simulation: The Evolution of Stellar and Gaseous Discs across Cosmic Time},
	urldate = {2025-08-17},
	volume = {490},
	year = {2019}}

@article{schayeEAGLEProjectSimulating2015,
	author = {Schaye, Joop and Crain, Robert A. and Bower, Richard G. and Furlong, Michelle and Schaller, Matthieu and Theuns, Tom and Dalla Vecchia, Claudio and Frenk, Carlos S. and McCarthy, I. G. and Helly, John C. and Jenkins, Adrian and {Rosas-Guevara}, Y. M. and White, Simon D. M. and Baes, Maarten and Booth, C. M. and Camps, Peter and Navarro, Julio F. and Qu, Yan and Rahmati, Alireza and Sawala, Till and Thomas, Peter A. and Trayford, James},
	doi = {10.1093/mnras/stu2058},
	issn = {0035-8711},
	journal = {Monthly Notices of the Royal Astronomical Society},
	langid = {english},
	month = jan,
	number = {1},
	pages = {521--554},
	shorttitle = {The {{EAGLE}} Project},
	title = {The {{EAGLE}} Project: Simulating the Evolution and Assembly of Galaxies and Their Environments},
	urldate = {2025-08-17},
	volume = {446},
	year = {2015}}

@article{balbus89,
	author = {Balbus, Steven A. and Soker, Noam},
	doi = {10.1086/167521},
	journal = {{\textbackslash} apj},
	pages = {611},
	title = {Theory of Local Thermal Instability in Spherical Systems},
	volume = {341},
	year = {1989}}

@article{Danovich15,
	author = {Danovich, Mark and Dekel, Avishai and Hahn, Oliver and Ceverino, Daniel and Primack, Joel},
	doi = {10.1093/mnras/stv270},
	eprint = {1407.7129},
	journal = {{\textbackslash} mnras},
	number = {2},
	pages = {2087--2111},
	title = {Four phases of angular-momentum buildup in high-z galaxies: from cosmic-web streams through an extended ring to disc and bulge},
	volume = {449},
	year = {2015}}

@article{lillyGasRegulationGalaxies2013,
	author = {Lilly, Simon J. and Carollo, C. Marcella and Pipino, Antonio and Renzini, Alvio and Peng, Yingjie},
	doi = {10.1088/0004-637X/772/2/119},
	issn = {0004-637X},
	journal = {The Astrophysical Journal, Volume 772, Issue 2, article id. 119, 19 pp. (2013).},
	langid = {english},
	month = aug,
	number = {2},
	pages = {119},
	shorttitle = {Gas {{Regulation}} of {{Galaxies}}},
	title = {Gas {{Regulation}} of {{Galaxies}}: {{The Evolution}} of the {{Cosmic Specific Star Formation Rate}}, the {{Metallicity-Mass-Star-formation Rate Relation}}, and the {{Stellar Content}} of {{Halos}}},
	urldate = {2025-09-10},
	volume = {772},
	year = {2013}}

@article{barbaniUnderstandingBaryonCycle2025,
	author = {Barbani, Filippo and Pascale, Raffaele and Marinacci, Federico and Torrey, Paul and Sales, Laura V. and Li, Hui and Vogelsberger, Mark},
	doi = {10.1051/0004-6361/202452608},
	issn = {0004-6361},
	journal = {Astronomy \&amp; Astrophysics, Volume 697, id.A121, 20 pp.},
	langid = {english},
	month = may,
	pages = {A121},
	shorttitle = {Understanding the Baryon Cycle},
	title = {Understanding the Baryon Cycle: {{Fueling}} Star Formation via Inflows in {{Milky Way-like}} Galaxies},
	urldate = {2025-09-14},
	volume = {697},
	year = {2025}}

@article{lehnerProjectAMIGAInner2025,
	archiveprefix = {arXiv},
	author = {Lehner, Nicolas and Howk, J. Christopher and Collins, Lucy and Sameer and Wakker, Bart P. and Augustin, Ramona and Barger, Kathleen A. and Berg, Michelle A. and Bordoloi, Rongmon and Brown, Thomas M. and Cashman, Frances H. and {Faucher-Gigu{\`e}re}, Claude-Andr{\'e} and Fox, Andrew J. and French, David M. and Gilbert, Karoline M. and Guhathakurta, Puragra and O'Meara, John M. and O'Shea, Brian W. and Peeples, Molly S. and Pisano, D. J. and Prochaska, J. Xavier and Stern, Jonathan and Tumlinson, Jason and Werk, Jessica K. and Williams, Benjamin F.},
	doi = {10.48550/arXiv.2506.16573},
	eprint = {2506.16573},
	journal = {eprint arXiv:2506.16573},
	langid = {english},
	month = jun,
	pages = {arXiv:2506.16573},
	shorttitle = {Project {{AMIGA}}},
	title = {Project {{AMIGA}}: {{The Inner Circumgalactic Medium}} of {{Andromeda}} from {{Thick Disk}} to {{Halo}}},
	urldate = {2025-09-14},
	year = {2025}}

@article{lyuDominantRoleCoplanar2025,
	author = {Lyu, Cheqiu and Wang, Enci and Zhang, Hongxin and Peng, Yingjie and Wang, Xin and Li, Haixin and Ma, Chengyu and Yu, Haoran and Chen, Zeyu and Jia, Cheng and Kong, Xu},
	doi = {10.3847/2041-8213/adb4ed},
	issn = {0004-637X},
	journal = {The Astrophysical Journal},
	langid = {english},
	month = mar,
	number = {1},
	pages = {L6},
	title = {Dominant {{Role}} of {{Coplanar Inflows}} in {{Driving Disk Evolution Revealed}} by {{Gas-phase Metallicity Gradients}}},
	urldate = {2025-09-14},
	volume = {981},
	year = {2025}}

@article{marascoSupernovadrivenGasAccretion2012,
	author = {Marasco, A. and Fraternali, F. and Binney, J. J.},
	doi = {10.1111/j.1365-2966.2011.19771.x},
	issn = {0035-8711},
	journal = {Monthly Notices of the Royal Astronomical Society},
	month = jan,
	pages = {1107--1120},
	title = {Supernova-Driven Gas Accretion in the {{Milky Way}}},
	urldate = {2022-08-03},
	volume = {419},
	year = {2012}}



\appendix
\red{\section{Angular Momentum Distribution}}
\label{appendix:AMD}

\newcommand{\jmean}{\langle j\rangle}
\red{
The angular momentum profile in our initial conditions (section~\ref{sec:ICs CGM} and top panel of Fig.~\ref{fig:theta_j}) is derived by assuming that baryons associated with the halo have an exponential angular momentum distribution (AMD). This implies that the mass of gas with specific angular momentum below some value $j$ can be written as
\begin{equation}
    M(<j) = \fb\Mvir\left(1-e^{-\frac{j}{\jmean}}\right)
\end{equation}
where $\fb=0.16$ is the cosmic baryon mass fraction, $\Mvir$ is the halo mass, and $\jmean$ is the mean specific angular momentum. This equation is equivalent to eqns.~(55) -- (56) in \cite{pezzulliAngularMomentumCosmological2017} in the limit $e^{-\xi}\rightarrow0$, where $\xi\sim2-6$ is an adjustable parameter they add to allow for truncation of the AMD at the virial radius.  We use the full form of the AMD with $\xi=3.5$ in our simulation initial conditions, though we note that $\xi$ mainly affects the angular momentum profile near $\rvir$. The exact value of $\xi$ is thus irrelevant for hot inflows in the simulations since they form at smaller radii. }

\red{Assuming the low-$j$ part of an exponential AMD is removed due to accretion and outflows (see arguments in section~6 of \citeauthor{pezzulliAngularMomentumCosmological2017}), and that the remaining hot CGM mass is a fraction $\fcgm$ of the halo baryon budget, we get
\begin{equation}
    M_{\rm hot}(<j) = \fb\Mvir\left(\fcgm-e^{-\frac{j}{\jmean}}\right) ~.
\end{equation}
In our initial conditions, it also holds that (see eqn.~\ref{e:rad solution})
\begin{equation}
    M_{\rm hot}(<r) = \fcgm\fb\Mvir\left(\frac{r}{\rvir}\right)^{3-a}
\end{equation}
where $a=1.5$ is the slope of the power-law density profile, so
\begin{equation}\label{eq:j0}
j(r) = -\jmean\left[\ln \fcgm+\ln \left(1- \left(\frac{r}{\rvir}\right)^{3-a}\right)\right]~.
\end{equation} 
Unless $\fcgm$ is close to unity, the implied angular momentum profile is rather flat at small radii and equal to
\begin{equation}\label{eq:low j}
    j(r\ll \rvir) \approx\jmean\ln\fcgm^{-1} ~.
\end{equation}}
\red{In our simulations we use $\fcgm=0.3$, as suggested by integrating eqn.~(\ref{e:rad solution}) out to $\rvir=250\kpc$ and assuming $\Mvir=10^{12}\,\msun$. We also assume $\jmean=2200\kpc\kmsmath$ consistent with typical spin parameters seen in simulations (see section~\ref{sec:hotinflows}). The hot inflows originate from radii $r<80\kpc$, so eqn.~(\ref{eq:j0}) implies $j$ has a narrow range of $2700-3000\kpc\kmsmath$, as can be seen in the initial conditions plotted in Fig.~\ref{fig:theta_j}. }

\red{In addition, we also ran our simulations with a constant angular momentum profile, with the normalisation set to equal that in the inner part of the cosmologically-motivated AMD used in the main text. The results of this simulation are almost identical to the fiducial simulations, indicating that warp formation via hot CGM inflows depends mainly on the properties of the inner CGM.}

\red{\section{Accretion rate}}\label{appendix:accretion}

\begin{figure}
    \centering
    \includegraphics[width=\linewidth]{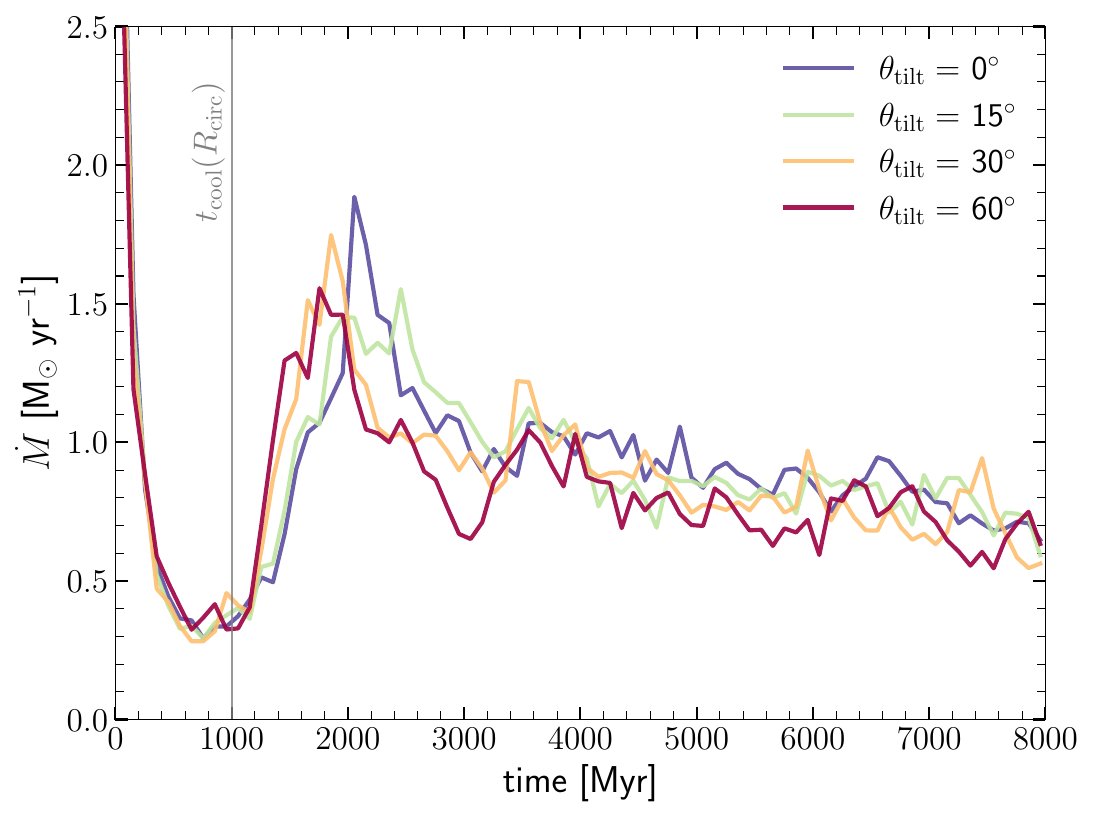}
    \caption{\red{Mass accretion rate versus time in the four simulations. The accretion rate is measured as the change in total mass of stars and cool gas ($<10^5\,{\rm K}$) per unit time. The cooling time at $\Rcirc$ is marked. }}
    \label{fig:accretion_rate}
\end{figure}

\red{
Figure~\ref{fig:accretion_rate} plots the accretion rate in the four simulations, measured as the change in total mass of stars and cool gas ($<10^5\,{\rm K}$) per unit time in the simulated domain. Almost all cooling and star formation occurs at radii $r\lesssim 20\kpc$. Note that the accretion rate drops and increases rapidly during the first $2\,{\rm Gyr}$, a timescale comparable to $\tcool(\Rcirc)\sim\,{\rm Gyr}$ (vertical line, see eqn.~\ref{eq:tcool}). To avoid this initial transient phase in our idealised simulations, we focus the analysis in the main text on snapshots at times $t>2\,{\rm Gyr}$. 
}

\section{Additional accretion tracks}
\label{appendix:tracks}
\begin{figure*}
\centering
    \includegraphics[width=\linewidth]{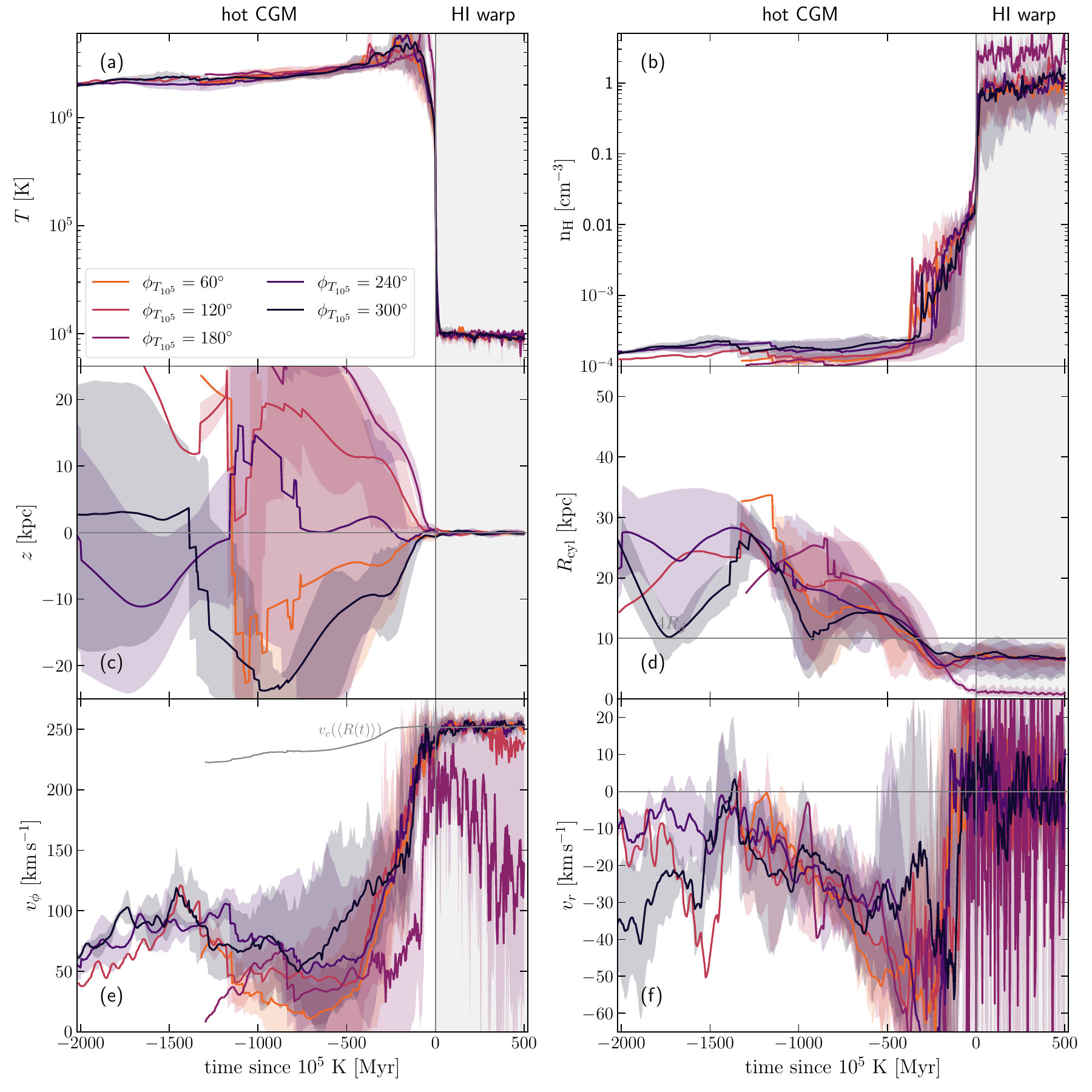} 
  \caption{Similar to Fig.~\ref{fig:track_props} for accreting gas that cools at $R_{\tfive} \leq 10\kpc$.}
  \label{fig:plane_track_props}
\end{figure*}

Figure \ref{fig:plane_track_props} is similar to Fig.~\ref{fig:track_props}, for fluid elements that cool at $R_{\tfive} \leq 10\kpc$ ($\approx 15$\%\ of all accreting gas, see Fig.~\ref{fig:R_hist}). These tracks are discussed in section~\ref{sec:tracks}. 

\bsp	
\label{lastpage}
\end{document}